\documentstyle[12pt,epsf,epsfig]{article}
\def\simle{\hspace*{0.2em}\raisebox{0.5ex}{$<$}
     \hspace{-0.8em}\raisebox{-0.3em}{$\sim$}\hspace*{0.2em}}

\newcommand{\saprox}{${\raisebox{-.6ex}{{$\stackrel{\textstyle
<}{\sim}$}}}$}
\newcommand{\hpinn}{h_{\pi NN}^1}

\begin{document}

\begin{titlepage}
\hfill{\normalsize Caltech MAP-299}\\
\vspace{0.75cm}

\begin{center}{\Large {\bf Nuclear Parity-Violation in Effective Field Theory}}

\vspace{0.75cm}

{\large Shi-Lin Zhu$^{a,b}$, C.M. Maekawa$^{b,c}$, B.R. Holstein$^{d,e}$,\\
M.J. Ramsey-Musolf$^{b,f,g}$, and U. van Kolck$^{h,i}$}

\vspace{0.5cm}
{\it
$^a$ Department of Physics, Peking University\\
Beijing 100871, China\\
$^b$ Kellogg Radiation Laboratory, California Institute of Technology\\
Pasadena, CA 91125, USA\\
$^c$ Departamento de F\'{\i}sica, Funda\c{c}\~ao
Universidade Federal do Rio Grande\\
Campus Carreiros, PO Box 474 96201,
Rio Grande, RS, Brazil\\
$^d$ Department of Physics-LGRT, University of Massachusetts\\
Amherst, MA 01003, USA\\
$^e$ Theory Group, Thomas Jefferson National Accelerator Facility\\
 Newport
News, VA 23606, USA\\
$^f$ Department of Physics, University of Connecticut\\
Storrs, CT 06269, USA\\
$^g$ Institute for Nuclear Theory, University of Washington\\
Seattle, WA 98195, USA\\
$^h$ Department of Physics, University of Arizona\\
Tucson, AZ 85721, USA\\
$^i$ RIKEN-BNL Research Center, Brookhaven National Laboratory\\
Upton, NY 11973, USA}

\end{center}

\vspace{0.25cm}

\begin{abstract}

We reformulate the analysis of nuclear parity-violation (PV)
within the framework of effective field theory (EFT). To ${\cal
O}(Q)$, the PV nucleon-nucleon ($NN$) interaction depends on five
{\em a priori} unknown constants that parameterize the
leading-order, short-range four-nucleon operators. When pions are
included as explicit degrees of freedom, the potential contains
additional medium- and long-range components parameterized by
PV $\pi NN$  coupling. We derive the form of the corresponding one- and
two-pion-exchange potentials.
We apply these considerations
to a set of existing and
prospective PV few-body measurements that may be used to determine the five
independent low-energy constants relevant to the pionless EFT and the additional
constants associated with dynamical pions. We also discuss
the relationship between the conventional meson-exchange framework and the EFT
formulation, and argue that the latter provides a more general and systematic
basis for analyzing nuclear PV.


\end{abstract}

\vfill
\end{titlepage}

\section{Introduction}
\label{sec1}

The cornerstone of traditional nuclear physics is the study of
nuclear forces and, over the years, phenomenological forms of the
nuclear potential have become increasingly sophisticated. In the
nucleon-nucleon ($NN$) system, where data abound, the present
state of the art is indicated, for example, by phenomenological
potentials such as AV18 that are able to fit phase shifts in the
energy region from threshold to 350 MeV in terms of $\sim$ 40
parameters. Progress has been made in the description of
few-nucleon systems \cite{few}, but such a purely phenomenological
approach is less efficient in dealing with the components of the
nuclear interaction that are not constrained by $NN$ data. At the
same time, in recent years a new technique ---effective field
theory (EFT)--- has been used in order to attack this problem by
exploiting the symmetries of QCD \cite{eft}. In this approach the
nuclear interaction is separated into long- and short-distance
components. In its original formulation \cite{wei}, designed for
processes with typical momenta comparable to the pion mass, $Q\sim
m_\pi$, the long-distance component is described fully quantum
mechanically in terms of pion exchange, while the short-distance
piece is described in terms of a number of
phenomenologically-determined contact couplings. The resulting
potential \cite{ray,3Npot} is approaching \cite{fits,chiralfew}
the degree of accuracy of purely-phenomenological potentials. Even
higher precision can be achieved at lower momenta, where all
interactions can be taken as short ranged, as has been
demonstrated not only in the $NN$ system \cite{aleph,crs}, but
also in the three-nucleon system \cite{3stooges,triton}. Precise
($\sim 1\%$) values have been generated also for low-energy,
astro-physically-important cross sections of reactions such as
$n+p\rightarrow d+\gamma$ \cite{npd}. Besides providing reliable
values for such quantities, the use of EFT techniques allows for a
realistic estimation of the size of possible corrections.

Over the past nearly half century there has also developed a
series of measurements attempting to illuminate the parity-{\it
violating} (PV) nuclear interaction.  Indeed the first
experimental paper was that of Tanner in 1957 \cite{tan}, shortly
after the experimental confirmation of parity violation in nuclear
beta decay by Wu et al. \cite{par}. Following the seminal
theoretical work by Michel in 1964 \cite{mic} and that of other
authors in the late 1960's \cite{ope,oth,pir}, the results of such
experiments have generally been analyzed in terms of a
meson-exchange picture, and in 1980 the work of Desplanques,
Donoghue, and Holstein (DDH) developed a comprehensive and general
meson-exchange framework for the analysis of such interactions in
terms of seven parameters representing weak parity-violating
meson-nucleon couplings \cite{ddh}. The DDH interaction has become
the standard setting by which hadronic and nuclear PV processes
are now analyzed theoretically.

It is important to observe, however, that the DDH framework is, at
heart, a {\em model} based on a meson-exchange picture. Provided
one is interested primarily in near-threshold phenomena, use of a
model is unnecessary, and one can instead represent the PV nuclear
interaction in a model-independent effective-field-theoretic
fashion.  The purpose of the present work is to formulate such a
systematic, model-independent treatment of PV $NN$ interactions.
We feel that this is a timely goal, since such PV interactions are
interesting not only in their own right but also as effects
entering atomic PV measurements \cite{ana} as well as experiments
that use parity violation in electromagnetic interactions to probe
nucleon structure \cite{str}.

In our reformulation of nuclear PV, we consider two versions of
EFT, one in which the pions have been \lq\lq integrated out" and
the other including the pion as an explicit degree of freedom. 
In
the pionless theory, the PV nuclear interaction is entirely
short-ranged, and the most general potential depends at leading
order on five independent operators parameterized by a set of five
{\em a priori} unknown low-energy constants (LECs). When applied
to low-energy ($E_{\rm cm}\simle 50$ MeV) two-nucleon PV
observables ---such as the neutron spin asymmetry in the capture
reaction ${\vec n}+p\to d+\gamma$--- it implies that there are
five independent PV amplitudes, which may be determined by an
appropriate set of measurements. 
We therefore recover previous results obtained without effective field theory
by Danilov \cite{dan} and Desplanques and Missimer \cite{dm}.
Making contact with these known results is an important motivation for
us to consider this pionless EFT. 
Going beyond this,
in next (non-vanishing) order in the EFT,
there are several additional independent operators. By contrast,
the DDH meson-exchange framework amounts to a model in which the
short-range physics is codified into six independent operators. On
one hand, the heavy-meson component of the DDH potential is a
redundant representation of the leading-order EFT. On the other,
it does not provide the most complete parameterization of the
short-ranged PV $NN$ force to subleading order, because it is
based on a truncation of the QCD spectrum after inclusion of the
lowest-lying octet of vector mesons. It may, therefore, not be
entirely physically realistic, and we feel that a more general
treatment using EFT is warranted.

When we are interested in observables at higher energies, we need
to account for pion propagation explicitly, simultaneously
removing its effects from the contact interactions. Inclusion of
explicit pions introduces a long-range component into the PV $NN$
interaction, whose strength is set at the lowest order by the PV
$\pi NN$ Yukawa coupling, $\hpinn$. This long-range component,
which is formally of lower-order than shorter-range interactions,
is identical to the long-range, one-pion-exchange (OPE) component
of the  DDH potential.  However, in addition, inclusion of pions
leads to several new effects that do not arise explicitly in the
DDH picture:

\begin{itemize}
\item A
medium-range, two-pion-exchange (TPE) component in the potential
that arises at the same order as the leading short-range
potential and that is also proportional to $\hpinn$. This
medium-range component was considered some time ago in Ref.
\cite{pir} but could not be systematically incorporated into the
treatment of nuclear PV until the advent of EFT. As a result, such
piece has not been previously included in the analysis of PV
observables. We find that the two-pion terms introduce a
qualitatively new aspect into the problem and speculate that their
inclusion may modify the $\hpinn$ sensitivity of various PV
observables.

\item Next-to-next-to-leading-order (NNLO) PV $\pi NN$ operators.
In principle, there exist several such operators that contribute
to the PV NN interaction at the same order as the leading short-range
potential. In practice, however, effects of all but one of the independent NNLO
PV $\pi NN$ operators can be absorbed via a suitable redefinition
of the short-range operator coefficients and $\hpinn$.  The coefficient of
the remaining, independent NNLO operator  -- $k_{\pi NN}^{1a}$ -- must be
determined from experiment.
Additional terms are generated in the potential at ${\cal O}(Q)$
by higher-order corrections to the strong $\pi NN$ coupling (here, $Q$ denotes a
small momentum or pion mass). These
terms have also not been included in previous treatments of the PV
$NN$ interaction. Their coefficients are  fixed  by either
reparameterization invariance or measurements of other
parity-conserving pion-nucleon observables.

\item A new electromagnetic operator. For PV observables involving photons,
the explicit incorporation of pions requires inclusion of a PV
$NN\pi\gamma$ operator that is entirely absent from the DDH
framework and whose strength is characterized by a constant ${\bar C}_\pi$.
\end{itemize}

In short, for the low-energy processes of interest here, the most
general EFT treatment of PV observables depends in practice on eight {\em a
priori} unknown constants when the pion is included as an explicit
degree of freedom: five independent combinations of ${\cal O}(Q)$
short-range constants and those associated with the effects of the
pion: $\hpinn$, ${\bar C}_\pi$, and the NNLO PV $\pi NN$ coupling $k_{\pi NN}^{1a}$.
In order to determine these PV
low-energy constants (LECs), one therefore requires a minimum of
five independent, low-energy observables for the pionless EFT and eight for the
EFT with dynamical pions.  Given the theoretical
ambiguities associated with interpreting many-body nuclear observables (see below),
one would ideally
 attempt to determine the PV LECs from measurements in few-body systems.
Indeed, the state
 of the art in few-body physics allows one perform {\em ab initio}
computations of few-body
 observables \cite{few}, thereby making the few-body system a
theoretically clean environment
 in which to study the effects of hadronic PV.
At present, however, there exist only two measurements
 of few-body PV observables:
$A_L^{pp}$, the longitudinal analyzing power in
polarized proton-proton scattering,
  and $A_L^{p\alpha}$, the longitudinal analyzing power
for ${\vec p}\alpha$ scattering. In what follows,
we outline a prospective program of additional measurements
that would afford a complete determination
of the PV LECs through ${\cal O}(Q)$.


Completion of this low-energy program would serve two additional
purposes. First, it would provide hadron structure theorists with a set of
benchmark numbers that are in principle calculable from first
principles. This situation would be analogous to what one
encounters in chiral perturbation theory for pseudoscalar mesons,
where the experimental determination of the ten LECs appearing in
the ${\cal O}(Q^4)$ Lagrangian presents a challenge to
hadron-structure theory. While many of the ${\cal O}(Q^4)$ LECs
are saturated by $t$-channel exchange of vector mesons, it is not
clear {\em a priori} that the analogous PV $NN$ constants are
similarly saturated (as is assumed implicitly in the DDH model).
Moreover, analysis of the PV $NN$ LECs involves the interplay of
weak and strong interactions in the strangeness-conserving sector.
A similar situation occurs in $\Delta S=1$ hadronic weak
interactions, and the interplay of strong and weak interactions in
this case is both subtle and only partially understood, as
evidenced, {\em e.g.}, by the well-known $\Delta I=1/2$ rule
enigma. The additional information in the $\Delta S=0$ sector
provided by a well-defined set of experimental numbers would
undoubtedly shed light on this fundamental problem.

The information derived from the low-energy few-nucleon PV program
could also provide a starting point for a reanalysis of PV effects
in many-body systems. Until now, one has attempted to use PV
observables obtained from both few- and many-body systems in order
to determine the seven PV meson-nucleon couplings entering the DDH
potential, and several inconsistencies have emerged. The most
blatant is the vastly different value for $\hpinn$ obtained from
the PV $\gamma$-decays of $^{18}$F and from the combination of the
${\vec p}p$ asymmetry and the Cesium anapole moment.
Although combinations of coupling constants can be found
that fit partial sets of experiments (see, {\em e.g.}, Ref.~\cite{wilburn}),
it seems difficult to describe all experiments consistently
with theory (see, {\em e.g.},
Ref.~\cite{ana} and references therein). The origin of
this clash could be due to any one of a number of factors. Using
the operator constraints derived from the few-body program as
input into the nuclear analysis  could help clarify the situation.
It may be, for example, that
the medium-range TPE potential or higher-order operators relevant
only to nuclear PV processes
play a more significant role in
nuclei than implicitly assumed by the DDH framework.
Alternatively, the treatment of the many-body system
---such as the truncation of the model space in shell-model approaches
to the Cesium anapole moment--- may be the culprit. 
(For an example of the relevance of nucleon-nucleon correlations
to parity violation in nuclei, see Ref.~\cite{jerry}.) 
In any
case, approaching the nuclear problem from a more systematic
perspective and drawing upon the results of few-body studies would
undoubtedly represent an advance for the field.

In the remainder of the paper, then, we describe in detail the EFT
reformulation of nuclear PV and the corresponding program of
study. In Section \ref{sec2a}, we briefly review the conventional, DDH
analysis and summarize the key differences with the EFT approach.
In particular, we write down the various components of the PV EFT
potential here, relegating its derivation to subsequent sections.
In Section \ref{fewbody}, we outline the phenomenology of the low-energy
few-body PV program, providing illustrative relationships between
various observables and the five relevant, independent
combinations of short-range LECs. We emphasize that the analysis
presented in Section \ref{fewbody} is intended to demonstrate {\em how} one
would go about carrying out the few-body program rather than to
give precise numerical formulas. Obtaining the latter will require
more sophisticated few-body calculations than we are able to
undertake here. Section \ref{pionless} contains the derivation of the PV
potential in the EFT without explicit pions. We then extend the
framework to include pions explicitly in Section \ref{sec3}.
In Section \ref{models}
we discuss the relationship between the PV LECs and the PV
meson-nucleon couplings entering the DDH framework, and illustrate
how this relationship depends on one's truncation of the QCD
spectrum. Section \ref{sec9} contains some final observations. Various
details pertaining to the calculations contained in the text
appear in the Appendices.


\section{Nuclear PV: Old and New}
\label{sec2a}

\begin{figure}
\begin{center}
\epsfig{file=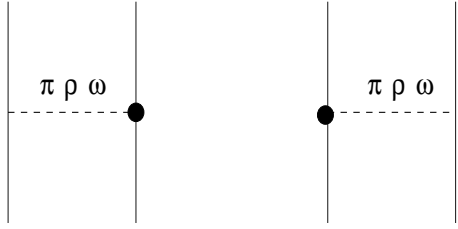,height=3cm,width=6cm}
\caption{Parity-violating $NN$ potential generated by meson exchange.}
\label{fig:DDH}
\end{center}
\end{figure}

The essential idea behind the conventional DDH framework relies
on the fairly successful representation of
the parity-conserving $NN$ interaction
in terms of a single meson-exchange approach.  Of course, this
requires the use of strong-interaction couplings of the lightest vector
($\rho$, $\omega$) and pseudoscalar ($\pi$) mesons($M$),
\begin{eqnarray}
{\cal H}_{\rm st}&=&
-ig_{\pi NN}\bar{N}\gamma_5\tau\cdot\pi N
-g_\rho\bar{N}\left(\gamma_\mu
                    +i{\chi_\rho\over 2m_N}\sigma_{\mu\nu}k^\nu\right)
\tau\cdot\rho^\mu N\nonumber\\
&&-g_\omega\bar{N}\left(\gamma_\mu
                        +i{\chi_\omega\over 2m_N}\sigma_{\mu\nu}k^\nu
\right)\omega^\mu N,
\label{eq:pch}
\end{eqnarray}
whose values are reasonably well determined. The DDH approach to
the parity-violating weak interaction utilizes a similar
meson-exchange picture, but now with one strong and one weak
vertex
---{\it cf.} Fig. \ref{fig:DDH}.

We require then a parity-violating $NNM$ Hamiltonian in
analogy to Eq. (\ref{eq:pch}).  The process is simplified somewhat
by Barton's theorem, which requires that in the CP-conserving
limit, which we employ, exchange of neutral pseudoscalars is
forbidden \cite{bar}.  From general arguments, the effective
Hamiltonian with fewest derivatives must take the form
\begin{eqnarray}
{\cal H}_{\rm wk} &=&{\hpinn\over
\sqrt{2}}\bar{N}(\tau\times\pi)_3N
-\bar{N}\left(h_\rho^0\tau\cdot\rho^\mu +h_\rho^1\rho_3^\mu
+{h_\rho^2\over 2\sqrt{6}}(3\tau_3\rho_3^\mu
-\tau\cdot\rho^\mu)\right)
\gamma_\mu\gamma_5N\nonumber\\
&&-\bar{N}
\left(h_\omega^0\omega^\mu+h_\omega^1\tau_3\omega^\mu\right)\gamma_\mu\gamma_5N
+h_\rho^{'1}\bar{N}(\tau\times\rho^\mu)_3{\sigma_{\mu\nu}k^\nu\over
2m_N}
\gamma_5N.
\end{eqnarray}
We see that there exist, in this model, seven unknown weak
couplings $\hpinn$, $h_\rho^0$, ...  However, quark model
calculations suggest that $h_\rho^{'1}$ is quite small \cite{bhh},
so this term is usually omitted, leaving parity-violating
observables described in terms of just six constants.  DDH
attempted to evaluate such PV couplings using basic quark-model
and symmetry techniques, but they encountered significant
theoretical uncertainties.  For this reason their results were
presented in terms of an allowable range for each, accompanied by
a ``best value'' representing their best guess for each coupling.
These ranges and best values are listed in Table \ref{tab0},
together with predictions generated by subsequent groups
\cite{dz,fcdh}.

\begin{table}
\begin{center}
\begin{tabular}{|c|c|c|c|c|}
\hline
\quad   & DDH\cite{ddh} & DDH\cite{ddh} & DZ\cite{dz} &
FCDH\cite{fcdh}\\
Coupling & Reasonable Range & ``Best" Value &  &  \\ \hline
$h_{\pi NN}^1$ & $0\rightarrow 30$ &+12&+3&+7\\
$h_\rho^0$& $30\rightarrow -81$&$-30$&$-22$&$-10$\\
$h_\rho^1$& $-1\rightarrow 0$& $-0.5$&+1&$-1$\\
$h_\rho^2$& $-20\rightarrow -29$&$-25$&$-18$&$-18$\\
$h_\omega^0$&$15\rightarrow -27$&$-5$&$-10$&$-13$\\
$h_\omega^1$&$-5\rightarrow -2$&$-3$&$-6$&$-6$\\ \hline
\end{tabular}
\caption{Weak $NNM$ couplings as calculated in Refs.
\cite{ddh,dz,fcdh}. All numbers are quoted in units of the ``sum
rule" value $g_\pi =3.8\cdot 10^{-8}$.}
\label{tab0}
\end{center}
\end{table}

Before making contact with experimental results, however,
it is necessary to
convert the $NNM$ couplings generated above into a
parity-violating $NN$ potential.  Inserting the strong and weak
couplings,
defined above,
into the meson-exchange diagrams shown in Fig.\ref{fig:DDH}
and taking the Fourier transform,
one finds the DDH parity-violating $NN$ potential
\begin{eqnarray}
V^{\rm PV}_{DDH}(\vec{r})
&=&i{\hpinn g_{A}m_N\over \sqrt{2}F_\pi}\left({\tau_1\times\tau_2\over
2}\right)_3
(\vec{\sigma}_1+\vec{\sigma}_2)\cdot
\left[{{\vec p}_1-{\vec p}_2\over 2m_N},w_\pi (r)\right]\nonumber\\
&&-g_\rho\left(h_\rho^0\tau_1\cdot\tau_2+h_\rho^1\left({\tau_1+\tau_2\over
2}
\right)_3+h_\rho^2{(3\tau_1^
3\tau_2^3-\tau_1\cdot\tau_2)\over 2\sqrt{6}}\right)
\nonumber\\
&&\quad \left((\vec{\sigma}_1-\vec{\sigma}_2)\cdot \left\{{{\vec
p}_1-{\vec p}_2\over 2m_N},w_\rho(r)\right\}
+i(1+\chi_\rho)\vec{\sigma}_1\times\vec{\sigma}_2\cdot \left[{{\vec
p}_1-{\vec p}_2\over 2m_N},w_\rho
(r)\right]\right)\nonumber\\
&&-g_\omega\left(h_\omega^0+h_\omega^1\left({\tau_1+\tau_2\over
2}\right)_3\right)\nonumber\\
&&\quad\left((\vec{\sigma}_1-\vec{\sigma}_2)\cdot \left\{{{\vec
p}_1-{\vec p}_2\over 2m_N},w_\omega (r)\right\}
+i(1+\chi_\omega)\vec{\sigma}_1\times\vec{\sigma}_2\cdot \left[{{\vec
p}_1-{\vec p}_2\over
2m_N},w_\omega(r)\right]\right)\nonumber\\
&&-\left(g_\omega h_\omega^1-g_\rho h_\rho^1\right)
\left({\tau_1-\tau_2\over 2}\right)_3
(\vec{\sigma}_1+\vec{\sigma}_2)\cdot \left\{{{\vec p}_1-{\vec
p}_2\over 2m_N},w_\rho(r)\right\}
\nonumber\\
&&-g_\rho h_\rho^{'1}i\left({\tau_1\times\tau_2\over 2}\right)_3
(\vec{\sigma}_1+\vec{\sigma}_2)\cdot \left[{{\vec p}_1-{\vec
p}_2\over 2m_N},w_\rho(r)\right],
\end{eqnarray}
where ${\vec p}_i=-i{\vec\nabla}_i$,
${\vec\nabla}_i$ denoting the gradient with respect to
the co-ordinate ${\vec x}_i$ of the $i$-th nucleon,
 $r=|{\vec x}_1 - {\vec x}_2|$ is the separation between the two
nucleons,
\begin{equation}
w_i(r)=\frac{\exp (-m_ir)}{4\pi r}
\label{wi(r)}
\end{equation}
is the usual Yukawa form,
and
the strong $\pi NN$ coupling $g_{\pi NN}$ has
been expressed in terms of the axial-current
coupling $g_A$ using the Goldberger-Treiman relation: $g_{\pi NN}=
g_A m_N/F_\pi$, with $F_\pi =92.4$ MeV being the pion decay constant.

Nearly all experimental results involving nuclear parity violation
have been analyzed using $V^{\rm PV}_{DDH}$ for the past
twenty-some years. At present, however, there appear to exist
discrepancies between the values extracted for the various DDH
couplings from experiment. In particular, the values of $\hpinn$
and $h_\rho^0$ extracted from $\vec{p}p$ scattering and the
$\gamma$ decay of $^{18}$F do not appear to agree with the
corresponding values implied by the anapole moment of $^{133}$Cs
measured in atomic parity violation \cite{atomic}.

These inconsistencies suggest that the DDH framework may not,
after all, adequately characterize the PV $NN$ interaction and provides
motivation
for our reformulation using EFT. The idea of using EFT methods in order to
study
parity-violating
hadronic $\Delta S=0$ interactions is not new \cite{kaplan}.
Recently, a flurry of activity
(see, for example, Refs. \cite{zhu,anapole1,anapole2,bs1,bs2,cj1,cj2})
has centered on PV processes involving a single nucleon,
such as
 $$
e p\rightarrow e' p,\quad
\gamma p\rightarrow\gamma p,\quad
\gamma p\rightarrow n\pi^+,\quad
 etc.$$
There has also been work on the $NN$ system,
with pion exchange treated perturbatively \cite{ss1,ss2}
or non-perturbatively \cite{radcappionful}.
However, a comprehensive analysis has yet to take place, and this
omission is rectified in the study described below, wherein we
generate a systematic framework within which to address PV
$NN$ reactions. We utilize the so-called Weinberg formulation \cite{wei},
wherein the pion, when included explicitly, is treated fully
quantum mechanically while shorter-distance phenomena ---as would be
produced by the exchange of heavier mesons such as $\rho$, $\omega$,
{\it etc.}--- are represented in terms of simple four-nucleon contact terms.
The justification for the non-perturbative treatment
of (parts of) pion exchange has been discussed
in a recent paper \cite{towards}.

Although a fully self-consistent procedure would involve use of
EFT to compute both the PV operators {\it and} few-body
wavefunctions, equally accurate results can be obtained by drawing
upon state-of-the art wave functions obtained from a
phenomenological, strong-interaction $NN$ potential, including PV
effects perturbatively, and using EFT to systematically organize
the relevant PV operators. Such a \lq\lq hybrid" approach has been
followed with some success in other contexts \cite{eft} and we
adopt it here. In so doing, we truncate our analysis of the PV
operators at order $Q/\Lambda_\chi$, where $Q$ is a small momentum
characteristic of the low-energy PV process and $\Lambda_\chi=4\pi
F_\pi\sim$ 1 GeV is the scale of chiral symmetry breaking. Since
realistic wave functions obtained from a phenomenological
potential effectively include strong-interaction contributions to
all orders in $Q/\Lambda_\chi$, the hybrid approach introduces
some inconsistency at higher orders in $Q/\Lambda_\chi$. For the
low-energy processes of interest here ($E_{\rm cm}\simle 50$ MeV),
however, we do not expect the impact of these higher-order
problems to be significant. We would not, however, attempt to
apply our analysis to higher-energy processes ({\em e.g.}, the
TRIUMF 221 MeV ${\vec p}p$ experiment \cite{pp}) where inclusion of
higher-order PV operators would be necessary.

With these considerations in mind, it is useful to compare $V^{\rm
PV}_{\rm DDH}$ with the leading-order PV $NN$ EFT potential. In the
pionless theory, this potential is entirely short ranged (SR) and has
co-ordinate space form
\begin{eqnarray}\label{3}\nonumber
V_{1,\ \rm SR}^{\rm PV} ({\vec r}) &=&
{2\over \Lambda_\chi^3} \left\{
\left[ C_1 + (C_2+C_4) \left({\tau_1 +\tau_2 \over 2}\right)_3
+ C_3 \tau_1 \cdot \tau_2 +{\cal I}_{ab} C_5 \tau_1^a \tau_2^b\right]
\right.\\
\nonumber &&
\left.
\qquad
\left( {\vec \sigma}_1 -{\vec \sigma}_2\right)\cdot \{-i\vec{\nabla},f_m(r)\}
\right.\\
\nonumber &&
\quad
\left. +  \left[ {\tilde C}_1
+ ({\tilde C}_2+{\tilde C}_4)\left({\tau_1 +\tau_2\over 2}\right)_3
+ {\tilde C}_3 \tau_1 \cdot \tau_2 +{\cal I}_{ab} {\tilde C}_5\tau_1^a \tau_2^b
\right]
\right.\\
\nonumber &&
\left.
\qquad
i \left( {\vec \sigma}_1\times {\vec \sigma}_2 \right) \cdot
[-i\vec{\nabla},f_m(r)]
\right.\\
\nonumber &&
\quad
\left. + \left( C_2 -C_4 \right)
\left({\tau_1-\tau_2\over 2}\right)_3
\left( {\vec \sigma}_1 +{\vec \sigma}_2\right) \cdot \{-i\vec{\nabla},f_m(r)\}
\right.\\
&& \quad \left. + C_6 i\epsilon^{ab3} \tau_1^a \tau_2^b
\left( {\vec \sigma}_1 +{\vec \sigma}_2\right)\cdot
[-i\vec{\nabla},f_m(r)] \right\}
\label{eq:vpvsr}
\end{eqnarray}
where the subscript \lq\lq 1" in the potential denotes the chiral
index of the operators\footnote{Roughly speaking, the chiral index
corresponds to the order of a given operator in the
$Q/\Lambda_\chi$ expansion. A precise definition is given in  Sec.
\ref{sec3} below.},
\begin{equation}
{\cal I}=
\left(\begin{array}{lll}1&0&0\\0&1&0\\0&0&-2
 \end{array} \right),\label{eq:chg}
\end{equation}
and $f_m({\vec r})$ is a function that \begin{itemize}
\item [i)]
is strongly peaked, with width $\sim 1/m$ about $r=0$, and
\item [ii)] approaches
$\delta^{(3)}({\vec r})$ in the zero-width  ($m\rightarrow
\infty$) limit.
\end{itemize}
A convenient form, for example, is the Yukawa-like function
\begin{equation}
f_m(r)={m^2\over 4\pi r}\exp{(-mr)}.\label{eq:yuk}
\end{equation}
Here $m$ is a mass chosen to reproduce the appropriate short-range effects
($m\sim \Lambda_\chi$ in the pionful theory,
but $m\sim m_\pi$ in the pionless theory).
Note that, since
the terms containing ${\tilde C}_2$ and ${\tilde C}_4$ are identical,
$V_{\rm SR}^{\rm PV}$
nominally contains ten independent operators. As we show below, however, only
five combinations of these operators are relevant at low-energies.

For the purpose of carrying out actual calculations, one could
just as easily use the momentum-space form of $V^{\rm PV}_{1,\ \rm
SR}$, thereby avoiding the use of $f_m({\vec r})$ altogether.
Nevertheless, the form of Eq. (\ref{eq:vpvsr}) is useful when
comparing with the DDH potential. For example, we observe that the
same set of spin-space and isospin structures appear in both
$V^{\rm PV}_{1,\ \rm SR}$ and the vector-meson exchange terms in
$V^{\rm PV}_{\rm DDH}$, though the relationship between the
various coefficients in $V^{\rm PV}_{1,\ \rm SR}$ is more general.
In particular, the DDH model is tantamount to taking $m\sim
m_\rho,m_\omega$ and assuming
\begin{equation}
{ {\tilde C}_1 \over C_1}= { {\tilde C}_2 \over C_2}= 1+\chi_\omega,
\end{equation}
\begin{equation}
{ {\tilde C}_3 \over C_3}= { {\tilde C}_4 \over C_4}={ {\tilde C}_5
\over C_5}=
 1+\chi_\rho,\label{eq:ddhr}
\end{equation}
assumptions which may not be
physically realistic. In Section \ref{models}, we give illustrative
mechanisms which may lead to a breakdown of these assumptions.

When pions are included explicitly,
one obtains
in addition the same long-range (LR)
component induced by OPE as in $V^{\rm PV}_{\rm DDH}$,
\begin{equation}
V^{\rm PV}_{-1,\ \rm LR} ({\vec r}) =
i{\hpinn g_{A}m_N\over \sqrt{2}F_\pi}\left({\tau_1\times\tau_2\over
2}\right)_3
(\vec{\sigma}_1+\vec{\sigma}_2)\cdot
\left[{{\vec p}_1-{\vec p}_2\over 2m_N},w_\pi (r)\right]
\label{10}
\end{equation}
where $w_\pi(r)$ is given by Eq. (\ref{wi(r)}). Note
that, as we will explain in Sect. \ref{sec3},
$V^{\rm PV}_{-1,\  \rm LR}$ is two orders lower than $V^{\rm
PV}_{1,\  \rm SR}$ ---in contrast to the strong potential where the
short- and long-range components first arise formally at the same
order.
(Even though Eq. (\ref{10}) has the same form as 
a term in Eq. (\ref{eq:vpvsr}),
it has no suppression by powers of $\Lambda_\chi$
or other heavy scales. Therefore, it appears at lower order.)

Furthermore,
two new types of contributions to the potential arise at the same order
as Eq. (\ref{eq:vpvsr}):
(a) a long-range component stemming from higher-order $\pi NN$ operators,
and
(b) a medium-range (MR), two-pion-exchange (TPE) contribution,
$V^{\rm PV}_{1,\  \rm MR}$.
At ${\cal O}(Q)$, the TPE potential is proportional to
$\hpinn$ and involves two terms having the same spin-isospin
structure as the terms in $V^{\rm PV}_{1,\  \rm SR}$ proportional to
${\tilde C_2}$ and ${C_6}$ but having a more complicated
spatial dependence.
In momentum space,
\begin{equation}\label{2}\nonumber
V_{1, \rm MR}^{PV} ({\vec q}) =  -{1\over \Lambda_\chi^3} \left\{
  {\tilde C}^{2\pi}_2 (q)  {\tau_1^z +\tau_2^z\over 2}
i \left( {\vec \sigma}_1 \times {\vec \sigma}_2 \right) \cdot
{\vec q}
\ + C^{2\pi}_6 (q) i\epsilon^{ab3} \tau_1^a
\tau_2^b \left( {\vec \sigma}_1 +{\vec \sigma}_2\right)\cdot {\vec
q} \right\},
\end{equation}
where the functions ${\tilde C}^{2\pi}_2 (q)$
and $C^{2\pi}_6 (q)$, defined below in Eq. (\ref{good}),
are determined by the leading-order $\pi NN$ couplings.
Again, it is more convenient to compute matrix
elements of $V^{\rm PV}_{1,\  \rm MR}$ using the momentum-space form,
and we defer a detailed discussion of the latter until Section \ref{sec3}
below. We emphasize, however, the presence of $V^{\rm PV}_{1,\  \rm MR}$
introduces a qualitatively new element into the treatment of
nuclear PV with pions not present in the DDH framework.

The NNLO long-range contribution to the potential generated by the
new PV $\pi NN$ operator is
\begin{equation}
\label{eq:vpvlrnnlo}
V_{1,\  \rm LR}^{\rm PV}  =
 - i{k_{\pi NN}^{1b} g_{A}\over 2\Lambda_\chi F_\pi^2}
 \left({\tau_1\times\tau_2\over 2}\right)_3\Biggl\{
 \epsilon_{abc}\sigma_1^c\sigma_2^e\Bigl\{\nabla_1^a,[\nabla_r^b
 \nabla_r^e,w_\pi(r)]\Bigr\}
 + (1\leftrightarrow 2)\Biggr\}+\cdots \ \ \ ,
 \end{equation}
 where $\nabla_r$ is the gradient with respect to the relative co-ordinate
${\vec x}_1-{\vec x}_2$ and where the $\cdots$ denote long-range, NNLO
contributions proportional to
 $\hpinn$ that are generated by NNLO effects at the strong $\pi NN$ vertex
(see Appendix B).

As we discuss in Section 3, a complete program of low-energy PV
measurements includes photo-reactions. In the DDH framework, PV
electromagnetic (EM) matrix elements receive two classes of
contributions: (a) those involving the standard, parity-conserving
EM operators in combination with parity-mixing in the nuclear
states, and (b) PV two-body EM operators derived from the
amplitudes of Fig. \ref{fig21}. Explicit expressions for these
operators in the DDH framework can be found in Ref. \cite{ana}. In
the case of EFT, the two-body PV EM operators associated with
heavy-meson exchange in DDH are replaced by operators obtained by
gauging the derivatives in $V^{\rm PV}_{1,\  \rm SR}$  as well as
by explicit photon insertions on external legs. The two-body
operators associated with $V^{\rm PV}_{-1,\  \rm LR}$ are
identical to those appearing in DDH, while the PV currents
associated with $V^{\rm PV}_{1,\  \rm MR}$ and $V^{\rm PV}_{1,\
\rm LR}$ are obtained by gauging the derivatives appearing in the
potential and by inserting the photon on all charged-particle
lines in the corresponding Feynman diagrams\footnote{The
derivation of the medium-range two-body operators involves an
enormously detailed computation, which we defer to a later
publication.}. The foregoing two-body currents introduce no new
unknown constants beyond those already appearing in the potential.
However, an additional, independent pion-exchange two-body
operator also appears at the same order as the short-range PV
two-body currents:
\begin{equation}\label{eq:mecnew}
{\vec J}({\vec x}_1, {\vec x}_2, {\vec q}) =
{\sqrt{2} g_A {\bar C}_\pi m_\pi^2\over
\Lambda_\chi^2 F_\pi} {\rm e}^{-i{\vec q}\cdot{\vec x}_1}\
\tau_1^+\tau_2^-\ {\vec\sigma}_1\times
{\vec q} \ {\vec\sigma}_2\cdot{\hat r} \ h_\pi(r) \
+ \ (1\leftrightarrow 2),
\end{equation}
where
\begin{equation}
h_\pi(r) = {\exp{(-m_\pi r)}\over m_\pi r}
\left(1+{1\over m_\pi r}\right),
\end{equation}
and ${\bar C}_\pi$ is an additional LEC parameterizing the leading PV
$NN\pi\gamma$ interaction.
Any photoreaction  sensitive to the short-range PV potential will also
depend on ${\bar C}_\pi$ when
pions are included explicitly.

Through ${\cal O}(Q)$, then, the phenomenology of nuclear PV
depends on five unknown constants in the pionless theory and eight when
pions are included explicitly ($\hpinn$, $k_{\pi NN}^{1a}$,  and ${\bar C}_\pi$
in addition to the contact interactions). As we
discuss below, an initial low-energy program will afford a determination of
the five constants in the pionless theory. Additional low-energy measurements in
few-body systems would provide a test of the self-consistency of the EFT at this
order. Any discrepancies could indicate the need to including pions as explicit
degrees of freedom, thereby necessitating the completion of additional
measurements in order to determine the pion contributions to ${\cal O}(Q)$. As
we discuss below, there exists a sufficient number of prospective measurements
that could be used for this purpose. Given the challenging nature of the
experiments, a sensible strategy would be to first test for the self-consistency
of the pionless EFT with a smaller set of measurements and then to complete the
additional measurements needed for EFT with pions if necessary.



\section{Parity Violation in Few-Body Systems}
\label{fewbody}

There exist
numerous low-energy experiments that have attempted to explore
hadronic parity violation.  Some, like the photon asymmetry in the decay
of a
polarized isomeric state of ${}^{180}$Hf,
\begin{equation}
A_\gamma=-(1.66\pm 0.18)\times 10^{-2}\cite{hf},
\end{equation}
or the asymmetry in longitudinally-polarized neutron
scattering on ${}^{139}$La,
\begin{equation}
A_z=(9.55\pm 0.35)\times 10^{-2}\cite{la},
\end{equation}
involve F-P shell nuclei wherein the effects of hadronic parity
violation are large and clearly observed, but where the difficulty
of performing a reliable wave function calculation precludes a
definitive interpretation.  For this reason, it is traditional to
restrict one's attention to S-D shell or lighter nuclei.  Here
too, there exist a number of experiments, such as the asymmetry in
the decay of the polarized first excited state of ${}^{19}$F,
\begin{eqnarray}
A_\gamma({1\over 2}^-,110\,\,{\rm keV})&=&-(8.5\pm 2.6)\times
10^{-5}\cite{f1}
\nonumber\\
        &=&-(6.8\pm 1.8)\times 10^{-5}\cite{f2},
\end{eqnarray}
wherein a clear parity-violating signal is observed, or those such as
the circular polarization in the decay of excited levels of ${}^{21}$Ne,
\begin{eqnarray}
P_\gamma({1\over 2}^-,2.789\,\,{\rm MeV})&=&(24\pm 24)\times
10^{-4}\cite{n1}
\nonumber\\
        &=&(3\pm 16)\times 10^{-4}\cite{n2},
\end{eqnarray}
or of ${}^{18}$F,
\begin{eqnarray}
P_\gamma(0^-, 1.081\,\,{\rm MeV})&=&(-7\pm 20)\times 10^{-4}\cite{f3}
\nonumber\\
        &=&(3\pm 6)\times 10^{-4}\cite{f4}\nonumber\\
        &=&(-10\pm 18)\times 10^{-4}\cite{f5}\nonumber\\
        &=&(2\pm 6)\times 10^{-4}\cite{f6},
\end{eqnarray}
wherein a nonzero signal has not been seen, but where the precision of
the
experiment is high enough that a significant limit can be placed on
the underlying parity-violating mechanism.

The reason that a
$10^{-4}$ experiment can reveal information about an effect which is on
the surface at the level
$$G_Fm_N^2\times (p_F/ m_N)\sim 10^{-6},$$
where $p_F\sim 270$ MeV is the Fermi momentum, is that the nucleus
can act as an PV-amplifier.  This occurs when there exist a pair
of close-by levels having the same spin but opposite parity,
$|J^\pm\rangle$.  In this case the parity mixing expected from
lowest-order perturbation theory, $|\psi_\pm\rangle\simeq
|J^\pm\rangle\pm \epsilon|J^\mp\rangle$, can become anomalously
large due to the smallness of the energy difference
$E_{J^+}-E_{J^-}$ in the mixing parameter
\begin{equation}
\epsilon\simeq{\langle J^-|H_{weak}|J^+\rangle\over
E_{J^+}-E_{J^-}}.
\end{equation}
Indeed, when compared with a typical level splitting of $\sim 1$
MeV, the energy differences exploited in ${}^{19}$F ($\Delta
E=110$ keV), ${}^{21}$Ne ($\Delta E=5.7$ keV), and ${}^{18}$F
($\Delta E=39$ keV) lead to expected enhancements at the level of
10, 100, and 25 respectively.  However, when interpreted in terms
of the best existing nuclear shell-model wave functions, there
exists a serious discrepancy between the values of the $\Delta
I$=1 pion coupling required in order to understand the ${}^{19}$F
or ${}^{21}$Ne experiments and the upper limit allowed by the
${}^{18}$F result.

Such matters have been extensively reviewed by previous authors
\cite{hol89,ah,hh}, and we do not intend to revisit these issues
here.  Instead we suggest that {\it at the present time} any
detailed attempt to understand the parity-violating $NN$
interaction must focus on experiments involving only the very
lightest
---$NN$, $Nd$, $N\alpha$--- systems,
wherein our ability to calculate the effects of a given
theoretical picture are under much better control. As we
demonstrate below, there exist a sufficient number of such
experiments, either in progress or planned, in order to accomplish
this task for either the pionless EFT or the EFT with dynamical pions.
Once a reliable set of low-energy constants are in hand, as obtained from
such very-light systems, theoretical work can proceed on at least
two additional fronts:
\begin{itemize}
\item[i)] experimental results from the heavier nuclear systems
---involving P, S-D, and F-P shells and higher levels--- can be
revisited and any discrepancies hopefully resolved with the confidence
that the
weak low-energy constants are correct, and
\item[ii)] one can attempt to evaluate the size of the
phenomenological weak constants starting from the fundamental
quark-quark weak interaction in the Standard Model.
\end{itemize}

This scheme mirrors the approach that has proven highly successful
in chiral perturbation theory (ChPT) \cite{wphysica}, wherein
phenomenological constants are extracted purely from experimental
results, using no theoretical prejudices other than the basic
(broken) chiral symmetry of QCD. In the meson sector \cite{gl},
the phenomenologically-determined counter-terms $L_1,L_2,...,
L_{10}$ have already become the focus of various theoretical
programs attempting to predict their size from fundamental theory.
Note that our approach to nuclear  parity violation is
similar in spirit to the one advocated in a prescient 1978 paper
by Desplanques and Missimer \cite{dm} that builds on ideas put
forward by Danilov \cite{dan}. In subsequent work, this approach
was superseded by the use of the DDH potential. In our study,
then, we are in a sense recasting the ideas of Refs. \cite{dm,dan}
in the modern and theoretically systematic framework of EFT.
\subsection{Amplitudes}

We now consider the first part of the program ---elucidation of
the basic weak couplings. We argue that, provided one is working
in a region of sufficiently low-energy, the parity-violating $NN$
interaction can be described in terms of just {\it five} real
numbers, which characterize S-P wave mixing in the spin singlet
and triplet
channels. 
The arguments in this section borrow heavily from the
work of Danilov \cite{dan} and Desplanques and Missimer \cite{dm}.
The following sections will show how to interpret this
phenomenology within EFT.

For simplicity we begin with a parity-conserving system of two
nucleons. Then the $NN$ scattering matrix can be expressed purely
in terms of S-wave scattering at low energies and has the
phenomenological form \cite{dan}
\begin{equation}
{\cal
M}(\vec{k}_f,\vec{k}_i)=
\langle\vec{k}_f|\hat{T}|\vec{k}_i\rangle=m_t(k)P_1+m_s(k)P_0,\label{eq:tmx}
\end{equation}
where
$$P_1={1\over 4}(3+\vec{\sigma}_1\cdot\vec{\sigma}_2),\qquad
P_0={1\over 4}(1-\vec{\sigma}_1\cdot\vec{\sigma}_2)$$ are
spin-triplet, -singlet projection operators. All other partial
waves can be neglected. We can determine the form of the functions
$m_i(k)$ by using the stricture of unitarity,
\begin{equation}
2{\rm Im}\hat{T}=\hat{T}^\dagger \hat{T}.
\end{equation}
In the S-wave sector this becomes
\begin{equation}
{\rm Im}m_i(k)=k|m_i(k)|^2,
\end{equation}
whose solution is of the familiar form
\begin{equation}
m_i(k)={1\over k}e^{i\delta_i(k)}\sin\delta_i(k).
\end{equation}
Since at zero energy
\begin{equation}
\lim_{k\rightarrow 0}m_i(k)=-a_i
\end{equation}
where $a_i$ is the scattering length, it is clear that unitarity can
 be enforced by the simple modification
\begin{equation}
m_i(k)={-a_i\over 1+ika_i},
\label{ert}
\end{equation}
which is the lowest-order effective-range result. The scattering
cross section is found via
\begin{equation}
{d\sigma\over d\Omega}={\rm Tr}{\cal M}^\dagger{\cal
M}={a_i^2\over 1+k^2a_i^2},
\end{equation}
so that at the lowest energy we have the familiar form
\begin{equation}
\lim_{k\rightarrow 0}{d\sigma_{s,t}\over d\Omega}=|a_{s,t}|^2.
\end{equation}
The associated scattering wave functions are given by
\begin{eqnarray}
\psi^{(+)}_{\vec{k}}(\vec{r})&=&\left[e^{i\vec{k}\cdot\vec{r}}-{m_N\over
4\pi}
\int d^3r'{e^{ik|\vec{r}-\vec{r}\,'|}\over |\vec{r}-\vec{r}\,'|}V(\vec{r}\,')
\psi^{(+)}_{\vec{k}}(\vec{r}\,')\right]\chi\nonumber\\
&\stackrel{r\rightarrow\infty}{\longrightarrow}&
\left[e^{i\vec{k}\cdot\vec{r}}+{\cal M}(-i\vec{\nabla},
\vec{k}){e^{ikr}\over r}\right]\chi, \label{eq:wfc}
\end{eqnarray}
where $\chi$ is the spin wave function. In the simple Born
approximation, then, we can represent the wave function in terms
of an effective local potential
\begin{equation}
V_{eff}^{(0)PC}(\vec{r})={4\pi\over
m_N}(a_tP_1+a_sP_0)\delta^3(\vec{r}), \label{eq:nrv}
\end{equation}
as can be confirmed by substitution into Eq. (\ref{eq:wfc}).

Parity mixing can be introduced into this simple representation,
as done by Danilov \cite{dan}, via generalization of the
scattering amplitude to include parity-violating structures.  Up
to laboratory momenta of 140 MeV or so, we can omit all but S- and
P-wave mixing, in which case there exist five independent such
amplitudes:
\begin{itemize}
\item[i)] $d_t(k)$, representing ${}^{3}S_1-{}^{1}P_1$ mixing;
\item[ii)] $d_s^{0,1,2}(k)$, representing ${}^{1}S_0-{}^{3}P_0$ mixing
with $\Delta I=0,1,2$ respectively; and
\item[iii)] $c_t(k)$, representing ${}^{3}S_1-{}^{3}P_1$ mixing.
\end{itemize}
and, after a little thought, it becomes clear that the low-energy
scattering matrix in the presence of parity violation can be
generalized to
\begin{eqnarray}
{\cal M}(\vec{k}_f,\vec{k}_i)
&=&m_t(k)P_1+m_s(k)P_0\nonumber\\
&&+\Bigl[\left(d_s^0(k)Q_1+d_s^1(k)Q_{1+}+d_s^2(k)Q_2\right)
   \left(\vec{k}_i\cdot(\vec{\sigma}_1-\vec{\sigma}_2)P_1
    +P_1\vec{k}_f\cdot(\vec{\sigma}_1-\vec{\sigma}_2)\right) \nonumber\\
&&\quad +d_t(k)\left(\vec{k}_i\cdot(\vec{\sigma}_1-\vec{\sigma}_2)P_0
     +P_0\vec{k}_f\cdot(\vec{\sigma}_1-\vec{\sigma}_2)\right)\Bigr]\nonumber\\
&&+c_t(k)Q_{1-}(\vec{\sigma}_1+\vec{\sigma}_2)\cdot
               \left(\vec{k}_iP_1+P_1\vec{k}_f\right),
    \label{eq:tpx}
\end{eqnarray}
where we have introduced the isovector and isotensor operators
\begin{equation}Q_{1-}={1\over 2}(\tau_1-\tau_2)_z,\qquad
Q_{1+}={1\over 2}(\tau_1+\tau_2)_z,\qquad
Q_2={1\over2\sqrt{6}}(3\tau_{1z}\tau_{2z}-\vec{\tau}_1\cdot\vec{\tau}_2),
\nonumber
\end{equation}
and isospin projection operators
\begin{equation}
Q_0={1\over 4}(1-{\vec\tau}_1\cdot{\vec\tau_2}), \qquad
Q_1={1\over 4}(3+{\vec\tau}_1\cdot{\vec\tau_2}).
\nonumber
\end{equation}
Each of the new pieces is indeed odd under spatial inversion
[$\vec{\sigma}_i\rightarrow\vec{\sigma}_i$ and
$\vec{k}_f,\vec{k}_i\rightarrow-\vec{k}_f,-\vec{k}_i$] and even
under time reversal [$\vec{\sigma}_i\rightarrow-\vec{\sigma}_i$
and
$\vec{k}_i\cdot(\vec{\sigma}_1-\vec{\sigma}_2)P_j\leftrightarrow
P_j \vec{k}_f\cdot(\vec{\sigma}_1-\vec{\sigma}_2)$].

Now consider what constraints can be placed on the forms
$d_i(k),c_i(k)$.
The requirement of unitarity reads
\begin{equation}
{\rm Im}\ d_i(k)=k[m_i^*(k)d_i(k)+d_i^*(k)m_p(k)]\label{eq:uni},
\end{equation}
where $m_i(k),m_p(k)$ are the scattering amplitudes in the S-, P-wave
channels connected by $d_i(k)$.  Eq. (\ref{eq:uni}) is satisfied by the
solution
\begin{equation}
d_i(k)=|d_i(k)|\exp i[\delta_i(k)+\delta_p(k)],
\end{equation}
{\it i.e.}, the phase of the transition amplitude is simply the sum of
the strong interaction phase shifts in the incoming and outgoing
channels.

Danilov \cite{dan} suggested that, on account of the short-range
of the weak interaction, the energy dependence of the weak
couplings $d_i(k)$ should be primarily determined, up to say 50
MeV or so solely by the strong interaction dynamics. Since at very
low energy the P-wave scattering can be neglected, he suggested
the use of the forms
\begin{equation}
c_t(k)=\rho_tm_t(k),\quad d_t(k)=\lambda_tm_t(k),\quad
d_s^i(k)=\lambda_s^im_s(k),\label{eq:eff}
\end{equation}
which provide the parity-mixing amplitudes in terms of the five
phenomenological constants: $\rho_t,\lambda_t,\lambda_s^i$.

We can understand the motivation behind Danilov's assertion by
writing down the simplest phenomenological form for a weak
low-energy parity-violating $NN$ potential. To do so, one may
start with the momentum-space form of $V^{\rm PV}_{1,\rm SR}$
given in Eq. (\ref{eq:vpvsr}):

\begin{eqnarray}\nonumber
V^{\rm PV}_{1,\  \rm SR} ({\vec q}, {\vec p})
&=&
-{1\over \Lambda_\chi^3}\Biggl\{
  -\Biggl[  (C_1 + C_3)Q_1 +(C_1-3C_3)Q_0
+(C_2+C_4)Q_{1+}
-\sqrt{\frac{8}{3}}C_5 Q_2\Biggr]\\\nonumber & &
\qquad \qquad \left( {\vec \sigma}_1 -{\vec \sigma}_2\right)\cdot {\vec p}
\nonumber \\ \nonumber
&&+ \Biggl[  ({\tilde C_1} + {\tilde C_3})Q_1 +({\tilde C_1}-3{\tilde C_3})Q_0
+({\tilde C_2}+{\tilde C_4})Q_{1+}
-\sqrt{\frac{8}{3}}
{\tilde C}_5 Q_2\Biggr]  \\\nonumber & &
\quad i \left( {\vec \sigma}_1
\times {\vec
\sigma}_2 \right) \cdot {\vec q}\\
&&+ \left[  C_2 -C_4 \right] Q_{1-} \left( {\vec \sigma}_1 +{\vec
\sigma}_2\right)
\cdot {\vec p}
+ C_6 i \left({\vec \tau_1}\times{\vec\tau_2}\right)_z
\left( {\vec \sigma}_1 +{\vec \sigma}_2\right)\cdot {\vec q} \Biggr\},
\label{eq:sht}
\end{eqnarray}
where $p = [(p_1-p_2) +(p_1^\prime -p_2^\prime)]/2$ and
$q=[(p_1-p_2)- (p_1^\prime -p_2^\prime)]/2$.

The change in the wave function generated by $V_{1,\ \rm SR}^{\rm
PV}(\vec{r})$ is understood to involve the full strong-interaction
Green's function and wave functions
\begin{equation}
\delta\psi^{(+)}(\vec{r}) =\int
d^3r'G_k(\vec{r},\vec{r}')V_{1,\ \rm SR}^{\rm PV}(\vec{r}')
\psi^{(+)}(\vec{r}')\label{eq:gnf}
\end{equation}
and the connection between the weak PV potential
$V_{1,\ \rm SR}^{\rm PV}(\vec{r})$ and the scattering matrix Eq.
(\ref{eq:tpx}) can be found via
\begin{equation}
d_i(k)\sim -{m_N\over 4\pi}( \langle\psi^{P(-)}_k|V_{1,\ \rm
SR}^{\rm PV}|\psi^{S(+)}_k\rangle +\langle\psi^{S(-)}_k|V_{1,\ \rm
SR}^{\rm PV}|\psi^{P(+)}_k\rangle). \label{eq:wf}
\end{equation}
Now, if we are at very low energy, we may
use the plane-wave approximation for the P wave,
\begin{equation}
\psi_k^{P-}(r)\simeq j_1(kr),
\end{equation}
and we can approximate the S wave by its asymptotic form
\begin{equation}
\psi_k^{S+}(r)\simeq {1\over kr}e^{i\delta_i(k)}\sin(kr+\delta_i(k))
\stackrel{k\rightarrow 0}{\longrightarrow}{1\over kr}e^{i\delta_i(k)}
\sin\delta_i(k).
\end{equation}
where we have used the experimental fact that $|\delta_i|\approx
|ka_i|>>k/m$ where $1/m$ is maximum range set by the integration.
Then, we can imagine calculating a generic parity-violating
amplitude $d_i(k)$ via Eq. (\ref{eq:wf}):
\begin{eqnarray}
d_i(k)&\sim& {4\pi\over k} C_i \int_0^\infty
drr^2j_1(kr)[{\partial\over \partial r},f_m(r)]{1\over kr}
e^{i\delta_i(k)}\sin\delta_i(k)\nonumber\\
&\equiv& \lambda_i {1\over
k}e^{i\delta_i}\sin\delta_i=\lambda_im_i(k)
\end{eqnarray}
with \lq\lq $C_i$" symbolically indicating the appropriate
combination of PV constants appearing in $V_{1,\ \rm SR}^{\rm PV}$
and
\begin{equation}
\lambda_i\sim {4\pi\over k}C_i\int_0^\infty
drrj_i(kr){df_m(r)\over dr} \approx {4\pi\over 3}C_i\int_0^\infty
drr^2{df_m(r)\over dr},\label{eq:sim}
\end{equation}
which is the basic form advocated by Danilov\footnote{Note that
this argument is not quite correct quantitatively.  Indeed, since
$1/m$ is much smaller than the range of the $NN$ interaction (in
the pionful theory), we should really use interior forms of the
wave functions. However, when this is done, the same qualitative
result is found, but the simple relationship in Eq. (\ref{eq:sim})
is somewhat modified. }. An analogous relationship holds for
$\rho_t$.

At low energy then it seems prudent to explicitly include the
appropriate S-wave scattering length in expressing the effective
weak potential, and we can define
\begin{equation}
\lim_{k\rightarrow 0}m_{s,t}(k)=-a_{s,t},\quad \lim_{k\rightarrow
0}c_t(k),
d_s(k),d_t(k)=-\rho_ta_t,-\lambda_s^ia_s,-\lambda_ta_t
\end{equation}
As emphasized above, the real numbers
$\rho_t,\lambda_s^i,\lambda_t$---which can in turn be related to
the effective parameters $C_i$---{\it completely} characterize the
low-energy parity-violating interaction and can be determined
experimentally, as we shall discuss below. Alternatively, we use
an isospin decomposition
\begin{eqnarray}
\lambda_s^{pp}&=&\lambda_s^0+\lambda_s^1+{1\over\sqrt{6}}\lambda_s^2,
                 \nonumber\\
\lambda_s^{np}&=&\lambda_s^0-{2\over \sqrt{6}}\lambda_s^2,
                 \nonumber\\
\lambda_s^{nn}&=&\lambda_s^0-\lambda_s^1+{1\over \sqrt{6}}\lambda_s^2.
\end{eqnarray}
to write things in terms of the appropriate $NN$ quantities. Now,
the explicit connection between the S-matrix elements
$\lambda_i,\rho_t$ and the weak interaction parameters $C_i,
\tilde{C}_i$ in our effective Lagrangian must be done carefully
using the Eq. (\ref{eq:wf}) and the best possible $NN$ wave
functions.  This work is underway, but is not yet
completed\cite{rocco}. In the meantime,
 we may obtain an indication of the connection
by using the following simple arguments:

When we restrict ourselves to a model-space containing only the
low-energy S,P amplitudes noted above, then several of the
operators in Eqs. (\ref{3},\ref{eq:sht}) become redundant. For
example, the $d_t$ amplitude involves a $T=0\to T=0$ transition,
so only the terms proportional to $Q_0$ contribute. In this case,
the spin-space operators $( {\vec \sigma}_1 -{\vec \sigma}_2)\cdot
{\vec p}$ and $i( {\vec \sigma}_1 \times {\vec\sigma}_2 ) \cdot
{\vec q}$ yield identical matrix elements up to an overall
constant of proportionality. This feature can be seen by
considering the co-ordinate space potential, which contains the
function $f_m(r)$ times derivatives acting on the initial and
final states. In the short range limit and in the absence of the
$NN$ repulsive core, both the P-wave and first derivative of the
S-wave vanish at the origin, whereas the product of the S-wave and
first derivative of the P-wave are non-zero. Thus, only the
components of $\vec q$ and $\vec p$ that yield derivatives of the
P-wave at the origin contribute, leading to identical matrix
elements of these two operators (up to an overall phase). Of
course, corrections to this statement occur when $f_m(r)$ is
smeared out over some short range $\sim 1/m$. Since $1/m<<
1/\mbox{typical momentum}\sim a$ where $a$ is the scattering
length, at low energy such corrections are higher-order in our
power counting, going as $K^2/m^2$, where
$K\sim\sqrt{M(E+\bar{V})}$ with $\bar{V}\sim 50$ MeV representing
some average depth of the $NN$ potential characterizing the
interior region.  Similarly, the operators $({\vec \sigma}_1
-{\vec \sigma}_2)_z$ and $( {\vec \sigma}_1 \times
{\vec\sigma}_2)_z$ each transform a spin triplet into a spin
singlet state, and vice versa. Hence, one may absorb the effect of
the term proportional to $({\tilde C_1}+{\tilde C_3})$ into the
corresponding term proportional to $(C_1+C_3)$ by a suitable
redefinition of the constants. Related arguments allow one to
absorb the remaining terms proportional to the ${\tilde C}_i$ --
as well as the term containing $C_6$ -- into the terms involving
$(C_1-3C_3)P_1$, $(C_2+C_4) Q_{1+}$, $(C_2-C_4)Q_{1-}$, and $C_5
Q_2$ for a net total of {\it five} independent operators, which in
turn generate the five independent low-energy PV amplitudes
$\lambda_s^0,\lambda_s^1,\lambda_s^2,\lambda_t,\rho_t$. In the
zero-range limit, then, we have
\begin{eqnarray}
\lambda_t & \propto & (C_1-3C_3) -({\tilde C_1}-3{\tilde C_3}) \nonumber \\
\lambda_s^0 & \propto & (C_1+C_3) +({\tilde C_1}+{\tilde C_3})\nonumber \\
\label{eq:lincomb}
\lambda_s^1 & \propto & (C_2+C_4) +({\tilde C_2}+{\tilde C_4}) \\
\lambda_s^2 & \propto & -\sqrt{8/3}(C_5 +{\tilde C_5})\nonumber\\
\rho_t & \propto & \frac{1}{2}(C_2-C_4) -C_6 \ \ \ .\nonumber
\end{eqnarray}
However, going away from strict threshold values and the use of
more realistic wave functions will modify these expectations
somewhat, as illustrated by a simple, didactic discussion in
Appendix E. We emphasize, however, that
what is needed at the present time is a purely empirical
evaluation in terms of five independent and precise experiments,
and that is what we shall discuss next.


\subsection{Relation to observables}

The next step of the program ---contact between this effective
parity-violating interaction and experimental observables--- was
initiated by Desplanques and Missimer \cite{dm}.  Before quoting
these results, we sketch the manner by which such a confrontation
is performed. In doing so, we emphasize that the following
analysis does {\it not} rely on definitive computations employing
state-of-the art few-body wave functions---carrying out such
calculations goes beyond the scope of the present study. Indeed,
obtaining precise values for the $\lambda_i$ and $\rho_t$ will
require a concerted effort on the part of both experiment and
few-body nuclear theory. What we provide below is intended,
rather, to serve as a qualitative roadmap for such a program,
setting the context for what we hope will be future experimental
and theoretical work.

For simplicity,  we begin with an illustrative example of $nn$
scattering, for which the Pauli principle demands that the initial
state at low energy must be purely ${}^1S_0$.  One can imagine
longitudinally polarizing a neutron of momentum $\vec{p}$ and
measuring the total scattering cross section from an unpolarized
target. Since $\vec{\sigma}\cdot \vec{p}$ is odd under spatial
inversion, the cross section can depend on helicity only if parity
is violated, and via trace techniques the helicity-correlated
cross section can easily be found.  Using
\begin{equation}
{\cal M}(\vec{k}_f,\vec{k}_i)=m_s(k)P_0+d_s^{nn}
[\vec{k}_i\cdot (\vec{\sigma}_1-\vec{\sigma}_2)P_0+P_0\vec{k}_f\cdot(\vec{\sigma
}_1-\vec{\sigma}_2)]
\end{equation}
we determine
\begin{eqnarray}
\sigma_\pm&=&\int d\Omega_f{\rm Tr}{\cal M}(\vec{k}_f,\vec{k}_i)
{1\over 2}(1\pm\vec{\sigma}_1\cdot\hat{k}_i){\cal
M}^\dagger(\vec{k}_f,
\vec{k}_i)\nonumber\\&=&4\pi |m_s(k)|^2{\rm Tr} P_0 + 8\pi{\rm Re}\ m_s^{\ast}(k
) d_s^{nn}(k){\rm Tr}\ P_0
({\vec\sigma}_1-{\vec\sigma}_2)\cdot{\hat k}_i(1\pm{\vec\sigma}_1\cdot{\hat k}_i
)+\cdots\nonumber\\
&=&4\pi |m_s(k)|^2\pm 16\pi {\rm Re}\ m_s^{\ast}(k) d_s^{nn}(k)+\cdots
\end{eqnarray}

Defining the asymmetry via the sum and difference of such helicity
cross sections and neglecting the tiny P-wave scattering, we have
then
\begin{equation}
A_L={\sigma_+-\sigma_-\over \sigma_++\sigma_-} ={4k\, {\rm
Re}[m_s^*(k)d_s^{nn}(k)] \over |m_s(k)|^2}\simeq
4k\lambda_s^{nn}.\label{eq:fre}
\end{equation}
Thus the helicity-correlated $nn$ asymmetry provides a direct measure
of the parity-violating parameter $\lambda_s^{nn}$.
Note that, since the total cross section is involved,
some investigators have opted to utilize the optical theorem
via \cite{bhk,oka,mil}
\begin{equation}
A_L=4k {{\rm Im} d_s^{nn}(k)\over {\rm Im} m_s(k)},
\end{equation}
which, using our unitarized forms, is completely equivalent to
Eq. (\ref{eq:fre}).

Of course, $nn$ scattering is currently just
a {\it gedanken} experiment, and we have discussed
it merely as a warm-up to the real problem: $pp$ scattering, which
introduces
the complications associated with the Coulomb interaction.  In spite of
this
complication, the calculation proceeds quite in parallel to the
discussion
above with obvious modifications.
We find
\begin{equation}
A_L={\sigma_+-\sigma_-\over \sigma_++\sigma_-} ={4k\, {\rm
Re}[m_s^*(k)d_s^{pp}(k)] \over |m_s(k)|^2}\simeq 4k\lambda_s^{pp}
\end{equation}
In the next section we show how this can be obtained
straightforwardly within an EFT approach.

On the experimental side, such asymmetries have been measured both
at low energies (13.6 and 45 MeV) as well as at higher energies (221 and
800 MeV). It is only the low-energy results\footnote{Note that the
13.6 MeV Bonn measurement is fully consistent with the earlier
but less precise number
\begin{equation}
A_L^{pp}(15\,\,{\rm MeV})=-(1.7\pm 0.8)\times 10^{-7}\cite{lan}
\end{equation}
determined at LANL.}
\begin{eqnarray}
A_L^{pp}(13.6\,\,{\rm MeV})&=&-(0.93\pm0.20\pm0.05)\times
10^{-7}\cite{bon},\nonumber\\
A_L^{pp}(45\,\,{\rm MeV})&=&-(1.50\pm 0.22)\times 10^{-7}\cite{psi},
\label{psi}
\end{eqnarray}
that are appropriate for our analysis, and from these results we
can extract the experimental number for the singlet mixing
parameter as
\begin{equation}
\left(\lambda_s^{pp}\right)^{expt} =-{A_L(45 {\rm MeV})\over
4k}=-(4.0\pm 0.6)\times 10^{-8}\,\,{\rm fm}.\label{eq:re}
\end{equation}
where $4k\approx 0.88 m_N$. Note that this Eq. (\ref{eq:re}) is
consistent with that of Desplanques and
Missimer\cite{dm}\begin{equation}
\left(\lambda_s^{pp}\right)^{expt} =-{A_L(45 {\rm MeV})\over
0.82m_N}=-(4.1\pm 0.6)\times 10^{-8}\,\,{\rm fm}, \label{eq:ppe}
\end{equation}

In a corresponding fashion, as described by Ref. \cite{dm},
contact can be made between other low-energy observables and the
effective parity-violating interaction. Clearly we require five
independent experiments in order to identify the five independent
S-P mixing amplitudes.  As emphasized above, we consider {\it
only} PV experiments on systems with $A=4$ or lower, in order that
nuclear-model dependence be minimized.  We utilize here the
results of Desplanques and Missimer \cite{dm}, but these forms
should certainly be updated using state-of-the art few-body
computations. There exist many such possible experiments and we
suggest the use of
\begin{itemize}
\item[i)] low-energy $pp$ scattering, for which
\begin{eqnarray}
pp(13.6\,\,{\rm MeV}):\quad A_L^{pp} &=&-0.48\lambda_s^{pp}m_N\nonumber\\
pp(45\,\,{\rm MeV}):\quad A_L^{pp} &=&-0.82\lambda_s^{pp}m_N
\end{eqnarray}
\item[ii)] low-energy $p\alpha$ scattering, for which
\begin{equation}
p\alpha(46\,\,{\rm MeV}):\quad A_L^{p\alpha} = [-0.48(\lambda_s^{pp}+{1\over
2}\lambda_s^{pn})-1.07(\rho_t+{1\over 2}\lambda_t)]m_N
\end{equation}
\item[iii)] threshold $np$ radiative capture for which there exist
two independent observables
\begin{eqnarray}
\mbox{circular polarization}:\quad
P_\gamma&=&(0.63\lambda_t-0.16\lambda_s^{pn})m_N\nonumber\\
\mbox{photon asymmetry}:\quad A_\gamma&=&-0.107\rho_tm_N
\end{eqnarray}
\item[iv)] neutron spin rotation from $^4$He
\begin{equation}
{d\phi^{n\alpha}\over dz}=[1.2(\lambda_s^{nn}+{1\over 2}
\lambda_s^{pn})-2.68(\rho_t-{1\over 2}\lambda_t)]M_n\,\,{\rm
rad/m}
\end{equation}
\end{itemize}
Inverting these results, we can determine the five S-P mixing
amplitudes via
\begin{eqnarray}
m_N\lambda_s^{pp}&=&-1.22\,A_L^{pp}(45\,\,{\rm MeV})\nonumber\\
m_N\rho_t&=&-9.35\,A_\gamma(np\rightarrow d\gamma)\\
m_N\lambda_s^{pn}&=&1.6\,A_L^{pp}(45\,{\rm
MeV})-3.7\,A_L^{p\alpha}(46\,{\rm MeV})+37\,A_\gamma(np\rightarrow
d\gamma)\nonumber\\
&&-2\,P_\gamma(np\rightarrow d\gamma)\nonumber\\
m_N\lambda_t&=&0.4\,A_L^{pp}(45\,{\rm
MeV})-0.7\,A_L^{p\alpha}(46\,{\rm MeV})+7\,A_\gamma(np\rightarrow
d\gamma)\nonumber\\
&&+\,P_\gamma(np\rightarrow d\gamma)\nonumber\\
m_N\lambda_s^{nn}&=&0.83\,{d\phi^{n\alpha}\over dz}
-33.3\,A_\gamma(np\rightarrow d\gamma)-0.69\,A_L^{pp}(45\,\,{\rm
MeV})\nonumber\\
&&+1.18\,A_L^{p\alpha}(46\,\,{\rm MeV})-1.08P_\gamma(np\rightarrow
d\gamma)
\end{eqnarray}

At the present time only two of these numbers are known
definitively ---the longitudinal asymmetry in $pp$, Eq. (\ref{psi}),
and in $p\alpha$ scattering,
\begin{equation}
A_L^{p\alpha}(46\,\,{\rm MeV})=-(3.3\pm 0.9)\times
10^{-7}\cite{moresin}.
\end{equation}
However, efforts are underway to measure the photon asymmetry in
radiative $np$ capture at LANSCE \cite{lansce} as well as the
neutron spin rotation on $^4$He at NIST \cite{nist}.  These
measurements are also proposed at the neutron beamline at the
Spallation Neutron Source (SNS) under construction at Oak Ridge
National Laboratory.   An additional, new measurement of the
circular polarization in $np$ radiative capture would complete the
above program, although this is very challenging because of the
difficulty of measuring the photon helicity.   Alternatively, one
could consider the inverse reaction
---the asymmetry in $\vec{\gamma}d\rightarrow np$--- and this is
being considered at Athens \cite{athens} and at HIGS at
Duke\cite{duke}.

To the extent that one can neglect inclusion of the $\pi$ as an
explicit degree of freedom, one could use this program of
measurements to perform a complete determination of the five
independent combinations of ${\cal O}(Q)$, PV LECs. Nonetheless,
in order to be confident in
 the results of such a series of
measurements, it is useful to note that other light systems can
and should also be used as a check of the consistency of the
extraction. There are various possibilities in this regard,
including
\begin{itemize}
\item[i)] $pd$ scattering
\begin{equation}
A_L^{pd} (15\,\,{\rm MeV})=(-0.21\rho_t-0.07\lambda_s^{pp}-0.13\lambda_t
                        -0.04\lambda_s^{pn})m_N
\end{equation}
\item[ii)] radiative $nd$ capture
\begin{equation}
A_\gamma=(1.42\rho_t+0.59\lambda_s^{nn}+1.18\lambda_t+0.51\lambda_s^{pn})m_N
\end{equation}
\item[iii)] neutron spin rotation on $H$
\begin{equation}
{d\phi^{np}\over dz}
=(1.26\rho_t-0.63\lambda_t+1.8\lambda_s^{np}+0.45\lambda_s^{pp}+0.45\lambda_s^{n
n})m_N\,{\rm
rad/m}
\end{equation}
\end{itemize}
Note that possible follow-ups of the LANSCE and NIST
experiments include the last three processes \cite{snow}.

We emphasize that the above results have been derived under the
assumption that the spin-conserving interaction $\rho_t$ is short
ranged -- an assumption applicable at energies well below the pion
mass. On the other hand, for the 46 MeV ${\vec p}\alpha$
measurement, the proton momentum is well above $m_\pi$, so
integrating out the pion may not be justified. In this case,
inclusion of the pion will lead to modification of the above
formulas, introducing a dependence on $\hpinn$, $k_{\pi
NN}^{1a}$, and  ${\bar C}_\pi$. Thus, a total of eight low-energy
few-body measurements would be needed to determine the relevant
set of low-energy constants. In the foregoing discussion, we have
identified eight few-body observables that could be used for this
purpose. Additional possibilities include the PV asymmetry in
near-threshold pion photo- or
electro-production\cite{cj1,cj2,riad} or deuteron
photodisintegration\cite{bogdan}. At present,  we are unable to
write down the complete dependence of the few-body PV observables
on $\hpinn$, $k_{\pi NN}^{1a}$, and ${\bar C}_\pi$, since only
the effects of the LO OPE PV potential (and associated two-body
currents) have been included in previous few-body computations.
Obtaining such expressions is a task requiring future theoretical
effort. In any case, it is evident from our discussion that there
exists ample motivation for several new few-body PV experiments
and that a complete determination of the relevant PV low-energy
constants is certainly a feasible prospect.

\section{EFT without explicit pions}
\label{pionless}

Although the foregoing analysis relied on traditional scattering
theory, it is entirely equivalent to an EFT approach. In the
following two sections, we present this EFT treatment in greater
detail, considering first only processes where the momenta $p$ of
all external particles are much smaller than the pion mass. In
this regime, the detailed dynamics underlying the $NN$ interaction
cannot be resolved, and interactions are represented by simple
delta-function potentials.  As with any EFT, this approximation is
justified by a separation of scales. In this case, one has scales
set by the $NN$ scattering lengths
---$a_s\sim-20$ fm, $a_t\sim 5$ fm---  that are
both much larger than the  $\sim 1/m_\pi$ range of the
pion-exchange component of the $NN$ strong interaction
\cite{aleph,crs}.
Because of this separation of scales, the deuteron can be described
within this pionless EFT. For example, one can calculate
the deuteron form factors at momenta up to the pion mass \cite{crs}.
This pionless EFT is limited in energy,
but it is very simple 
(since all interactions among nucleons are of contact character)
and high-order calculations
can be carried out. Therefore, although its expansion parameter
is not particularly small, high precision can be reached easily.

In this very-low-energy regime the EFT of
the two-nucleon problem is not much more than a reformulation of
the analysis in Section \ref{fewbody}. The full benefits of an EFT
framework will, however, be evident when we consider the regime of
momenta comparable to the pion mass in the next section.

\subsection{Effective Lagrangian}

Nucleons with momenta much smaller than the pion mass are
non-relativistic, and in this case, it is convenient to redefine
the nucleon fields so as to eliminate the term proportional to
$m_N$ from the Lagrangian. In so doing, one obtains an infinite
tower of operators proportional to powers of $p/m_N << 1$. This
widely-used heavy-fermion formalism \cite{georgi,jenkins}, is
nothing but a Galilean-covariant expression of the usual
non-relativistic expansion. Since the non-relativistic EFT must
match the relativistic theory for $p\sim m_N$, Lorentz invariance
relates various terms in the tower of $(p/m_N)^k$-suppressed
effective operators. Thus, one way to construct the effective
Lagrangian is to write the most general rotational-invariant
non-relativistic Lagrangian, then to relate parameters by imposing
this matching condition, or ``reparameterization'' invariance
\cite{reparam}. Alternatively, we can simply write a  relativistic
Lagrangian and then take the non-relativistic limit.

The most general Lagrangian involving two nucleon fields
$N,\bar{N}$ and a photon field $A_\mu$ that is invariant under
Lorentz, parity, time reversal and $U(1)$ gauge symmetries is
\begin{eqnarray}\label{pcnopi1}
{\cal L}_{N,PC} &=& \bar N \Bigl\{ iv\cdot D +{1\over 2m_N} \left(
(v\cdot D)^2- D^2 \right) +[S_\mu, S_\nu][{D}^\mu, {
D}^\nu]\\\nonumber & &+ {\kappa_0+\kappa_1\tau_3\over
m_N}\epsilon_{\mu\nu\alpha\beta} v^\alpha S^{\beta} F^{\mu\nu}
+\ldots \Bigr\} N,
\end{eqnarray}
where $\kappa_0 (\kappa_1)$ is the isoscalar (isovector) anomalous
magnetic moment, $v^\mu$ and $S^\mu$ are the nucleon velocity and
spin [$v^\mu=(1, \vec{0})$ and $S^\mu=(0, \vec{\sigma}/2)$ in the
nucleon rest frame], $D_\mu=
\partial_\mu +ie Q_N A_\mu$ is the electromagnetic covariant
derivative, with $Q_N=(1+\tau_z)/2$ the nucleon charge matrix, and
$F^{\mu\nu}=
\partial^\mu A^\nu -\partial^\nu A^\mu$. Here, as in the following
Lagrangians, ``$\ldots$'' denote terms with more derivatives,
which give rise to other nucleon properties such as
polarizabilities.

When we relax the restriction of parity invariance, we can write
additional terms, such as
\begin{equation}\label{pvnopi}
{\cal L}_{N,PV}
= \frac{2}{m_N^2}\bar N (a_0 + a_1 \tau_z)S_\mu N \partial_\nu
F^{\mu\nu}
+\ldots,
\end{equation}
where $a_0$ ($a_1$) is the isoscalar (isovector) anapole moment
of the nucleon. These terms were discussed in Refs.
\cite{anapole1,anapole2};
they appear in PV electron scattering but
not in the processes we focus on here.

For the two-nucleon system, we need to consider contact terms with
{\it four} nucleon fields. The simplest parity-conserving (PC)
interactions are
\begin{equation}
\label{pcnopi2}
{\cal L}_{PC,NN} =
-\frac{1}{2}C_{S}\bar{N}N\bar{N}N
+2C_{T}\bar{N}S^{\mu }N\bar{N}S_{\mu }N
+\ldots,
\end{equation}
where $C_S$, $C_T$ are dimensional coupling constants first introduced
in
Ref. \cite{wei}.
Their projections onto the two $NN$ S waves are
\begin{eqnarray}
C_{0s}&=& C_S -3C_T    \nonumber \\
C_{0t}&=& C_S+C_T\ \ \ .
\end{eqnarray}
These parameters are related to the respective scattering lengths,
while higher-derivative operators give rise to additional
parameters, such as S-wave effective ranges and P-wave scattering
volumes \cite{aleph}.

For future reference, it is also useful to write down the
first-quantized $NN$ potential arising from ${\cal L}_{PC,NN}$. To
order ${\cal O}(Q)$, we have
\begin{equation}\label{pcconpot}
V^{CT}_{PC} ({\vec q}, {\vec p})
= C_S + C_T {\vec \sigma}_1 \cdot {\vec \sigma}_2,
\end{equation}

Similarly, we can construct PV two-nucleon contact interactions. A
detailed derivation appears in Appendix A and leads at ${\cal
O}(Q)$ to
\begin{eqnarray}\label{lllprime}\nonumber
{\cal L}_{PV,NN}&=& {1\over \Lambda_\chi^3}
\left\{
-C_1 N^\dag  N \, N^\dag  {\vec \sigma} \cdot i{\vec D}_- N
+ C_1 N^\dag  iD^i_- N \,  N^\dag  \sigma^i  N
\right.\\ \nonumber
&&\left. \quad  - {\tilde C}_1
i\epsilon^{ijk} \, N^\dag  iD_+^i \sigma_j N \, N^\dag  \sigma^k N
\right.\\ \nonumber
&&\left. \quad  -C_2\ N^\dag  N \, N^\dag \tau_3 {\vec \sigma}
\cdot i{\vec D}_- N
 + C_2 N^\dag  iD^i_- N  \, N^\dag \tau_3  \sigma^i  N
\right.\\ \nonumber
&&\left. \quad  - {\tilde C}_2
i\epsilon^{ijk} \, N^\dag  iD_+^i \sigma_j N \, N^\dag \tau_3  \sigma^k N
\right.\\ \nonumber
&&\left. \quad  -C_3 N^\dag \tau^a N \, N^\dag \tau^a {\vec \sigma}
\cdot i{\vec D}_- N
 + C_3 N^\dag\tau^a  iD^i_- N  \, N^\dag \tau^a \sigma^i  N \right.\\ \nonumber
&& \left. \quad  - {\tilde C}_3
i\epsilon^{ijk} \, N^\dag \tau^a iD_+^i \sigma_j N \, N^\dag \tau^a \sigma^k N
\right.\\ \nonumber
&&\left. \quad  -C_4 N^\dag \tau_3 N \, N^\dag  {\vec \sigma} \cdot i{\vec
D}_- N
 + C_4 N^\dag \tau_3 iD^i_- N  \, N^\dag  \sigma^i  N
\right.\\ \nonumber
&&\left. \quad  - {\tilde C}_4
i\epsilon^{ijk} \, N^\dag \tau_3 iD_+^i \sigma_j N \, N^\dag  \sigma^k N
\right.\\ \nonumber
&&\left. \quad  -C_5 {\cal I}_{ab}\, N^\dag \tau^a N \, N^\dag \tau^b {\vec
\sigma}
\cdot i{\vec D}_- N
 +C_5{\cal I}_{ab} \, N^\dag \tau^a iD^i_- N \, N^\dag\tau^b
\sigma^i N
\right.\\ \nonumber
&&\left.  \quad -{\tilde C}_5 {\cal I}_{ab}
i\epsilon^{ijk} \, N^\dag \tau^a iD_+^i \sigma_j N \, N^\dag \tau^b \sigma^k N
\right.\\
&&\left. \quad -C_6 i\epsilon^{ab3} \, N^\dag \tau^a N \, N^\dag \tau^b {\vec
\sigma}
\cdot i{\vec D}_+ N \right\}
+\ldots,
\label{PVcontacts}
\end{eqnarray}
where we have introduced the short-hand notation
\begin{equation}
N^\dag iD^\mu_\pm N \equiv \left( iD^\mu N \right)^{\dag} N \pm N^\dag
\left( iD^\mu N \right).
\label{Dpm}
\end{equation}
(in momentum space, $iD^\mu_{+}$ and $iD^\mu_{-}$ give rise to the
difference and sum, respectively, of the initial and final nucleon
momenta). The effects of the weak interaction are represented by
the LECs $C_i$. We have normalized the operators to a scale
$\Lambda_\chi=4\pi F_\pi\sim$ 1 GeV, as would appear natural in a
pionful theory. One might then anticipate that the $ C_i$ are of
order $G_F\Lambda_\chi^2\sim 10^{-5}$. In fact, as discussed in
Section 5, naive dimensional analysis (NDA) suggests that these
quantities have the magnitude $C_i\sim (\Lambda_\chi/F_\pi)^2
g_\pi$, where $g_\pi= 3.8\times 10^{-8}$ sets the scale for
non-leptonic weak interactions. One may also attempt to predict
such constants using models (see Section \ref{models}) and compare
with the experimentally determined linear combinations discussed
above.

The ${\cal O}(Q)$ Lagrangian ${\cal L}_{PV,NN}$ gives rise to the
potential in Eq. (\ref{3}), which generates energy-independent S-P
wave mixing as discussed earlier. Higher-derivative PV operators
lead both to energy-dependence in the S-P mixing amplitudes as
well as mixing involving higher partial waves. Given the level of
complexity already appearing at ${\cal O}(Q)$, we will {\it not}
consider these higher-order terms.

\subsection{Amplitudes}

In processes involving a single nucleon, amplitudes can be
expanded in loops.  The situation is more subtle when two or more
nucleons are present \cite{wei}. This is due to the fact that
intermediate states that differ from initial states only by nucleon
kinetic energies receive $O(m_N/p)$ enhancements. A resummation
then must be performed, leading, {\it e.g.}, to nuclear bound
states, and it is not immediately obvious that such resummations
can be done while maintaining the derivative expansion necessary
to retain predictive power order by order. The large values for
the $NN$ scattering lengths, however, provide justification for
such a procedure \cite{aleph,crs}. Before considering PV effects,
it is helpful to review what resummation technique yields for the
case of the strong $NN$ interaction.

In lowest order,
the S-wave $NN$ interaction can be represented via a contact term
\begin{equation}
T_{0i}=C_{0i}(\mu).
\end{equation}
Including the rescattering corrections, the full $T$-matrix is found to
be
\begin{eqnarray}
T_i(k)&=&C_{0i}(\mu)+C_{0i}(\mu)G_0(k)C_{0i}(\mu)+\ldots\nonumber\\
&=&{C_{0i}(\mu)\over 1-C_{0i}(\mu)G_0(k)}=-{4\pi\over m_N}{1\over
-{4\pi\over m_N C_{0i}(\mu)}-\mu-ik},
\label{pionlessresum}
\end{eqnarray}
where
\begin{equation}
G_0(k)=\lim_{\vec{r},\vec{r}'\rightarrow
0}G_0(\vec{r},\vec{r}')=\int{d^3s\over (2\pi)^3} {1\over {k^2\over
m_N}-{s^2\over m_N}+i\epsilon}=-{m_N\over 4\pi}(\mu+ik)
\end{equation}
is the zero-range Green's function, which displays
the large nucleon mass in the numerator.
Identifying the scattering length
 via
\begin{equation}
-{1\over a_i}=-{4\pi\over m_NC_{0i}(\mu)}-\mu\label{eq:sca}
\end{equation}
and using the relation
\begin{equation}
m_i(k)=-{m_N\over 4\pi}T_i(k)
\end{equation}
connecting the scattering and $T$-matrices, we find
\begin{equation}
m_i(k)={1\over -{1\over a_i}-ik}=-{a_i\over 1+ika_i}.
\end{equation}
It is important to note here that since $a_i$ is a physical quantity,
 it cannot depend on the scale parameter $\mu$ and this invariance is observed in
 Eq. (\ref{eq:sca}), wherein the $\mu$ dependence of the Green's
 function is canceled by the corresponding scale dependence in
 $-4\pi/m_N C_{0i}(\mu)$.

We observe that the resummation is at this order completely
equivalent to the unitarization that lead to Eq. (\ref{ert}), and
one can show similarly that in the $NN$ system inclusion of
higher-derivative operators reproduces higher powers of energy in
the effective-range expansion \cite{aleph,crs}.

It is straightforward to generalize the above calculation
to account for electromagnetic interactions.
As shown in Ref. \cite{hol} (see also Ref. \cite{kong})
the unitarized $pp$ scattering amplitude has the form
\begin{equation}
m_s(k)=-{m_N\over
4\pi}{C_{0s}(\mu)C_\eta^2(\eta_+(k))\exp2i\sigma_0\over
1-C_{0s}(\mu)G_C(k)},
\end{equation}
where $\eta_+(k)=M\alpha/2k$,
\begin{equation}
C_\eta^2(x)={2\pi x\over e^{2\pi x}-1}
\end{equation}
is the usual Sommerfeld factor, $\sigma_0=
{\rm arg}\Gamma(\ell+1+i\eta_+(k))$ is the Coulomb phase shift, and
the free Green's function $G_0(k)$ has also been replaced by its Coulomb
analog
\begin{equation}
G_C(k)=\int{d^3s\over (2\pi)^3}{C^2(\eta_+(k))\over
{k^2\over m_N}-{s^2\over m_N}+i\epsilon}.
\end{equation}
Remarkably, this integral can be performed analytically, yielding

\begin{equation}
G_C(k)=-{m_N\over
4\pi}\left[\mu+m_N\alpha\left(H(i\eta_+(k))-\log{\mu\over
m_N\pi\alpha}-\zeta\right)\right].
\end{equation}
Here $\zeta$ is defined in terms of the Euler constant $\gamma_E$ via
$\zeta=2\pi-\gamma_E$ and
\begin{equation}
H(x)=\psi(x)+{1\over 2x}-\log x.
\end{equation}
The resultant scattering amplitude has the form
\begin{eqnarray}
m_s(k)&=&{C_\eta^2(\eta_+(k))e^{2i\sigma_0}\over -{4\pi\over
m_NC_{0s}(\mu)}
-\mu-m_N\alpha\left[H(i\eta_+(k))-\log{\mu\over m_N\pi\alpha}
-\zeta\right]}\nonumber\\
&=&{C_\eta^2(\eta_+(k))e^{2i\sigma_0}\over -{1\over a_{0s}}
-m_N\alpha\left[h(\eta_+(k))-\log{\mu\over m_N\pi\alpha}
-\zeta\right]-ikC_\eta^2(\eta_+(k))},
\end{eqnarray}
where we have defined, as before,
\begin{equation}
-{1\over a_{0s}(\mu)}=-{4\pi\over MC_{0s}(\mu)}-\mu,
\end{equation}
and
\begin{equation}
h(\eta_+(k))
={\rm Re}H(i\eta_+(k)).
\end{equation}

The experimental scattering length $a_{Cs}$ in the presence of the
Coulomb interaction is defined via
\begin{equation}
-{1\over a_{Cs}}=-{1\over
a_{0s}(\mu)}+m_N\alpha\left(\log{\mu\over m_N\pi\alpha}
-\zeta\right),\label{eq:sca1}
\end{equation}
in which case the scattering amplitude takes its traditional
lowest-order
form
\begin{equation}
m_s(k)={C_\eta^2(\eta_+(k))e^{2i\sigma_0}\over -{1\over a_{Cs}}
-m_N\alpha H(i\eta_+(k))}\label{eq:cou}.
\end{equation}
Of course, Eq. (\ref{eq:cou}) requires that the Coulomb-corrected
scattering length be
different from its non-Coulomb partner, and comparison of the
experimental
$pp$ scattering length ---$a_{pp}=-7.82$ fm--- with its $nn$ analog
---$a_{nn}= -18.8$ fm--- is roughly consistent with
Eq. (\ref{eq:sca1}) if a reasonable
cutoff is chosen ({\em e.g.}, $\mu\sim 1$ GeV).

Having unitarized the strong scattering amplitude, we can now
proceed analogously for its parity-violating analog. The
lowest-order S-P mixing amplitude is
\begin{equation}
T_{0SP}=W_{0SP}(\mu).
\end{equation}
Inclusion of S-wave rescattering effects while neglecting P-wave
scattering and Coulomb contributions yields the result
\begin{equation}
T_{SP}(k)=W_{0SP}(\mu)+W_{0SP}(\mu)G_0(k)C_{0i}(\mu)+\ldots={W_{0SP}(\mu)\over
1-G_0(k)C_{0i}(\mu)}.\label{eq:ho}
\end{equation}
Writing Eq. (\ref{eq:ho}) in the form
\begin{eqnarray}
d_i(k)&=&-{m_N\over 4\pi}T_{SP}(k)={{W_{0SP}(\mu)\over
C_{0i}(\mu)}\over
-{4\pi\over m_NC_{0i}(\mu)}-\mu-ik}\nonumber\\
&=&{\lambda_i\over -{1\over a_i}-ik}=\lambda_im_i(k),
\end{eqnarray}
we identify the {\it physical} ($\mu$-independent) S-P wave mixing
amplitude
via
\begin{equation}
\lambda_i=\frac{W_{0SP}(\mu)}{C_{0i}(\mu)}.
\end{equation}
Similarly, including the Coulomb interaction, we find for the
unitarized weak amplitude
\begin{equation}
T_{0SP}={W_{0SP}(\mu)C_\eta^2(\eta_+(k))e^{i(\sigma_0+\sigma_1)}
\over (1-C_{0s}(\mu)G_C(k))}
\equiv
{\lambda_{SP}^{pp}C_\eta^2(\eta_+(k))e^{i(\sigma_0+\sigma_1)}\over
-{1\over a_{Cs}}-m_N\alpha a_{Cs}H(i\eta_+(k))},
\end{equation}
where we have again neglected the P-wave scattering, and have
identified
\begin{equation}
\lambda_{SP}^{pp}={W_{0SP}(\mu)\over C_{0s}(\mu)}
\end{equation}
as the physical mixing parameter.

Having obtained fully unitarized forms, we can now proceed to
evaluate the helicity-correlated cross sections, finding, as
before, at the very lowest energies,
\begin{equation}
A_L={\sigma_+-\sigma_-\over \sigma_++\sigma_-} ={4k\, {\rm
Re}(d_s^*(k)m_s^{pp}(k)) \over |m_s(k)|^2}\simeq
4k\lambda_s^{pp}.
\end{equation}

%

Somewhat more involved, of course, are processes involving more
than two nucleons. Besides the inherent calculational difficulty,
interesting new physics arises when three nucleons can overlap.
When pions are integrated out of the theory, three-nucleon
interactions become significant. In fact, it has been shown
\cite{triton} that its strong running requires that the
non-derivative contact three-body force be included at {\it
leading} order in the EFT, together with the non-derivative
contact two-body forces considered above. This three-nucleon force
acts only on the S$_{1/2}$ channel, and provides a mechanism for
triton saturation. The existence of one three-body parameter in
leading order is the reason behind the phenomenological Phillips
line. Note that most three-nucleon channels are free of a
three-nucleon force up to high order, and can therefore be
predicted to high accuracy with two-nucleon input only
\cite{3stooges}. Similar renormalization might also take place in
the four-nucleon system. It remains to be seen whether the same
phenomenon also enhances PV few-body forces. We defer a detailed
treatment of $A\geq 3$ PV forces and related renormalization
issues to a future study.

\section{EFT with explicit pions}
\label{sec3}

For processes in which $p\sim m_\pi$, it is no longer sufficient to
integrate the pions out of the effective theory. Incorporation of
the pion as an explicit degree of freedom requires use of consistent PV chiral
Lagrangian, which we develop in this section.

\subsection{Effective Lagrangian}

Chiral perturbation theory ($\chi$PT) provides a systematic
expansion of physical observables in powers of small momenta and
pion mass for systems with at most one nucleon \cite{gl,jenkins}.
The interactions obtained from $\chi$PT can be used to build
four-nucleon operators arising from pion exchange, though care
must be taken to avoid double-counting the effects of multi-pion
exchange in both operators and wave functions (see below). In the
approach we follow here, pionic effects are generally included
non-perturbatively. Strong $\pi N$ interactions are derivative in
nature, and thus scale as powers of $p/\Lambda_\chi$. As a result,
one can include them while maintaining a systematic derivative
expansion \cite{wphysica}. By contrast, weak $\pi N$ interactions
need not involve derivatives, but the small scale associated with
hadronic weak interactions ($g_\pi$) implies that one needs at
most one weak vertex. In addition, explicit chiral
symmetry-breaking effects associated with the up- and down-quark
masses also enter perturbatively, since $m_\pi <<\Lambda_\chi$. To
incorporate all these effects, we require the most general
effective Lagrangian to a given order in $p$ containing local
interactions parameterized by {\em a priori} unknown low-energy
constants (LECs). The corrections from quark masses and loops are
then included order by order.

We give here the basic ingredients to our
discussion. (For a review, see Ref. \cite{ijmpe}.)
The nucleon mass $m_N$ is much larger than the pion mass $m_\pi$,
so we continue to employ a heavy-nucleon field.
The pion fields $\pi^a$, $a=1, 2, 3$, enter through
\begin{equation}
\xi  = \exp \left( i\pi^a \tau^a\over 2 F_\pi  \right),
\end{equation}
where $F_\pi =92.4$ MeV is the pion decay constant \cite{bho}.
This quantity allows us to construct chiral vector and axial-vector
currents
given by
\begin{eqnarray}\nonumber
V_\mu &=&{1\over 2}(\xi D_\mu \xi^\dag +\xi^\dag D_\mu \xi),\\ \nonumber
A_\mu &=& -{i\over 2}(\xi D_\mu \xi^\dag -\xi^\dag D_\mu
\xi)=-{D_\mu\pi\over F_\pi} +O(\pi^3),
\end{eqnarray}
respectively.

Chirally-symmetric strong interaction pionic effects
can be incorporated into the pionless Lagrangian
by substituting $D_\mu \rightarrow {\cal D}_\mu$, where
the chiral covariant derivative is
\begin{equation}
{\cal D}_\mu =  D_\mu +V_\mu,
\end{equation}
and by adding interactions involving $A_\mu$.
On the other hand, the
quark mass matrix ${\cal M}={\rm diag}(m_u, m_d)$
generates
chiral-symmetry breaking that
can be incorporated via
\begin{equation}
\chi_{\pm}  =  \xi^{\dagger}\chi \xi^{\dagger}\pm \xi\chi^{\dagger}\xi,
\end{equation}
where
\begin{equation}
\chi  =  2B(s+ip),
\end{equation}
with $B$ a constant with dimensions of mass, and $s, p$
representing scalar and pseudoscalar source fields. In the present
application $s={\cal M}$ and $p=0$, and in the following, we work
in the isospin-symmetric limit, $m_u=m_d={\hat m}$.
Isospin-breaking effects will generate small ($\simle 10^{-2})$
multiplicative corrections to the tiny PV effects of interest
here, so we safely neglect them. In this case, to  leading order
in the chiral expansion we have
\begin{eqnarray}\nonumber
\chi_{+} &=& 2B{\hat m} + O(\pi^2)\\
\chi_{-} &=&2B{\hat m}{i\pi\over F_\pi} + O(\pi^3).
\end{eqnarray}
The building blocks for including a $\Delta$ field in the
Lagrangian can be found in Ref. \cite{barrysdeltas}. For
simplicity we here integrate out $\Delta$ isobars. It is
straightforward but tedious to use these building blocks to extend
the results of our paper by including explicit $\Delta$ effects.

We group terms in Lagrangians ${\cal L}^{(\nu)}$ labeled by the
chiral index $\nu = d + f/2 -2$, where $d$ is the number of
derivatives and powers of the pion mass and $f$ the number of
fermion fields. We only display terms that are relevant for the
arguments that follow.

\begin{itemize}

\item Parity-conserving $\pi N$ Lagrangian

We then arrive at
\begin{equation}\label{pc1}
{\cal L}_{\pi N, PC}^{(0)} = \bar N [iv\cdot {\cal D} +2g_A^0 S\cdot A]N,
\end{equation}
with the lowest index.
Similarly, we have for the next to leading order (NLO) Lagrangian
\begin{eqnarray}\label{pc2}
{\cal L}_{\pi N, PC}^{(1)}& =& {1\over 2m_N} \bar N \Bigl\{ (v\cdot {\cal
D})^2
-{\cal D}^2
+[S_\mu, S_\nu][{\cal D}^\mu, {\cal D}^\nu]\\&-&2ig_A^0(S\cdot {\cal D} v\cdot A
 +v\cdot A S\cdot {\cal D})
+2(\kappa_0+\kappa_1\tau_3)\epsilon_{\mu\nu\alpha\beta}v^\alpha
S^\beta F^{\mu\nu}\Bigr\}N +\ldots\nonumber
\end{eqnarray}
and the next to next to leading order (NNLO) terms
\begin{eqnarray}\label{pc3}\nonumber
{\cal L}_{\pi N, PC}^{(2)}& =& \bar{N}
\left\{
\frac{g_A^0}{4m_N^2}\,[{\cal D}^\mu,[{\cal D}_\mu,S\!\cdot\! A]]
- \frac{g_A^0}{2m_N^2}\,v\cdot\!\!\stackrel{\leftarrow}{{\cal D}}
S\!\cdot\!A\,v\!\cdot\! {\cal D}
\right. \\ \nonumber
&& \left.
- \frac{g_A^0}{2m_N^2}\left(\{S\!\cdot\! {\cal D},v\!\cdot\!
A\}\,v\!\cdot\!
{\cal D} + \mbox{h.c.}\right)
-  \frac{g_A^0}{4m_N^2}\left(S\!\cdot\! A\,{\cal D}^2 + \mbox{h.c.}\right)
\right. \\ \nonumber
&&\left.
-\frac{g_A^0}{2m_N^2}\left(S\cdot\!\!\stackrel{\leftarrow}{{\cal D}} A\!
\cdot\! {\cal D} + \mbox{h.c.}\right)
+2\widehat{d}_{16}S\cdot A \langle\chi_{+}\rangle
\right. \\
&&\left.
+2\widehat{d}_{17}\langle S\cdot A\chi_{+}\rangle
+i\widehat{d}_{18} [S\cdot {\cal D},\chi_{-}]
+i\widehat{d}_{19}[S\cdot {\cal D},\langle\chi_{-}\rangle]
\right\} N+\ldots
\end{eqnarray}
Here the ellipses denote counter-terms not relevant in our present
calculation, a complete list of which is given in Ref.
\cite{nadia}.  The superscript \lq\lq 0" in $g_A $ and
$\mu_N$ indicates that these quantities must be appended by the
corresponding loop contributions in order to obtain the physical
(renormalized) axial coupling
 and nucleon
magnetic moment.

\item Parity-conserving $NN$ Lagrangian

For nuclear systems,
we require the PC Lagrangian involving more than two nucleon fields.
Here we will only need
the lowest index ($\nu=0$) terms, containing four nucleon fields.
The relevant Lagrangian has the same form as Eq. (\ref{pcnopi2}),
\begin{eqnarray}
\label{pc4}
{\cal L}_{NN,PC}^{(0)} =
-\frac{1}{2}C_{S}\bar{N}N\bar{N}N+2C_{T}\bar{N}S^{\mu
}N\bar{N}
S_{\mu }N+\ldots,
\end{eqnarray}
but here $C_S, C_T$ are constants whose numerical values are
different from the ones in the pionless theory. This is because we
are now removing soft-pion contributions from the counter-terms,
and including them explicitly.

The NLO four-nucleon corrections occur at $\nu=2$,
which will not be used since in this work we truncate the chiral
expansion of
the PV potential at ${\cal O}(Q)$.
Likewise, six-nucleon PC interactions first appear at $\nu=1$
so their contribution to PV observables will be at higher order
in loop diagrams.

\item Parity-violating $\pi N$ Lagrangian

The lowest-index ($\nu=-1$) PV interaction arises from the
$\pi NN$ Yukawa interaction,
\begin{eqnarray}\label{pv1}
{\cal L}^{(-1)}_{\pi N,PV}&=&-{\hpinn\over 2\sqrt{2} }{\bar N}
X_{-}^3 N\nonumber\\
&=&-i\hpinn (\bar p n \pi^+ -\bar n p \pi^-) +\ldots
\end{eqnarray}
where
\begin{equation}
X_{-}^3 = \xi^{\dag}\tau^3\xi - \xi\tau^3\xi^{\dag}
\end{equation}
and the ``$\ldots$'' denote the terms in this operator containing
additional (odd) numbers of pions. At $\nu=0$ there exist also PV
vector and axial-vector $\pi NN$ interactions, detailed
expressions for which can be found in Refs. \cite{kaplan,zhu}.
However, as discussed in Ref. \cite{cj2}, the effects of the
vector operators can be eliminated through ${\cal O}(Q)$ by using
the equations of motion and by suitably re-defining the constant
${\bar C}_\pi$ (defined below). The PV axial-vector couplings
involve two or more pions, and, as pointed out in Ref. \cite{zhu},
such couplings renormalize $\hpinn$ at ${\cal O}(Q^3)$.
Consequently, their contribution to the PV $NN$ potential appears
at ${\cal O}(Q^2)$, via loop effects.

At NNLO ($\nu=1$) we find
several new PV $\pi NN$ operators that will contribute to the PV
$NN$ potential at ${\cal O}(Q)$. In principle, these operators can
be expressed in terms of the quantities $X^a_{L,R}$ defined in
Ref. \cite{kaplan}, thereby allowing one to determine the full,
non-linear dependence on the pion fields. For our purposes,
however, it is sufficient to truncate the expansion of these
operators at one power of the pion field, since terms containing
additional pion fields only contribute to the PV $NN$ interaction
beyond ${\cal O}(Q)$. After implementing the strictures of reparameterization
invariance, we
obtain the Lagrangian
\begin{eqnarray}
\nonumber
{\cal L}^1_{\pi NN,\ PV} & = &{2ik_{\pi NN}^{1a}\over  \Lambda_\chi F_\pi}
\epsilon_{\mu\nu\alpha\beta} {\bar N}
{\overleftarrow D}^\mu
({\vec \tau}\times {\vec \pi})_3
{\overrightarrow D}^\nu v^\alpha S^\beta N
+{k_{\pi NN}^{1b}\over  \Lambda_\chi F_\pi} {\bar N} \left[
D^\lambda D_\lambda,
 ({\vec \tau}\times {\vec \pi})_3 \right]  N\\
\label{eq:nnlolropa}
&& + {k_{\pi NN}^{1c} m_\pi^2 \over  \Lambda_\chi F_\pi}
{\bar N}({\vec \tau}\times {\vec \pi})_3 N  +\ldots \  ,
\end{eqnarray}
where  we have chosen a normalization such that the constants
$k_{\pi NN}^{1a-c}$ ought to be of order a few times $g_\pi$
according to naive dimensional analysis (see below) and where the
``$\ldots$'' indicate terms involving more than one pion field.

Nominally, then, there exist three new, independent operators that contribute
to the PV NN potential at ${\cal O}(Q)$. A proof of their independence
under reparameterization invariance, following the arguments of Ref. \cite{luke92},
will appear in a forthcoming publication and we do not reproduce the full
arguments here. Heuristically, however, the existence of these operators
can be seen from their correspondence with the independent ${\cal O}(Q^2)$ scalars
that can be formed from the independent momenta, nucleon spin, and pion
mass\footnote{The presence of the single pion field leads to a pseudoscalar
interaction with the nucleon.}: ${\vec p}\cdot{\vec p}^{\prime}$,
${\vec \sigma}\cdot{\vec p}\times{\vec p}^{\prime}$, $({\vec p}-{\vec
p}^{\prime})^2$,
and $m_\pi^2$. Naively, then, one would have expected four independent ${\cal
O}(Q^2)$
operators, rather than just three as given in Eq. (\ref{eq:nnlolropa}), with the
operator corresponding to ${\vec p}^{\prime}\cdot{\vec p}$ given by
\begin{equation}
{\bar N }{\overleftarrow
D}^\lambda
 ({\vec \tau}\times {\vec \pi})_3  {\overrightarrow D}_\lambda N\  .
\end{equation}
However, in a relativistic formulation of the theory, the corresponding operator
can be rewritten in terms of ${\bar N}({\vec \tau}\times {\vec \pi})_3 N$ and
${\bar N} \left[
D^\lambda D_\lambda, ({\vec \tau}\times {\vec \pi})_3 \right]  N$ through
suitable integrations by parts and application of the LO equations of
motion\footnote{We thank Vincenzo Cirigliano for discussions on this point.};
consequently, it cannot exist as an independent operator in the heavy baryon
formulation. Indeed, similar arguments eliminate an analogous term, ${\bar N}
{\overleftarrow D}^\lambda S\cdot A  {\overrightarrow D}_\lambda N$, from the
parity conserving Lagrangian. In contrast, the remaining terms in ${\cal
L}^1_{\pi NN,\ PV}$ cannot be eliminated in the relativistic theory via such
arguments and, thus, must exist as independent terms in the non-relativistic case.

We also note that in order for EFT with non-relativistic nucleon fields
to match the fully relativistic theory, the coefficients $k_{\pi NN}^{1a-c}$
in general receive contributions proportional to $\hpinn$ that arise from a
non-relativistic reduction of  the LO PV $\pi NN$ Yukawa interaction in
addition to contributions that represent {\em bona fide} ${\cal O}(Q^2)$ effects.
This situation is analogous to what occurs for the ${\cal O}(Q^2)$ nucleon
magnetic moment operator, whose coefficient $\mu_N=Q_N+\kappa_N$ receives a
contribution (the Dirac term) that is dictated by relativity and that is
proportional to the ${\cal O}(Q)$ constant ($Q_N$) and a genuine, {\em a priori}
unknown ${\cal O}(Q^2)$ contribution (the Pauli term) parameterized by the anomalous
magnetic moment. In the present case, only $k_{\pi NN}^{1a,b}$ receive contributions
proportional to $\hpinn$ as dictated by relativity:
\begin{equation}
k_{\pi NN}^{1a}= {\hpinn\over 4\sqrt{2}}{\Lambda_\chi F_\pi\over M_N^2}+\ldots
\ , \qquad
k_{\pi NN}^{1b}= -{\hpinn\over 8\sqrt{2}}{\Lambda_\chi F_\pi\over M_N^2}
+\ldots\ ,
\end{equation}
where the ``$\ldots$''
indicate the unconstrained ${\cal O}(Q^2)$ contributions.

In practical terms, only two of the operators in Eq. (\ref{eq:nnlolropa}) are
likely to be experimentally distinguishable. In momentum space, the
second and third terms can be written as independent linear
combinations of $({\vec p}-{\vec p}^{\prime})^2+m_\pi^2$ and $m_\pi^2$.
The latter acts like a chiral
correction to $\hpinn$, so to ${\cal O}(Q)$ in the EFT, it cannot be resolved
experimentally. When inserted into the PV NN potential, the former cancels the
pion propagator, leading effectively to an ${\cal O}(Q)$ contact operator that
is indistinguishable from the SR operator proportional to $C_6$. In contrast,
the effects of the remaining operator involving
$k_{\pi NN}^{1a}$  cannot
be absorbed into the LO $\pi$ exchange potential or any of the
short-range ${\cal O}(Q)$ operators. Its contribution to the
potential has been given in Eq. (\ref{eq:vpvlrnnlo}).

\item Parity-violating $\gamma \pi N$ Lagrangian

Finally, there exists also a contact $\pi\gamma NN$ interaction at
$\nu =1$ \cite{cj2},
 \begin{equation}\label{pv2}
 {\cal L}^{(1)}_{\pi \gamma N,PV}=
-i e {{\bar C}_\pi\over \Lambda_\chi F_\pi}\bar p
               \sigma^{\mu \nu}F_{\mu\nu}n\pi^++{\mbox{H.c.}}
 \end{equation}



\item Parity-violating $NN$ Lagrangian and $\gamma \pi N N$ Lagrangian

The $\nu=1$ PV four-nucleon terms have the same form as in Eq.
(\ref{PVcontacts}) but with the gauge covariant derivatives
$D_\mu$ replaced by ${\cal D}_\mu$, the gauge {\em and} chiral
covariant derivatives. In this case, the coefficients $C_i$,
${\tilde C}_i$ will differ numerically from those appropriate to
the pionless theory, since in the latter case, the effects of pion
exchange are incorporated into the operator coefficients.

\end{itemize}

\subsection{Power counting}
\label{sec4}

Throughout this work we use power-counting arguments to
guide us in the task of identifying the most significant
contributions to PV observables.
Power counting is carried out under an implicit assumption
about the size of the couplings of the EFT.
It is assumed that the couplings are neither particularly small nor
particularly large compared with
``naive dimensional analysis'' (NDA) \cite{gm},
in which LECs scale with $F_\pi$ and $\Lambda_\chi$ as
\begin{equation}
\label{eq:georgi}
\left({D_\mu \over\Lambda_\chi}\right)^d
\left({\pi\over F_\pi}\right)^p
\left({{\bar N} N\over\Lambda_\chi F_\pi^2}\right)^{f/2}
\times (\Lambda_\chi F_\pi)^2
\times (g_\pi)^n \ ,
\end{equation}
where $d$, $p$, $f=2k$, $k$ and $n$ are positive integers and
\begin{equation}
\label{eq:gpi}
g_\pi\sim {G_F F_\pi^2\over 2\sqrt{2}} \sim 3.8\times 10^{-8}\ \ \ ,
\end{equation}

In the absence of weak interactions ($n=0$), the LECs scale
with a large mass scale as
$(\Lambda_\chi)^{-\nu}$, where $\nu= d +f/2 -2$ is
the chiral index defined earlier. Hence, one obtains
the ordering of operators in $Q/\Lambda_\chi$ described
earlier. At energies that are small compared with the mass of
$W$ and $Z$ bosons, weak interactions have a strength
given by the Fermi constant
$G_F=1.16639 \times 10^{-5}$ GeV$^{-2}$.
The effective operators they entail are proportional to (powers
of) the Fermi constant times the square of a mass scale.
A natural scale is the pion decay constant, so we assume that
these operators have coefficients of order of
$G_F F_\pi^2 \sim 10^{-7}$.
In Eq. (\ref{eq:gpi}), we use $g_\pi=3.8\times 10^{-8}$  because
this scale appears naturally in quark-model estimates
as in Ref. \cite{ddh}.

Here we  limit ourselves to $n=1$. Up to two derivatives, then, we
have one  PV $\pi NN$ Yukawa coupling $\hpinn$, three NNLO PV
$\pi NN$ couplings $k_ {\pi NN}^{1a-c}$, ten
short-distance LECS $C_i, {\tilde C}_i$, and one additional
independent PV LEC $\bar C_\pi$ if we consider PV photo-reactions.
As emphasized earlier, only five independent combinations of $C_i$
and ${\tilde C}_i$ are relevant to low-energy PV observables in
few-body systems, while the effects of all but one of the NNLO PV $\pi NN$
operators can be absorbed into other terms in the potential. In
practice, then,  the inclusion of pions leads to
 a total of eight independent LECs. From Eq. (\ref{eq:georgi}), the expected size
 of the
relevant PV couplings is
\begin{eqnarray}
\label{hpiNDA}
\hpinn &\sim& \left(\frac{\Lambda_\chi}{F_\pi}\right) g_\pi\\
\label{CiNDA}
C_i, {\tilde C}_i &\sim&
\left(\frac{\Lambda_\chi}{F_\pi}\right)^2  g_\pi\\
\label{CbarNDA}
k_{\pi NN}^{1a-c},\ {\bar C}_\pi & = & g_\pi \ \ \  .
\end{eqnarray}

The most challenging part of the power counting
is to order the strong-interaction effects.
Here we count powers of $Q$, where as above $Q$ denotes a small
quantity such as
the pion mass $m_\pi$, an external momentum $p$,
or the
electric charge, $e$. For example, the strong $\pi NN$ vertex is
counted as ${\cal O}(Q)$, the PV Yukawa vertex
is ${\cal O}(Q^0)$, the pion propagator is ${\cal O}(Q^{-2})$,
and the four-nucleon vertices proportional to $C_{S,T}$ are also counted as ${\cal
O}(Q^0)$.

\begin{figure}
\begin{center}
\epsfig{file=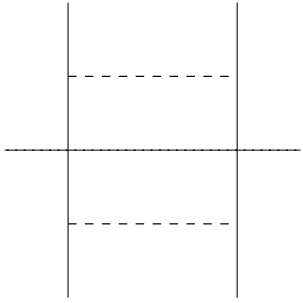,height=4cm,width=4cm}
\caption{Parity-conserving iterated one-pion-exchange diagram.
A solid (dashed) line
represents a nucleon (pion).
The dotted line
indicates the cut line
which picks out the two-nucleon intermediate state.}
\label{fig2}
\end{center}
\end{figure}

In the one-nucleon system, a loop integral $\int d^4 k$ can be
simply counted as ${\cal O}(Q^4)$. If there are two or more
nucleons, this naive counting breaks down. The reason is that
within nuclei nucleons are nearly on-shell. Thus, instead of being
${\cal O}(Q)$, the $q^0$ component of the pion four-momentum in
the one-pion-exchange (OPE) diagram shown in Fig. \ref{fig2} is
${\cal O}(Q^2)$, since
\begin{equation}
q^0=p^0_f-p^0_i\simeq {{\vec p}^2_f-{\vec p}^2_i\over 2m_N},
\end{equation}
where $i, f$ label initial and final states.
This simply means that in first approximation OPE is static.
Now consider the loop diagram generated by the exchange of two pions
between two nucleons,
and focus on
the $dq^0$ integral, which is, schematically,
\begin{eqnarray}
&&\int \frac{d q^0}{2\pi}
{i\over E/2 +q^0 -{\vec q}^2/2m_N+i\epsilon }{i\over E/2
-q^0
-{\vec q}^2/2m_N+i\epsilon}
\left({i\over (q^0)^2 -{\vec q}^2 -m_\pi^2+i\epsilon}\right)^2 \nonumber\\
& & \sim {i\over E-{\vec q}^2/m_N+i\epsilon}
\left({1\over {\vec q}^2 +m_\pi^2}\right)^2
+\ldots,
\end{eqnarray}
where $E\sim {\cal O}(p^2/m_N)$ is the nucleon kinetic energy.
The ``$\ldots$'' are contributions from the pion poles,
which scale according to naive power counting, and other small
terms.
Yet, the term shown explicitly, stemming from the nucleon pole,
represents an ${\cal O}(m_N/Q)$ enhancement
over naive counting.

This enhancement is more general than the specific diagram considered
above.
It is present in any diagram that represents a time ordering
displaying an intermediate state with nucleons only.
Such an intermediate state differs from the initial state only
by a difference of nucleon kinetic energies,
which is small because of the heavy nature of the nucleons.
This type of intermediate state already appeared in the pionless EFT,
and led to the resummation (\ref{pionlessresum}),
which is equivalent to unitarization of the potential,
{\it i.e.} to the solution of the Schr\"odinger equation.

To carry out the resummation in the presence of explicit pions,
two approaches have been
proposed,
which differ in the treatment of pion effects
relative to the contact interactions.
In the simplest approach \cite{KSW},
pion interactions are assumed to be small compared to
the non-derivative contact interactions,
and only the latter are resummed.
Unfortunately, this assumption does not converge for all $NN$ channels
at momenta of the order of the pion mass \cite{FMS}.
In the other approach \cite{wei},
non-derivative contact interactions are assumed to be
comparable to OPE, and both interactions are resummed.
In its original form, Weinberg's approach was proposed
as an expansion of the potential.
This approach appears to be successful
in accounting for
a broad array of nuclear observables \cite{eft},
but it, too, has problems:
iteration of the chiral-symmetry-breaking piece of
OPE leads to inconsistent renormalization
\cite{KSW,towards}.

Progress has been made recently in the understanding
of the power counting relevant for $NN$ scattering at $Q\sim m_\pi$
\cite{towards}.
If an expansion is made around the chiral limit, the aforementioned problems are
 in principle resolved,
 and one obtains an  expansion that is both consistent
and converges.
More work is necessary to test the new power counting,
but at this stage we can see the reason for its success.
The iteration of OPE in the chiral limit, together with the
non-derivative contact interactions, makes
the $NN$ amplitude numerically similar to Weinberg's original
proposal.
Therefore, while unnecessarily resumming higher-order terms,
Weinberg's power counting can still be used to organize
the potential.

With this scheme, we separate
Feynman diagrams into two classes: two-particle reducible (2PR) and
two-particle irreducible (2PI). Only 2PR diagrams lead to the anomalous
enhancement
factor after loop integration discussed above. The 2PI diagrams, in
contrast, do not contain
shallow poles, so they have the same power counting as the
one-nucleon system.
With this classification in hand,  one can use effective field theory
to organize
the calculation order by order. The sum of 2PI diagrams yields the
potential,
which
is the kernel for the Lippman-Schwinger (LS) equation. Through
iterations
the 2PR
diagrams are generated. Solving the LS equation, or equivalently
the Schr\"odinger equation, one arrives at the amplitude
from which scattering can be calculated, and whose poles are the
$NN$ bound states.

In this work we will follow Weinberg's formalism and derive the PV
$NN$ potential up to ${\cal O}(Q)$.
Only the 2PI PV diagrams are included
in the PV potential. All 2PR diagrams can be generated when the PV
potential
is inserted in the LS equation. In practice, the PV potential is much
smaller than the strong potential so it can be treated as a perturbation.
One can treat it as a PV operator and calculate the PV matrix element
using the wave function from LS equation with the strong potential.
The connection with the expansion of Ref. \cite{towards}
is easily made by further expanding in powers of $m_\pi^2$.

%
%
%
%

\subsection{The PV $NN$ Potential}

Using the above power counting we construct the PV potential,
classifying terms according to their size.
We truncate the chiral expansion of PV potential
at ${\cal O}(Q)$, although
the procedure can be carried out to higher orders in similar fashion.
The PC potential has been derived to ${\cal O}(Q^4)$ in Ref. \cite{ray}.

\begin{figure}
\begin{center}
\epsfig{file=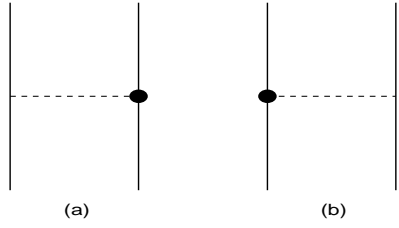,height=5cm,width=6cm}
\vspace{-1.75cm}
\caption{OPE diagram that contributes to the
long-range part of the PV potential.
The filled circle indicates the PV $\pi NN$ Yukawa coupling.}
\label{fig3}
\end{center}
\end{figure}

At ${\cal O}(Q^{-1})$, the only contribution comes from OPE
diagrams of Fig. \ref{fig3}, where the PV vertex is the LO Yukawa
interaction and the strong vertex arises from the operator in Eq.
(\ref{pc1}). These diagrams give rise to a long-range potential,
$V^{\rm PV}_{\rm LR}(\vec{k})$:
\begin{equation}\label{long}
V^{PV}_{(-1, LR)} ({\vec k})= -{g_A \hpinn\over 2\sqrt{2} F_\pi}
i[{\vec\tau}_1\times {\vec \tau}_2]_3 { \left( {\vec \sigma}_1
+{\vec \sigma}_2\right)\cdot {\vec k} \over {\vec k}^2 +m_\pi^2}
\end{equation}
where the $-1$ subscript denotes the chiral index of the corresponding
amplitude and where $k=p_1-p_1'=p_2'-p_2$.

Subleading corrections arise from several sources. First, there
are corrections to the long-range potential from corrections at
the PC vertex (see Fig. \ref{fig10}).  As discussed in Appendix B,
the corrections involving ${\hat d}_{16,18,19}$ amount to a
renormalization of the bare coupling $g_A^0$ while the term
containing ${\hat d}_{17}$ does not contribute. The remaining
terms in Eqs. (\ref{pc2},\ref{pc3}) are proportional to $g_A$ and
do not introduce any new unknown constants into the PV potential.
Since their contributions are discussed in Appendix B, we do not
reproduce them here.

\begin{figure}
\begin{center}
\epsfig{file=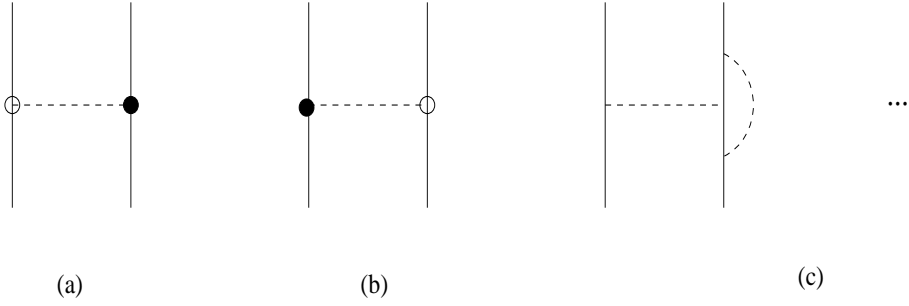,height=4cm,width=12cm} \caption{Corrections to
the long-range PV $NN$ potential from insertions of
(a,b) higher-order PC $\pi NN$ terms, which are denoted by the unfilled circle,
and (c) loops.}
\label{fig10}
\end{center}
\end{figure}

\begin{figure}
\begin{center}
\epsfig{file=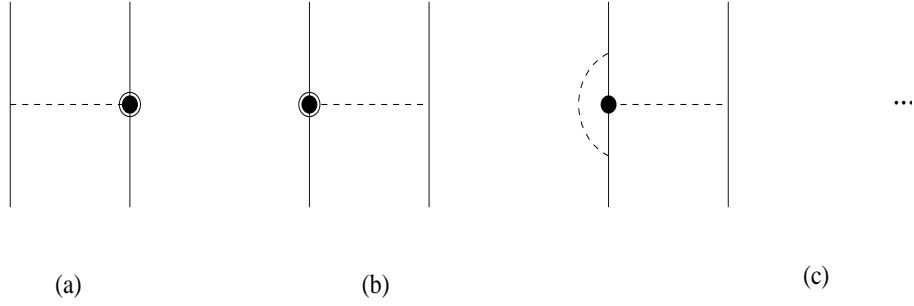,height=4cm,width=12cm}
\caption{Corrections to
the long-range PV $NN$ potential from insertions of
(a,b) higher-order PV $\pi NN$ terms, which are denoted by the
circled filled circle,
and (c) loops.}
\label{fig10-prime}
\end{center}
\end{figure}

Qualitatively new corrections arise at ${\cal O}(Q)$ from long-,
medium-,  and short-range effects, $V^{\rm PV}_{1,\ \rm LR}$,
$V^{\rm PV}_{1,\ \rm  MR}$ and $V^{\rm PV}_{1,\ \rm SR}$,
respectively. The NNLO long-range contributions arise from
inserting the operators in Eq. (\ref{eq:nnlolropa}) in the OPE
diagrams (see Fig. \ref{fig10-prime}). As noted in above, the
effects of the operators proportional to $k_{\pi NN}^{1b,c}$ can
be absorbed in the potential through a suitable redefinition of
$\hpinn$ and $C_6$. The momentum space form associated with the
remaining operator is
\begin{equation}
\label{eq:vpvlrnnlomom}
V^{\rm PV}_{1,\ \rm LR}({\vec p}_1, \cdots, {\vec p}_2^\prime)  =
{ g_{A}k_{\pi NN}^{1a}\over \Lambda_\chi F_\pi^2}\left({{\vec \tau}_1\times
{\vec\tau}_2\over 2}\right)_3\Biggl[
 {{\vec\sigma}_1\cdot{\vec p}_1^\prime\times{\vec p}_1 {\vec\sigma}_2\cdot
{\vec q}_1\over {\vec q}_1^2+m_\pi^2}+(1\leftrightarrow 2)\Biggr]+\ldots \ ,
\end{equation}
where ${\vec p}_i$ (${\vec p}_i^\prime$) is the initial (final)
momentum of the $i$th nucleon, ${\vec q}_i={\vec p}_i^\prime
-{\vec p}_i$, and the ``$\ldots$'' denote the ${\cal O}(Q)$
contributions generated by corrections to the strong $\pi NN$
vertex through NNLO [see Eqs. (\ref{pc2},\ref{pc3}) ]. Taking the
Fourier transform of Eq. (\ref{eq:vpvlrnnlomom}) leads, after some
algebra, to the co-ordinate space potential in Eq.
(\ref{eq:vpvlrnnlo}). In a similar way, one may evaluate the
contributions to $V^{\rm PV}_{1,\ \rm LR} $ generated by order
$Q^3$ contributions to the parity conserving $\pi NN$ vertex.

\begin{figure}
\begin{center}
\epsfig{file=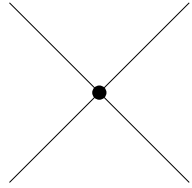,height=3cm,width=3cm}
\caption{PV $NN$ contact interactions that contribute to the PV
short-range potential.}
\label{fig7}
\end{center}
\end{figure}

\begin{figure}
\begin{center}
\epsfig{file=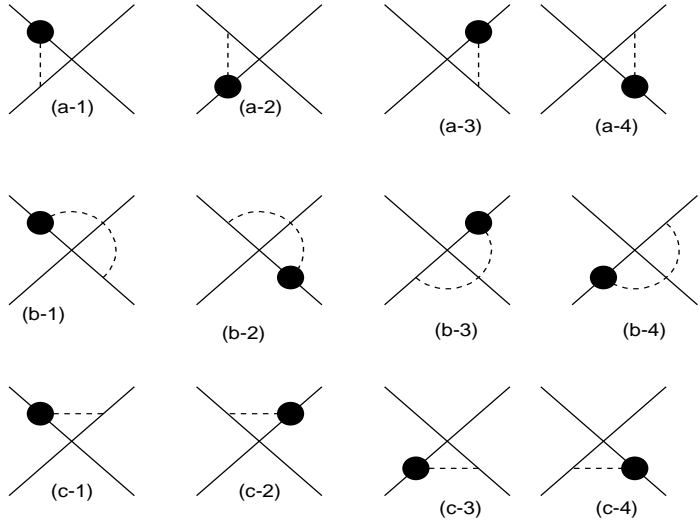,height=8cm,width=10cm}
\caption{Possible PV chiral corrections to PC $NN$ couplings $C_{S,T}$.}
\label{fig9}
\end{center}
\end{figure}

The short-range part $V^{\rm PV}_{\rm SR}$ arises from
\begin{itemize}
\item [i)] the PV $NN$ contact interactions in Fig. \ref{fig7} and
\item [ii)] possible chiral corrections to PC $NN$ operators $C_{S,T}$,
as shown in Fig. \ref{fig9}.

\end{itemize}

The contact interactions have exactly the same form as Eq.
(\ref{3}), so we do not reproduce the expression here. Of course,
the values of the $C_i, {\tilde C_i}$ differ from those in the
pionless theory, where they effectively account for the effects of
low-energy pion exchanges. In principle, one would expect these
couplings to be renormalized by $\pi$ loop effects, as in the case
of $\hpinn$. As we show in Appendix C, however, such loop effects
vanish to ${\cal O}(Q)$. Similarly, PV loop corrections to the
leading-order PC operators---illustrated in
Fig.\ref{fig9}---generate no corrections to the short-range
couplings at this order.

\begin{figure}
\begin{center}
\epsfig{file=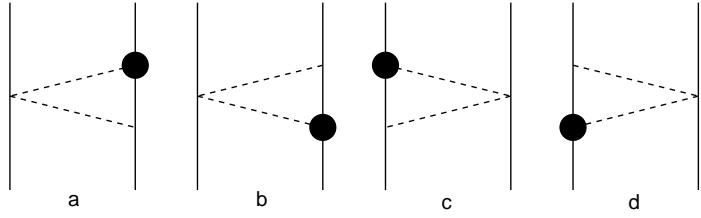,height=5cm,width=10cm}
\vspace{-1.0cm}
\caption{PV TPE triangle diagrams that contribute to the
medium-range PV $NN$ interaction at ${\cal O}(Q)$.}
\label{fig4}
\end{center}
\end{figure}

\begin{figure}
\begin{center}
\epsfig{file=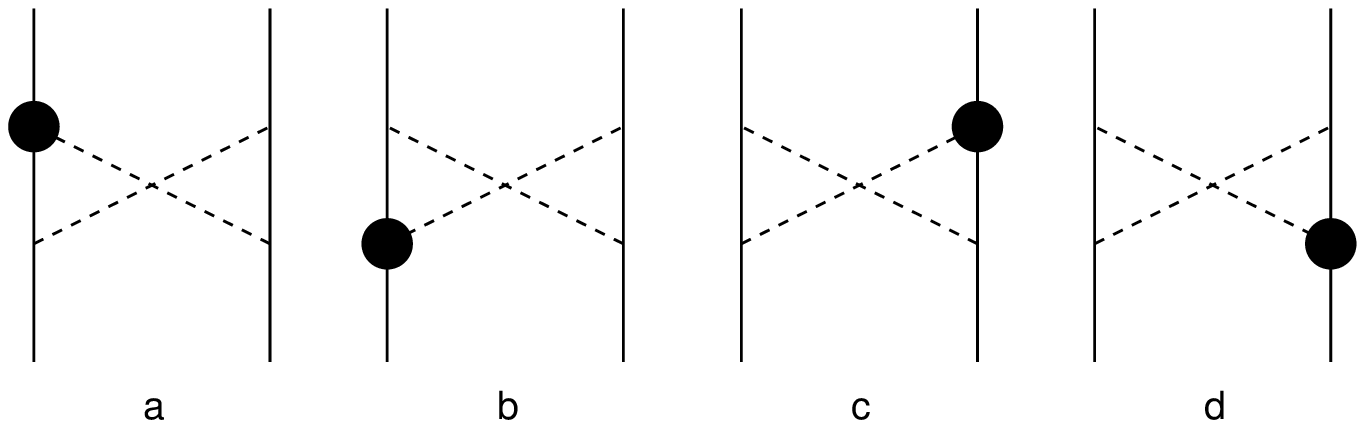,height=5cm,width=10cm}
\vspace{-1.0cm}
\caption{PV TPE crossed diagrams that contribute to the
medium-range PV $NN$ interaction at ${\cal O}(Q)$.}
\label{fig5}
\end{center}
\end{figure}

\begin{figure}
\begin{center}
\epsfig{file=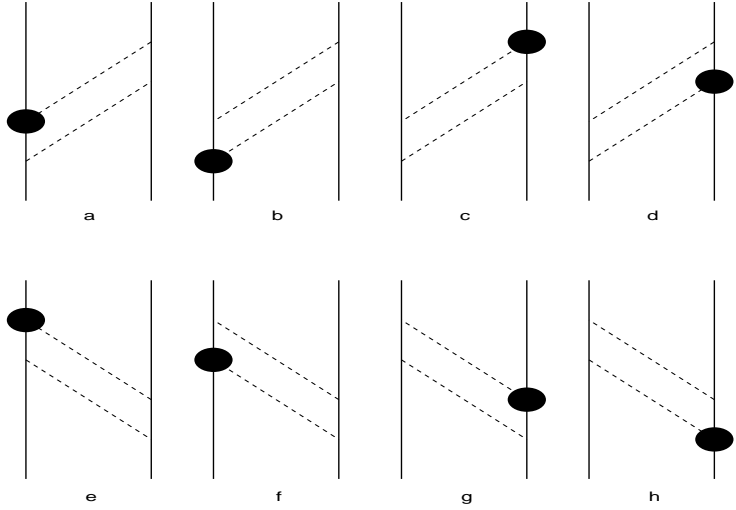,height=9cm,width=10cm}
\vspace{-1.5cm}
\caption{PV TPE box diagrams that contribute to the
medium-range PV $NN$ interaction at ${\cal O}(Q)$.}
\label{fig6}
\end{center}
\end{figure}

The medium-range part $V^{PV}_{MR}$ arises from the
two-pion-exchange (TPE)
diagrams, including
\begin{itemize}
\item [i)] the triangle diagrams in Fig. \ref{fig4},
\item [ii)] the crossed diagrams in Fig. \ref{fig5}, and
\item [iii)] the box diagrams in Fig. \ref{fig6}.
\end{itemize}

The evaluation of these diagrams is somewhat involved, and we give
a detailed discussion in Appendix D. Here, however, we note a few
salient features of the calculation. First, 
the explicit form of the TPE potential is linked to the definition
of OPE and the procedure to subtract the iterated OPE from the box
diagrams. 
The slanted box diagrams are meant here as a representation 
of the full box diagram
with the iterated static OPE subtracted (according to the procedure
explained in Appendix D).
Relativistic corrections (beyond those in OPE)
appear at higher orders. 
Next, 
we regulate the loop
integrals using dimensional regularization. The
regulator-dependence is removed by the appropriate counter-terms,
which in general have the form given in Eq. (\ref{3}). The
remaining, finite parts of the integrals contain terms \lq\lq
regular" -- or polynomial -- in momenta and $m_\pi$ and \lq\lq
irregular", or nonanalytic, terms. The former are
indistinguishable from operators appearing in Eq. (\ref{3}) (and
higher-order parts of the potential), whereas the latter are
uniquely identified with the loop integrals. In principle, one may
choose to retain explicitly any portion of the regular terms and
absorb the remainder into the short-range LECs appearing in Eq.
(\ref{3}). The meaning of the $C_i, {\tilde C}_i$ is, thus,
scheme-dependent. Here, we adopt a scheme in which all of the
regular terms are absorbed into the corresponding $C_i, {\tilde
C}_i$, leaving only the irregular contributions explicitly in
$V^{PV}_{1,\ \rm MR}$:
\begin{eqnarray}\label{v3}\nonumber
V_{(1,\ \rm  MR)}^{PV} ({\vec q})  &=&
-{1\over \Lambda_\chi^3} \left\{
  {\tilde C}^{2\pi}_2(q)  {\tau_1^z +\tau_2^z\over 2}
i \left( {\vec \sigma}_1 \times {\vec \sigma}_2 \right) \cdot {\vec q}
\right.
\nonumber \\
&& \quad
\left.
+ C^{2\pi}_6(q) i\epsilon^{ab3} [{\vec\tau_1} \times{\vec\tau_2}]_3
\left( {\vec \sigma}_1 +{\vec \sigma}_2\right)\cdot {\vec q}
\right\},
\end{eqnarray}
where
\begin{eqnarray}\label{good}\nonumber
& {\tilde C}^{2\pi}_2(q) =4\sqrt{2} \pi g_A^3 \hpinn L(q) \\
& C^{2\pi}_6 (q)= -\sqrt{2}\pi g_A \hpinn L(q) + {3\sqrt{2}\over
2}\pi
\left[ 3L(q) -H (q)\right] g_A^3 h_\pi^1,
\end{eqnarray}
and
\begin{eqnarray}\label{LandH}
L(q) &=& {\sqrt{4m_\pi^2 +{\vec q}^2} \over |{\vec q}|}
\ln \left( {\sqrt{4m_\pi^2 +{\vec q}^2}+|{\vec q}| \over 2m_\pi}\right),
\nonumber\\\
H(q) &=& {4m_\pi^2 \over 4m_\pi^2 +{\vec q}^2} L(q).
\end{eqnarray}

Thus, the PV TPE amplitudes produce contributions with the same
spin-isospin structure as the contact interactions ${\tilde C}_2,
C_6$. In fact, since the regulator-dependent and regular parts of
the amplitudes can be absorbed into $V^{\rm PV}_{\rm SR}$, we would not
expect any new spin-isospin dependence to emerge from the
divergent TPE amplitudes. The spatial-dependence of the finite,
non-analytic part, however, is qualitatively different. We discuss
this difference below.

Finally, we observe that there is no PV three-nucleon
force to ${\cal O}(Q)$. In connecting a third nucleon via a
pion-exchange interaction, one increases the order of a given
diagram by the same amount as if one added an additional loop.
Consequently, the ingredients given in Section 5.1 allow at ${\cal
O}(Q)$ only tree-level three-nucleon diagrams that involve the
leading
 order PC vertices
and the PV Yukawa coupling. However, these diagrams cancel against recoil terms
in the
iteration of the two-nucleon potential.
In fact, the situation here is analogous
 to the ${\cal O}(Q^2)$
PC three-nucleon force, where a similar cancellation
occurs\cite{wei,3Npot}. As a result, if one employs an
energy-independent potential (as is usually more convenient in
few-body calculations), one may omit these three-nucleon diagrams.
Non-trivial three-nucleon PV effects should appear only at ${\cal
O}(Q^2)$, which is beyond the order of our truncation here.

\subsection{EFT PV Potential: Qualitative Features}

As shown above, the PV $NN$ potential to ${\cal O}(Q)$  is given
by Eqs. (\ref{eq:sht},\ref{long},\ref{eq:vpvlrnnlomom},\ref{v3}). The
corresponding coordinate-space $V^{\rm PV}(r)$ can be obtained
straightforwardly by taking the Fourier transform of these
expressions. On the basis of the power counting, one would expect
the OPE potential $V^{\rm PV}_{-1,\ \rm LR}$ to dominate in those
channels where it contributes, unless $\hpinn$ is anomalously
small compared with the NDA estimate in Eq. (\ref{hpiNDA}). This
potential is, of course, not new \cite{ope}. Several contributions
arise with chiral index $\nu=1$. Although they are all formally of
the same order in power-counting, their effects may nevertheless
be distinct due to the different operator structures and spatial
ranges.  The SR potential has already been discussed extensively
in the treatment of the pionless EFT. Qualitatively, the only
impact on the SR potential of including the pion as an explicit
degree is that the numerical values of relevant combinations of
the $C_i$ and ${\tilde C}_i$ will differ for the theory with
pions.

The two-pion exchange contribution $V^{\rm PV}_{1,\ \rm MR}$ also appears at
${\cal O}(Q)$.
 The result in Eq. (\ref{v3}) appears to be the first analytic expression
for the PV TPE potential that is model-independent and consistent
with the symmetries of QCD. Although studies of PV TPE effects
have appeared previously in the literature (see, {\it e.g.}, Ref
\cite{pir}), direct comparison with our treatment is difficult.
First, we have not been able to find an analytic expression in the
literature. Second,  two terms in the PV TPE amplitudes that
depend strongly on the cutoff would have appeared explicitly had
we not used dimensional regularization. This regulator, or cutoff,
dependence requires inclusion of short-range counter-terms in
order to guarantee that physical observables are
regulator-independent. In the analysis of Ref. \cite{pir},
however, no mention is made of the counter-terms, and we suspect
that the corresponding TPE potential is not cutoff-independent.
Third, the component of the TPE amplitude unique to the loop
diagrams is determined by chiral symmetry, and it is notoriously
difficult to maintain this symmetry without using $\chi$PT (see
Ref. \cite{3Npot} for an illustrative example in the parity
conserving three-nucleon sector).
The
situation for PV interactions closely mirrors the developments in
the PC TPE $NN$ potential, whose first derivation in accordance
with chiral symmetry was given within EFT \cite{ray}, and whose
form was recently clearly identified in a phase-shift analysis of
$NN$ data \cite{nijmegen}.

\begin{figure}
\begin{center}
\epsfig{file=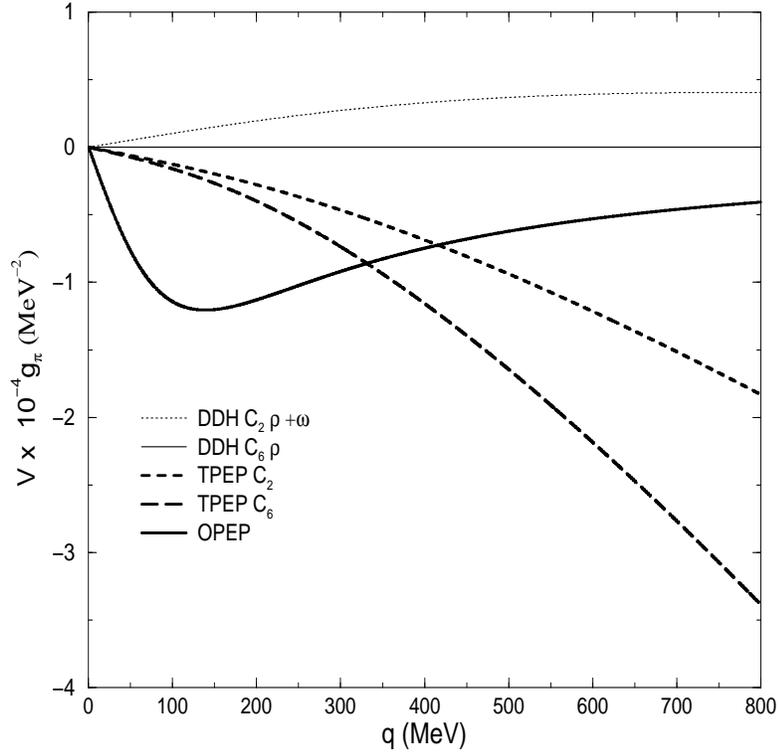,height=12cm,width=12cm}
\vspace{-0.75cm} \caption{Components of the PV $NN$ potential (in
units of $g_\pi 10^{-4}$ MeV$^{-2}$) as function of the momentum
transferred (in MeV): OPE (thick solid line); $C_6$ component of
TPE (long-dash line); ${\tilde C}_2$ component of TPE (short-dash
line); $C_6$ component of DDH (thin solid line); ${\tilde C}_2$
component of DDH (dotted line).} \label{TPEplot}
\end{center}
\end{figure}

The PV TPE contributes two spin-isospin operators.
One,
\begin{equation}
O_6 = i[{\vec\tau_1}\times{\vec\tau_2}]_3
\left( {\vec \sigma}_1 +{\vec \sigma}_2\right)\cdot {\hat q},
\label{op6}
\end{equation}
appears also in $V^{\rm PV}_{-1,\ \rm LR}$,
but $C^{2\pi, Loop}_6 (q)$ is not a simple Yukawa function.
The structure is also the same
as the $h_\rho^{1'}$ term in the DDH potential, where it
is usually neglected.
The other spin-isospin structure,
\begin{equation}
{\tilde O}_2= {\tau_1^z +\tau_2^z\over 2}
i \left( {\vec \sigma}_1 \times {\vec \sigma}_2 \right) \cdot {\hat q},
\label{op2}
\end{equation}
has the structure of the $h_\omega^1$-term in the DDH potential.
In Fig. \ref{TPEplot} we plot the momentum-dependence of the coefficients
of the operators
$O_6$ [Eq. (\ref{op6})] and ${\tilde O}_2$ [Eq. (\ref{op2})] for
$V^{\rm PV}_{1,\ \rm MR}$, in comparison with
the corresponding components of $V^{\rm PV}_{-1, {\rm LR}}$
and the DDH potential using DDH best values from Table \ref{tab0}.

As expected on the basis of power counting,
the OPE potential gives the largest effect for  $q\sim m_\pi$.
As $q$ increases, the TPE potential grows and eventually overcomes OPE.
This feature can be understood simply from the more singular nature of TPE:
while $V^{\rm PV}_{-1,\ \rm LR}$ scales as $q^{-1}$ at large $q$ (or $r^{-2}$ at
 small $r$),
$V^{\rm PV}_{1,\ \rm MR}$ scales as $q^1$ (or $r^{-4}$).
In comparison with isovector $\omega$-exchange term in DDH, the ${\tilde O}_2$
component of  $V^{\rm PV}_{1,\ \rm MR}$ has qualitatively similar behavior at
low-$q$ (up
 to an
overall phase).
The rise with $q$ is more rapid, however, indicating a longer effective
range than for
$\omega$-exchange.
As pointed out above, the $O_6$ component at
distances $r\saprox 1/m_\pi$ is missing in DDH,
while it is not particularly small in $V^{\rm PV}_{1,\ \rm MR}$.
This component will generate an additional energy-dependence
in the same channels OPE contributes.
Presumably, the conventional practice of neglecting the TPE component
leads to inconsistency in the analysis of experiments that
probe the ${\tilde O}_6$ operator at different scales.
We see no theoretical justification to neglect TPE.

It may, perhaps, be surprising that the TPE contributions to
${\tilde O}_{2,6}$ become numerically non-negligible compared to
the OPE effect at relatively low-momentum. For example, when $q$
is of the order of typical Fermi momentum for nuclei ($\sim 200$
MeV), the TPE contribution to ${\tilde O}_6$ is roughly one third
the OPE contribution. One may wonder, therefore, whether the EFT
converges too slowly to justify truncation at ${\cal O}(Q)$. One
should keep in mind, however, that TPE effects always appear in
tandem with short-range components of the same order and that the
latter properly compensate for the most singular part of the TPE
contribution.

New long-range, single pion-exchange terms also
arise at subleading order. The structure of the operator associated with $k_{\pi
NN}^{1a}$  -- shown
in Eq. (\ref{eq:vpvlrnnlo}) -- is distinct from those appearing in
$V^{\rm PV}_{-1,\ \rm LR}$, $V^{\rm PV}_{1,\ \rm MR}$ and $V^{\rm
PV}_{1,\ \rm S R}$ as well as from the operators appearing in the
DDH potential. Additional structures  are
induced by relativistic corrections to the PV $\pi NN$
Yukawa interaction, which are neglected in the DDH approach (see Appendix B). A
consistent power counting, however, requires that one include them
along with the SR and MR operators.


Finally, one might also worry that we have not included $\Delta$ isobar
contributions explicitly since $m_\Delta-m_N$ is comparable to $m_\pi$.
Indeed, in our treatment, $\Delta$ effects are implicit in the LECs.
Had we kept the $\Delta$ as an explicit degree of freedom, it would
contribute to the two-body PV $NN$ interaction solely via loops. Because
the PV $\pi N \Delta$ interaction vertices are of D-wave
character, loops that contain this new PV interaction are generically two orders
 higher than the
corresponding $\pi N$ loops containing the PV Yukawa coupling. Similarly, there
 would also be $\Delta$ contributions
to the renormalization of $h_\pi^1$ appearing in $V^{\rm PV}_{-1,\ \rm
LR}$. Since experiments are sensitive only to the renormalized
Yukawa couplings, the treatment of $\Delta$ loops will only affect
the interpretation of $h_\pi^1$ and not its extraction from
experiment (see the last article in Ref. \cite{cj2}). The only new
contributions from the $\Delta$ to the PV $NN$ interaction would
be in the TPE potential where the $\Delta$ appears between two PC
$\pi N\Delta$ vertices\footnote{In the two-nucleon PV interaction,
these are diagrams analogous to those in Figs. \ref{fig5},
\ref{fig6} but with the $\Delta$ substituted for a nucleon on the
line without a filled circle.}.  There would also be three-nucleon
diagrams that are the PV version of the leading PC three-nucleon
force \cite{3Npot}.  The calculation of these effects is
straightforward, and they introduce no new , {\em a priori}
unknown PV couplings.  We leave the \lq\lq improved" version of
the PV EFT containing these effects for the future when it may be required by
phenomenological considerations.

\subsection{Currents}

\begin{figure}
\begin{center}
\epsfig{file=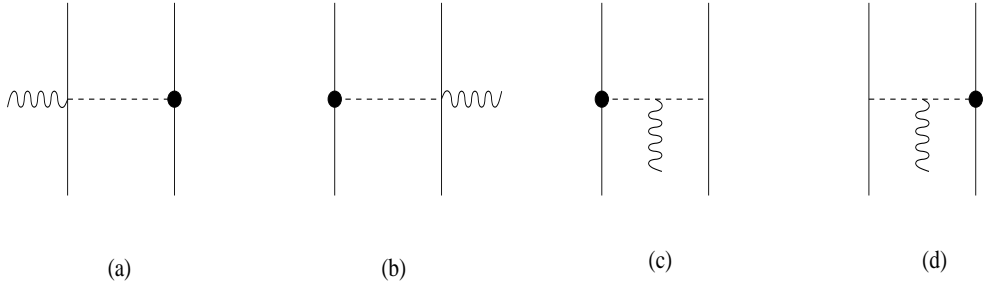,height=3.75cm,width=13cm}
\caption{Long-range PV meson-exchange currents in leading order.
A wavy lines represents a photon.}
\label{fig20}
\end{center}
\end{figure}

\begin{figure}
\begin{center}
\epsfig{file=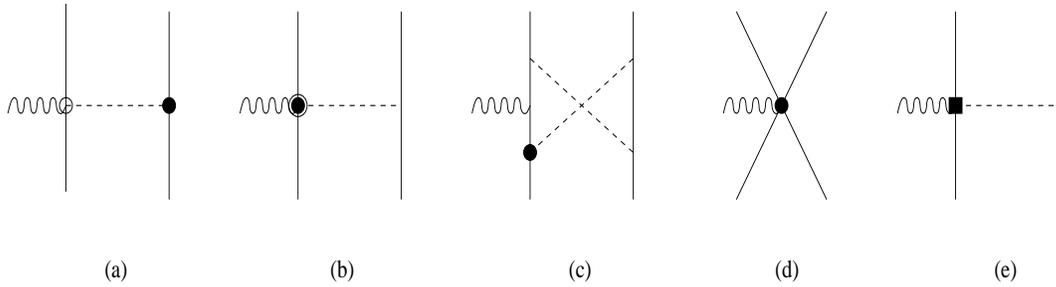,height=3.75cm,width=14cm}
\caption{Corrections to PV meson-exchange currents:
OPE from minimal substitution in the sub-leading (a) PC and (b) PV $\pi NN$
vertices,
(c) TPE,
(d) short-range contribution from minimal substitution in the PV
contact interaction,
and  (e)
OPE from new $\gamma \pi NN$ vertex.
Not all ordering and topologies are displayed.}
\label{fig21}
\end{center}
\end{figure}

As discussed earlier, any experimental program aimed at
determining the PV low-energy constants will likely include
electromagnetic processes. In order to maintain gauge invariance,
one must include the appropriate set of meson-exchange-current
operators. Typically in nuclear physics, one expresses
 the requirements of gauge invariance
through the continuity equation
\begin{equation}
\label{eq:cont}
{\vec\nabla}\cdot {\vec J} = [{\hat H}, \rho],
\end{equation}
where $J^\mu=(\rho, {\vec J})$.
For the long- and medium-range components of the
 potential, a minimal set
of current operators satisfying Eq. (\ref{eq:cont})
can be obtained by inserting
 the photon on all charged lines
in one- and two-pion-exchange diagrams. The meson exchange
current (MEC) operator corresponding to $V^{\rm PV}_{-1,\ \rm LR}$,
Fig. \ref{fig20},
is given in Ref. \cite{ana}.
The operators associated with $V^{\rm PV}_{1,\ \rm LR}$ and
$V^{\rm PV}_{1,\ \rm MR}$
are more involved [see Fig. \ref{fig21}(a-d)].
In particular, construction of the MEC operator associated with
$V^{\rm PV}_{1,\ \rm MR}$ is
technically arduous,  as one must evaluate a large number of Feynman diagrams
---a task which goes beyond the scope
of the present study. Thus, we defer a derivation of these
MEC operators to a future publication.

The foregoing set of MECs constitute a minimal, model-independent set required
to ensure
that Eq. (\ref{eq:cont}) is satisfied.
In addition, one may consider MECs that
 satisfy Eq. (\ref{eq:cont})
independently from the terms in the potential.
At ${\cal O}(Q)$, we find that there exists one such MEC that is
 not determined from $V^{\rm PV}$ by gauge invariance.
This operator is obtained by OPE
with an insertion of the operator
from Eq. (\ref{pv2})  [Fig.  \ref{fig21}(e)], leading to
the momentum space two-body current
\begin{equation}
\label{eq:mecnewcoord} {\vec J}  =  -i\left[{\sqrt{2} g_{\pi NN}
{\bar C}_\pi\over m_N \Lambda_\chi F_\pi}\right]\tau_1^+
{{\vec\sigma}_2\cdot{\vec q}_2 \ ({\vec q}_1+{\vec
q}_2)\times{\vec\sigma}_1\over {\vec q}_2^2+ m_\pi^2}+
(1\leftrightarrow 2)\ \ \
\end{equation}
and to Eq. (\ref{eq:mecnew}) in co-ordinate space.

\section{Short-Distance Archeology: Correspondence with DDH and Beyond}

\label{models}

In the ideal situation, a systematic EFT treatment would use
experimental low energy measurements in order to determine the
counter-terms $\lambda_t$, $\lambda_s^{0,1,2}$, $\rho_t$,
$\hpinn$, $k_{\pi NN}^{1a}$, and ${\bar C}_\pi$ entirely from
data. As emphasized earlier, there exists in principle a program
of low-energy few-body measurements which will yield at least five
linear combinations of these constants. Alternately, one would
ultimately hope to gain a theoretical understanding of the values
of these constants (and their linear combinations) probed by
experiment. However, obtaining reliable theoretical predictions is
complicated, since the PV $NN$ interaction involves a non-trivial
interplay of weak and non-perturbative strong interactions.
Indeed, carrying out a first-principles calculation of the PV LECs
is not yet possible, since lattice QCD techniques are not yet
sufficiently advanced to address this problem. Consequently, in
order to say anything about the LECs beyond NDA estimates,
theorists have of necessity relied on model approaches. In this
section we illustrate how the PV LECs can in principle be
estimated from details of the short-range dynamics.

Before proceeding further, it is useful to comment on the
correspondence with, and difference from, the conventional DDH formalism
in the treatment of short-distance PV physics.
\begin{itemize}

\item [(i)] The EFT approach is systematic and model-independent.
No assumption is made about the dynamics underlying the
short-range interactions in the EFT, whereas the
DDH formalism relies on a light pseudoscalar- and
vector-meson-exchange
picture as indicated in Fig. \ref{fig:DDH}.

\item [(ii)] The LECs $C_{1-5}$ have a straightforward correspondence
with the DDH PV meson-nucleon couplings $h_\omega^{0,1},
h_\rho^{0,1,2}$. In the EFT framework, however, $C_{1-5}$ could deviate
strongly from the DDH values, as we illustrate below.

\item [(iii)] In terms of the DDH meson-exchange language we have the
constraints
\begin{equation}
{ {\tilde C}_1^{DDH} \over C_1^{DDH}}= { {\tilde C}_2^{DDH} \over
C_2^{DDH}}= 1+\chi_\omega,
\end{equation}
\begin{equation}
{ {\tilde C}_3^{DDH} \over C_3^{DDH}}= { {\tilde C}_4^{DDH} \over
C_4^{DDH}}={ {\tilde
C}_5^{DDH}
\over C_5^{DDH}}=
 1+\chi_\rho,
\end{equation}
where $\chi_{\rho, \omega}$ denotes the ratio between tensor and vector
couplings of $\rho ,\omega$ meson-nucleon interaction.
In our EFT approach, however, ${\tilde C}_{1-5}$ constitute five
LECs whose values need not be related to $C_{1-5}$ as in
the DDH picture.

\item [(iv)] The DDH parameter $h_\rho^{ {\prime} 1}$ is generally
discarded
since its ``best value'' is tiny. In EFT, on the
other hand, $h_\rho^{'1}$ contributes to $C_6$, but
$C_6$ need not be small since it can receive a contribution, {\it
e.g.}, from $a_0$ meson exchange. Moreover, although the operator
accompanying $C_6$ has the same spin-isospin structure as PV pion
exchange,
these interactions have different ranges and may in principle  be
distinguished as long as a sufficient range of energies
is probed.

\end{itemize}

\subsection{Resonance saturation}

One popular model approach ---which we adopt here for purely
illustrative purposes--- assumes that the short-distance dynamics
is governed by the exchange of light meson resonances. This \lq\lq
resonance saturation" approach has some theoretical justification
from the standpoint of the large-$N_c$ expansion, where $N_c$
denotes the number of colors in QCD \cite{largeN}. It is also
supported by several phenomenological studies. It is well known,
for example, that in the ${\cal O}(Q^4)$ chiral Lagrangian
describing pseudoscalar interactions, the low-energy constants are
well-described by the exchange of heavy mesons \cite{egpr}. In
particular, the charge radius of the pion receives roughly a 7\%
long-distance loop contribution, while the remaining 93\% is
saturated by $t$-channel exchange of the $\rho^0$. Similarly, in
the baryon sector, dispersion-relation analyses of the isovector
and isoscalar nucleon electromagnetic form factors indicate
important contributions from the lightest vector mesons
\cite{Hoh76}. Finally, the primary features of the $NN$ PC
potential seem to be well described in such a picture
\cite{Stoks}.  Thus, it seems reasonable to assume that
low-lying-meson exchange may play an important role in the
short-distance physics associated with the PV LECs.

With these observations in mind, we invoke resonance saturation to
arrive at illustrative estimates for the PV LECs. The relevant
Feynman diagram is the same as in Fig. \ref{fig:DDH}, where the
exchanged bosons include all possible heavy mesons with
appropriate quantum numbers.  Here, parity violation enters
through one of the  meson-nucleon interaction vertices. While the
DDH framework includes only the lowest-lying vector mesons to
describe the short-distance PV $NN$ interaction, we also consider
the exchange of $a_0(980)$, $a_1(1260)$ and $f_1(1285)$, as well
as the radial excitations of these systems.  Of course, the PV
LECs receive additional contributions from higher resonances,
correlated meson exchange, {\em etc}. However, we limit our
consideration to this set, as it already suffices to illustrate to
what extent the short-distance PV $NN$ interaction can differ from
the predictions of the DDH model.

In order to estimate specific values of LECs in the framework of
the meson-exchange model we require the corresponding PC and PV
meson-nucleon Lagrangians:

\begin{itemize}
\item Vector-meson exchange

The parity-conserving vector-meson-nucleon interaction Lagrangian reads
\begin{eqnarray}
{\cal L}^{PC}_{\rho NN}& =& g_{\rho NN} \bar N
\left[\gamma_\mu +{\chi_\rho \over 2m_N} i\sigma_{\mu\nu} q^\nu\right]
{\tau \cdot \rho^\mu} N, \\
{\cal L}^{PC}_{\omega NN} &=&g_{\omega NN} \bar N
\left[\gamma_\mu +{\chi_\omega \over 2m_N}i\sigma_{\mu\nu} q^\nu\right]
 \omega^\mu N, \\
{\cal L}^{PC}_{\phi NN} &=&g_{\phi NN} \bar N
\left[\gamma_\mu +{\chi_\phi \over 2m_N}i\sigma_{\mu\nu} q^\nu\right]
 \phi^\mu N.
\end{eqnarray}

The parity-violating vector-meson-nucleon ($VNN$)
interaction Lagrangian is given
in Ref. \cite{ddh}:
\begin{eqnarray} \nonumber
{\cal L}^{PV}_{\rho NN} &=& \bar N
\gamma^\mu \gamma_5  \left[ h^0_{\rho }\tau\cdot \rho_\mu
+h^1_{\rho }\rho^0_\mu +
{h^2_{\rho }\over 2\sqrt{6}} \left(3\tau_3\rho_\mu^0 -\tau\cdot
\rho_\mu\right) \right] N  \\
&&-{h_\rho^{\prime 1} \over 2m_N}
\bar N \left( {\vec \tau}\times {\vec \rho}_\mu\right)_3
\sigma^{\mu\nu}q_\nu \gamma_5 N, \\
{\cal L}^{PV}_{\omega NN} &=& \bar N
\gamma^\mu \gamma_5 \omega_\mu
\left[ h^0_{\omega }+h^1_{\omega }\tau_3\right] N, \\
{\cal L}^{PV}_{\phi NN} &=& \bar N
\gamma^\mu \gamma_5 \phi_\mu \left[ h^0_{\phi } +h^1_{\phi
}\tau_3\right] N.
\end{eqnarray}
Note that we have adopted the convention for $\gamma_5$ following
Ref. \cite{BjD}, which is {\it different} from that used in
Ref. \cite{ddh}.

To our knowledge, the following contributions to the PV $NN$
short-distance interaction have not been discussed elsewhere in
the literature.

\item $a_0 (980)$-meson exchange
\begin{eqnarray}
{\cal L}^{PC}_{a_0 NN}& =& g_{a_0 NN} \bar N
{\tau \cdot a_0} N,
\end{eqnarray}
\begin{eqnarray}
{\cal L}^{PV}_{a_0 NN}& =& h_{a_0} \bar N i\gamma_5
\left( {\vec \tau}\times {\vec  a}_0\right)_3 N.
\end{eqnarray}

\item $a_1 (1260)$-meson exchange
\begin{eqnarray}
{\cal L}^{PC}_{a_1 NN}& =& g_{a_1 NN} \bar N \gamma_\mu\gamma_5
{\tau \cdot a^\mu_1} N.
\end{eqnarray}

Note that the structure
$\bar N i \sigma_{\mu\nu}q^\nu \left({\vec\tau}\cdot {\vec a}_1^\mu
\right) \gamma_5 N$ is analogous to the weak-electricity form
factor of nuclear beta decay; it is parity conserving but CP
violating and hence is not included. The PV Lagrangian is
\begin{eqnarray} \nonumber
{\cal L}^{PV}_{a_1 NN} &=& \bar N
\gamma_\mu  \left[ h^0_{a_1 }\tau\cdot a_1^\mu +h^1_{a_1 }{a_1}_0^\mu +
{h^2_{a_1 }\over 2\sqrt{6}} \left(3\tau_3 {a_1}_0^\mu-\tau\cdot
a_1^\mu\right) \right] N \\
&& +
\bar N {i\sigma_{\mu\nu} q^\nu\over 2m_N}
\left[ h^3_{a_1 }\tau\cdot a_1^\mu +h^4_{a_1 }{a_1}_0^\mu +
{h^5_{a_1 }\over 2\sqrt{6}} \left(3\tau_3 {a_1}_0^\mu-\tau\cdot
a_1^\mu\right) \right] N,
\end{eqnarray}
where $a_1^0$ is the neutral component of $a_1$ meson.

\item $f_1 (1285)$-meson exchange
\begin{eqnarray}
{\cal L}^{PC}_{f_1 NN}& =& g_{f_1 NN} \bar N \gamma_\mu\gamma_5 f^\mu_1
N,
\end{eqnarray}
\begin{eqnarray}
{\cal L}^{PV}_{f_1 NN} &=& \bar N
\gamma_\mu  \left[ h^0_{f_1 } f_1^\mu +h^1_{f_1 }{f_1}^\mu \tau^3
\right] N.
\end{eqnarray}
\end{itemize}

In principle, one may also include in such a model exchange of the radial
excitations of $\rho, \omega, \phi, a_0, a_1, f_1$
mesons is also allowed.

\subsection{LECs with DDH framework and beyond}

With these couplings in hand, we can identify our predictions for the
various low-energy constants.

If we consider only vector-meson exchange \`a la DDH, we have
\begin{eqnarray}\nonumber
 {{\tilde C}_{i}^{DDH}\over C_{i}^{DDH}}&=&1+\chi_\omega\ \ \ i=1,2,\\
\nonumber
 {{\tilde C}_{i}^{DDH}\over C_{i}^{DDH}}&=&1+\chi_\rho\ \ \ i=3-5,\\ \nonumber
 C_1^{DDH} &=&-\Lambda_\omega
h_\omega^0,\\ \nonumber
 C_2^{DDH}& =&-\Lambda_\omega
h_\omega^1,\\ \nonumber
 C_3^{DDH} &=&-\Lambda_\rho
h_\rho^0,  \\
\nonumber
 C_4^{DDH} &=&-\Lambda_\rho
h_\rho^1, \\
\nonumber
 C_5^{DDH} &=&{ \Lambda_\rho\over 2\sqrt{6}}
h_\rho^2, \\ \nonumber
 C_6^{DDH}&=& -\Lambda_\rho g_{\rho NN}
h_\rho^{\prime 1}.
\end{eqnarray}

where we have defined

\begin{equation}
\Lambda_M={\Lambda_\chi^3\over 2m_Nm_M^3}
\end{equation}

However, within the context of resonance saturation, these LECs
could also receive contributions from radial excitations of rho
and omega mesons, and from $a_0$-, $a_1$-, $f_1$-meson exchange.
Hence we have, more generally,
\begin{eqnarray}\nonumber
C_{1,2}&=&C_{1,2}^{DDH}+ C_{1,2}^{Radial}+ C_{1,2}^{f_1}, \\
\nonumber {\tilde C}_{1,2} &=&{\tilde C}_{1,2}^{DDH} +{\tilde
C}_{1,2}^{Radial}, \nonumber\\
 C_{3-5} &= &C_{3-5}^{DDH}+
C_{3-5}^{Radial} + C_{3-5}^{a_1},\\ \nonumber {\tilde C}_{3-5}& =&
{\tilde C}_{3-5}^{DDH} +{\tilde C}_{3-5}^{Radial}+{\tilde
C}_{3-5}^{a_1},
\\ \nonumber
C_6 & = & C_6^{DDH}+ C_6^{Radial} +C_6^{a_0},
\end{eqnarray}
where
\begin{eqnarray}\nonumber
 C_1^{f_1}& =&-\Lambda_{f_1} g_{f_1 NN}
h_{f_1}^0,\\
\nonumber C_2^{f_1}& =& - \Lambda_{f_1}g_{f_1 NN}
h_{f_1}^1,\\
\nonumber C_3^{a_1} & = & -\Lambda_{a_1} g_{a_1 NN}
h_{a_1}^0,\\
\nonumber {\tilde C}_3^{a_1} &=&-\Lambda_{a_1} g_{a_1 NN}
\left( h_{a_1}^0 + h_{a_1}^3 \right),\\
\nonumber C_4^{a_1} & =& -\Lambda_{a_1} g_{a_1 NN}
h_{a_1}^1, \\
\nonumber {\tilde C}_4^{a_1} &=& -\Lambda_{a_1} g_{a_1 NN}
\left( h_{a_1}^1 + h_{a_1}^4\right),\\
\nonumber C_5^{a_1}& =&{\Lambda_{a_1}\over 2\sqrt{6}} g_{a_1 NN}
h_{a_1}^2,
\\\label{eq;ressat} {\tilde C}_5^{a_1}& =& -{\Lambda_{a_1}\over 2\sqrt{6}}
g_{a_1 NN}
\left( h_{a_1}^2 + h_{a_1}^5\right),\\
\nonumber C_6^{a_0} &=& -\Lambda_{a_0} h_{a_0}.
\end{eqnarray}
Similar relations will hold for the radial excitations.

\subsection{Estimates for $C_{1-6}, {\tilde C}_{1-5}$}

As noted above, arriving at reliable theoretical predictions for
the PV LECs, even within the context of a model framework, is a
formidable task, and one which certainly goes beyond the scope of
the present work. Nevertheless, it is useful to have in hand
educated guesses for their magnitudes and signs, if for no other
reason than to provide benchmarks for comparison with experiment.
To that end, we quote below both expectations based on naive
dimensional analysis and values obtained from correspondence with
the DDH model. Future work could include, for example, computing
the weak couplings entering Eq. (\ref{eq;ressat}), thereby
providing model estimates for the departures of the $C_i$ and
${\tilde C}_i$ from their NDA or DDH values.

There exist various values for the parity-conserving couplings
$g_{\rho NN}$, $\chi_\rho$, $g_{\omega NN}$, and $\chi_\omega $
quoted in the literature \cite{bonn,speth,zhu5}. Fortunately, the
combination $g_{\rho NN}(1+\chi_\rho)$ takes roughly the same
value in different approaches: $g_{\rho NN}(1+\chi_\rho)\approx
21$. Likewise various approaches consistently yield a very small
value for $\chi_\omega$. It is thus reasonable to use the values $
\chi_\rho=6$, $g_{\rho NN}=3$ or $ \chi_\rho=3.7$, $g_{\rho
NN}=4.5$. A word of caution is in order here. The proper
accounting of chiral symmetry in multi-pion contributions might
affect the extractions of strong couplings. For example, the
effect of $\omega$ exchange in $NN$ scattering is significantly
reduced when correct TPE is considered \cite{nijmegen}. As we
emphasized earlier, the estimates here should be considered to
yield only an educated guess for the order of magnitude of the
LECs. The theoretical uncertainty from this exchange model is much
larger than the choice of $ \chi_\rho$ and $g_{\rho NN}$. Here, we
simply use $ \chi_\rho=3.7$, $g_{\rho NN}=4.5$,
$\chi_\omega=-0.12$, and $g_{\omega NN}=14$ to make our best
guess.   Results are given in Table \ref{tab2} and should be used with due
caution.

\begin{table}
\begin{center}~
\begin{tabular}{|c||c|c|c|}\hline
\hbox{LECs} & Naive Dimensional Analysis &\hbox{Best
values}&\hbox{Range}
\\\hline\hline
$C_1$  & $\pm 158$ & 32 & $-95 \to 172$ \\
${\tilde C}_1$  &$\pm 158$ & 28 & $-84\to 151$ \\
$C_2$  & $\pm 158$ & 17 & $ 13 \to 32$ \\
${\tilde C}_2$  &$\pm 158$ & 15 & $ 11\to 28$ \\
$C_3$  &$\pm 158$ & 63 & $ -63\to 171$ \\
${\tilde C}_3$ &$\pm 158$  & 296 & $-296\to 803$ \\
$C_4$  & $\pm 158$ & 95 & $-289 \to 520$ \\
${\tilde C}_4$  &$\pm 158$ & 1 & $0\to 1$ \\
$C_5$  &$\pm 158$  & $-11$ & $-13 \to -8 $ \\
${\tilde C}_5$ &$\pm 158$  & $-51$  & $ -61 \to -28$ \\
$C_6$ & $\pm 158$ & --- & --- \\
\hline
\end{tabular}
\end{center}
\caption{\label{tab2}
Estimates of ranges and best values for PV coupling constants $C_{1-6},
{\tilde
C}_{1-5}$
(in units of $g_\pi= 3.8 \times 10^{-8}$).}
\end{table}

\section{Conclusions}
\label{sec9}

In summary, we have performed a systematic study of the
parity-nonconserving nucleon-nucleon potential, and have suggested
ways by which the present confused experimental situation can be
resolved. We have proposed breaking this program into two separate
pieces:
\begin{itemize}
\item[i)] Since the low-energy parity-violating potential involves
five S-P wave
mixing amplitudes, we have constructed a simple local
effective potential
in order to reliably extract such quantities from
experiments involving only the $NN$, $Nd$, or $N\alpha$ systems.  We
have
also suggested the critical experiments that are needed in order to
successfully complete this task and have given explicit formulas which
will express the mixing amplitudes in terms of experimental
observables. We have also suggested a two-phase experimental program, where
phase one would include six (or possibly seven) measurements needed to test the
consistency of the pionless EFT and phase two would involve additional
measurements needed to determine the pion-related parameters if necessitated by
the results of phase one.

\item[ii)] A second important facet of this
program is to confront the extracted phenomenological potential
with theoretical expectations.  For this task, we have
systematically constructed a parity-violating nucleon-nucleon
potential $V^{\rm PV}(\vec{r})$ within the framework of
effective field theory using the Weinberg counting scheme up to
the order ${\cal O}(Q)$. The correspondence with, and difference
from, the conventional DDH potential were discussed.
\end{itemize}
In order for this scheme to come to fruition additional work
is required on several fronts.  Experimentally it
is critical to complete the key experiments, resulting in a confirmed and
reliable set of low-energy phenomenological parameters.  Once
these parameters are known, it is important to use them in order to analyze
the heavier nuclear systems and resolve the various existing
conflicts. Doing so could have important implications for the
applicability of EFT to other electroweak processes in heavy
nuclei, such as neutrinoless double $\beta$-decay\cite{prezeau}.
Future work is also needed to understand the relation between the
underlying effective weak potential $V^{\rm PV}(\vec{r})$
and the effective phenomenological parameters
$\rho_t,\lambda_t,\lambda_s^i$ and should involve the best
available nuclear wave functions.  What should result from this
program is the resolution of the presently confusing experimental
situation and a reliable form for the parity-violating nuclear
potential, which we hope will set the standard for future work in
this field.

\vspace{0.5cm}
\begin{center}
{\bf Acknowledgments}
\end{center}

We thank Paulo Bedaque, Vincenzo Cirigliano, Xiangdong Ji, Roxanne Springer, and
Mike
Snow for useful discussions. BRH and MJRM are grateful to the
Institute for Nuclear Theory and to the theory group at JLab (BRH)
for hospitality. UvK thanks the Kellogg Radiation Laboratory at
Caltech and the Nuclear Theory Group at the University of
Washington for hospitality, and RIKEN, Brookhaven National
Laboratory and the U.S. Department of Energy [contract
DE-AC02-98CH10886] for providing the facilities essential for the
completion of this work. This work was supported in part by the
National Natural Science Foundation of China under Grant 10375003,
Ministry of Education of China, FANEDD and SRF for ROCS, SEM
(SLZ); by the U.S. National Science Foundation under awards
PHY-0071856 (SLZ, MJRM) and PHY-02-44801 (BRH); by Brazil's
FAPERGS under award PROADE2 02/1266-6 (CMM); by the U.S.
Department of Energy under contracts DE-FG03-02ER41215,
DE-FG03-88ER40397, and DE-FG02-00ER41132 (MJRM) and an Outstanding
Junior Investigator award (UvK); and by the Alfred P. Sloan
Foundation (UvK).

\section*{Appendix A: The PV $NN$ Contact Lagrangian}
\label{sec5}

Since we employ the heavy-fermion formalism, one can build the
most general PV operators by using heavy-baryon fields directly.
This approach, however, yields redundant operators, which then
have to be eliminated by imposing reparameterization invariance.
Alternatively, we can obtain the relevant operators starting from
the relativistic theory, then performing a non-relativistic
expansion. We use $\psi_N$, ${\bar\psi}_N$ for the relativistic
nucleon field and $N$, $N^\dag$ to denote the nucleon field after
non-relativistic reduction. In general, there exist twelve
possible PV and CP conserving $NN$-interaction terms up to one
derivative, which we write as
\begin{eqnarray}\nonumber
{\cal O}_1&=&{g_1\over \Lambda_\chi^2}{\bar \psi}_N 1 \gamma_\mu  \psi_N
{\bar \psi}_N 1 \gamma^\mu\gamma_5 \psi_N\\
\nonumber
{\tilde {\cal O}}_1&=&{{\tilde g}_1\over \Lambda_\chi^3}{\bar \psi}_N 1
i\sigma_{\mu\nu}q^\nu
\psi_N {\bar \psi}_N 1
\gamma^\mu\gamma_5 \psi_N\\
\nonumber
{\cal O}_2&=&{g_2\over \Lambda_\chi^2}{\bar \psi}_N 1 \gamma_\mu
\psi_N {\bar \psi}_N\tau_3 \gamma_\mu \gamma_5\psi_N\\
\nonumber
{\tilde {\cal O}}_2&=&{{\tilde g}_2\over \Lambda_\chi^3}{\bar \psi}_N 1
i\sigma_{\mu\nu}q^\nu
\psi_N {\bar \psi}_N\tau_3 \gamma_\mu \gamma_5\psi_N\\
{\cal O}_3&=&{g_3\over \Lambda_\chi^2}{\bar \psi}_N \tau^a \gamma_\mu
\psi_N {\bar \psi}_N\tau^a \gamma_\mu \gamma_5\psi_N\\
\nonumber
{\tilde {\cal O}}_3&=&{{\tilde g}_3\over \Lambda_\chi^3}{\bar \psi}_N
\tau^a
i\sigma_{\mu\nu}q^\nu
\psi_N {\bar \psi}_N\tau^a \gamma_\mu \gamma_5\psi_N\\
\nonumber
{\cal O}_4&=&{g_4\over \Lambda_\chi^2}{\bar \psi}_N \tau_3 \gamma_\mu
\psi_N {\bar \psi}_N 1 \gamma_\mu \gamma_5\psi_N\\
\nonumber
{\tilde {\cal O}}_4&=&{{\tilde g}_4\over \Lambda_\chi^3}{\bar \psi}_N
\tau_3
i\sigma_{\mu\nu}q^\nu
\psi_N {\bar \psi}_N 1 \gamma_\mu \gamma_5\psi_N\\
\nonumber
{\cal O}_5&=&{g_5\over \Lambda_\chi^2}{\cal I}^{ab}{\bar \psi}_N \tau_a
\gamma_\mu
\psi_N {\bar \psi}_N\tau_b \gamma_\mu \gamma_5\psi_N\\
\nonumber
{\tilde {\cal O}}_5&=&{{\tilde g}_5\over \Lambda_\chi^3}{\cal I}^{ab}{\bar
\psi}_N \tau_a i\sigma_{\mu\nu}q^\nu
\psi_N {\bar \psi}_N\tau_b \gamma_\mu \gamma_5\psi_N \\
\nonumber
{\cal O}_6&=&{g_6\over \Lambda_\chi^2}\epsilon^{ab3}{\bar \psi}_N \tau_a
\psi_N {\bar \psi}_N \tau_b i\gamma_5 \psi_N \\
\nonumber
{\tilde {\cal O}}_6&=&{{\tilde g}_6\over
\Lambda_\chi^3}\epsilon^{ab3}{\bar
\psi}_N \tau_a \gamma_\mu
\psi_N {\bar \psi}_N \tau_b
 i\sigma_{\mu\nu}q^\nu\gamma_5 \psi_N\ \ \ ,
\end{eqnarray}
where ${\cal I}^{ab}$ is defined in Eq. (\ref{eq:chg}).
In writing down these PV operators, we have assumed that all the
isospin violation arises from the weak interaction, thus neglecting
isospin violation from up and down quark mass
difference and electromagnetic interactions, since corrections from
such effects are typically around a few percent and negligible for our
purposes.  The isospin content of the above terms is transparent:
the $1\cdot 1, \tau\cdot \tau$ terms conserve isospin,
the piece with ${\cal I}^{ab}$ carries $\Delta I=2$, and
all remaining pieces change isospin by one unit.

In order to understand the constraints that relativity imposes, we
consider a simple example ---the expansion of ${\cal O}_1$ and
${\tilde {\cal O}}_1$.  Up to ${\cal O}(Q)$ we have
\begin{eqnarray}\label{o1}\nonumber
{\cal O}_1 =&{g_1\over \Lambda_\chi^2}{1\over 2m_N}[ -N^\dag 1 N N^\dag
1
{\vec \sigma} \cdot i{\vec D}_- N
+ N^\dag 1 iD^i_- N  N^\dag 1 \sigma^i  N \\
& -i\epsilon^{ijk} N^\dag 1 iD_+^i \sigma_j N N^\dag 1 \sigma^k N ]
,
\end{eqnarray}
\begin{eqnarray}\nonumber
{\tilde {\cal O}}_1=-{{\tilde g}_1\over \Lambda_\chi^3}
i\epsilon^{ijk} N^\dag 1 iD_+^i \sigma_j N N^\dag 1 \sigma^k N,
\end{eqnarray}
Note that the {\it two} relativistic structures ${\cal O}_1$ and
${\tilde{\cal O}}_1$ together yield {\it three} distinct
non-relativistic spacetime forms.  However, only two linear
combinations of these forms are independent according to the
strictures of relativity. On the other hand, if we had started
from the non-relativistic theory and tried to write the most
general effective Lagrangian, we would have naively identified
each of these three structures as being independent and would have
mistakenly postulated three, rather than two, LECs.  The
requirements imposed on the independence of various
non-relativistic operators which follow from consistency with the
relativistic theory is known as reparameterization invariance.
Physically, this invariance amounts to stating that the
non-relativistic theory should not contain more physics ({\em
e.g.}, LECs) than the relativistic one. Analogous situations occur
in heavy-quark EFT and in the non-relativistic expansion of the
nucleon kinetic operator in heavy-nucleon EFT.

Similar results follow for the operators ${\cal O}_{2-5}$ and
${\tilde {\cal O}}_{2-5}$ in that each set $\{{\cal O}_i,
{\tilde{\cal O}}_i\}$ generates two {\it independent} combinations
of non-relativistic operators and, thus, two independent LECs.  On
the other hand, after non-relativistic reduction, ${\cal O}_6$ and
${\tilde {\cal O}}_6$ yield exactly the {\it same} form up to
${\cal O}(Q)$.  Hence, these structures yield only {\it one}
independent LEC in the non-relativistic theory even though  there
are two different LECs in the original, relativistic theory. The
new LEC is a linear combination of $g_6$ and ${\tilde g}_6$.

The full PV contact heavy-nucleon Lagrangian at ${\cal O}(Q)$ in
the $NN$ sector can then be written in the form given in Eq.
(\ref{PVcontacts}), where the LECs $C_i$ are related to the
relativistic couplings $g_i$ via

\begin{equation}
C_{1-5}= {\Lambda_\chi\over 2m_N} g_{1-5} \; ,
\end{equation}
\begin{equation}
{\tilde C}_{1-5}= {\tilde g}_{1-5}+ {\Lambda_\chi\over 2m_N} g_{1-5} \;
,
\end{equation}
\begin{equation}
C_6= {\tilde g}_6- {\Lambda_\chi\over 2m_N} g_6.
\end{equation}
We thus have a total of ten PV LECs describing PV short-distance
$NN$ physics.  For the purpose of characterizing PV operators in
the $NN$ system through ${\cal O}(Q)$, these ten constants are
sufficient.

\section*{Appendix B: Corrections to $V^{\rm PV}_{-1,\ \rm LR}$}

There are subleading corrections to OPE that arise from Fig.
\ref{fig10}, where the strong vertex comes from subleading PC
operators in Eqs. (\ref{pc2}-\ref{pc3}). In fact, when the $\nu=1$
operators from Eq. (\ref{pc2}) are inserted in Fig. \ref{fig10},
the resulting PV potential is naively of ${\cal O}(Q^0)$.
However, with
\begin{equation}
v\cdot q = q_0 = {{\vec p}_{1i}^2-{\vec p}_{1f}^2\over 2m_N} \sim
{\cal O}(Q^2/m_N),
\end{equation}
with $i$ ($f$) denoting the initial (final) nucleon, we have
\begin{equation}
V^{\rm PV}_{1a,\ \rm LR} = i {g_A \hpinn\over 4\sqrt{2}m^2_N
F_\pi} \left({\tau_1\times \tau_2\over 2} \right)_3 { ({\vec
p}_{1i}^2-{\vec p}_{1f}^2){\vec \sigma}_1 \cdot ({\vec p}_{1f}
+{\vec p}_{1i})-\left(1\leftrightarrow 2 \right) \over {\vec q}^2
+m_\pi^2},
\end{equation}
where $q=p_{2f}-p_{2i}=p_{1i}-p_{1f}$. Thus, this contribution
enters at ${\cal O}(Q)$ and must be included for consistency.

Similarly, the second and third  operators from Eq.
(\ref{pc3}) are nominally $\nu=2$ but lead to corrections that are  ${\cal O}(Q^3)$,
since they contain two kinetic operators from $v\cdot D$ or
$v\cdot A$. However, the insertion of the first, fourth and fifth
$\nu=2$ operators from Eq. (\ref{pc3}) in Fig. \ref{fig10}
lead to contributions at ${\cal O}(Q)$. We list these terms
below. The contribution from the first operator in Eq. (\ref{pc3})
reads
\begin{equation}
V^{\rm PV}_{1b,\ \rm LR} =  {{\vec q}^2 \over 8 m^2_N }V^{\rm
PV}_{-1,\ \rm LR} .
\end{equation}
The contribution from the fourth operator in Eq. (\ref{pc3}) is
\begin{equation}
V^{\rm PV}_{1c,\ \rm LR} = -i {g_A \hpinn\over 8\sqrt{2}m^2_N
F_\pi} \left({\tau_1\times \tau_2\over 2} \right)_3 { ({\vec
p}_{1i}^2+{\vec p}_{1f}^2){\vec \sigma}_1 \cdot {\vec q}
+\left(1\leftrightarrow 2 \right) \over {\vec q}^2 +m_\pi^2}.
\end{equation}
Finally, the contribution from the fifth operator in Eq. (\ref{pc3}) reads
\begin{equation}
V^{\rm PV}_{1d,\ \rm LR} = i {g_A \hpinn\over 4\sqrt{2}m^2_N
F_\pi} \left({\tau_1\times \tau_2\over 2} \right)_3 { ({\vec
\sigma}_1 \cdot {\vec p}_{1f})({\vec q}\cdot {\vec p}_{1i})
+({\vec \sigma}_1 \cdot {\vec p}_{1i})({\vec q}\cdot {\vec
p}_{1f}) + \left(1\leftrightarrow 2 \right) \over {\vec q}^2
+m_\pi^2}.
\end{equation}

Now consider the operators associated with $\widehat{d}_{16-19}$.
In the isospin-symmetric limit, to  leading order in the chiral
expansion we have
\begin{eqnarray}\nonumber
& S\cdot A \langle\chi_{+}\rangle \sim {m_\pi^2\over F_\pi^2} S\cdot A
\\
\nonumber
&\langle S\cdot A\chi_{+}\rangle \sim 0 \\ \nonumber
&[S\cdot {\cal D},\chi_{-}]\sim  {m_\pi^2\over F_\pi^2}S\cdot A \\
&[S\cdot {\cal D},\langle\chi_{-}\rangle]\sim {m_\pi^2\over F_\pi^2}
S\cdot A.
\end{eqnarray}
As a consequence, the LECs $\widehat{d}_{16, 18, 19}$ simply
renormalize the bare $\pi NN$ coupling constant $g_A^0$ at order
${\cal O}(Q^2)$ while LEC $\widehat{d}_{17}$ does not contribute.
Up to the truncation order ${\cal O}(Q)$ the corrections from
these LECs to PV $NN$ potential are automatically taken into
account as long as we use the renormalized (or physical) $g_A$ in
Eq. (\ref{long}).

Another possible source of corrections to the long-range PV
potential is the insertion of subleading PV operators in Fig.
\ref{fig3}. As pointed out in the Section \ref{sec3}, the PV
vector operator does not contribute due to vector-current
conservation. The axial-vector operator involves two pions and
leads to loop corrections at ${\cal O}(Q^2)$. The contribution from the
remaining PV operator proportional
to $k_{\pi NN}^{1a}$ has been discussed extensively in the main body of the paper.

Many  chiral loops exist at this order, from self-energy and PC-
and PV-vertex corrections.  The chiral loops do yield
contributions ${\cal O}(Q)$. However, these effects are included
in the renormalization of $g_A$ \cite{ijmpe} and of $\hpinn$
\cite{zhu}.

\section*{Appendix C: Loop Corrections to $V^{\rm PV}_{1,\ \rm SR}$}

\begin{figure}
\begin{center}
\epsfig{file=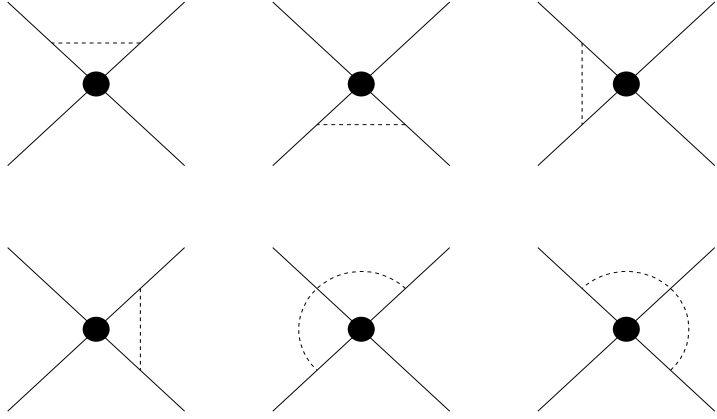,height=6cm,width=10cm}
\caption{Possible chiral corrections to PV $NN$ contact interactions.}
\label{fig8}
\end{center}
\end{figure}

The contact PV interactions in Fig. \ref{fig7} appear at ${\cal
O}(Q)$. From a simple counting of the chiral order of vertices ,
propagators, and loops, it is clear that loop corrections to these
PV LECs, shown in Fig. \ref{fig8}, first appear at ${\cal
O}(Q^3)$, which is beyond our truncation order.

Potentially more important are the loop corrections to the contact
PC interactions, where one vertex is the PV Yukawa coupling of the
pion to the nucleon. The relevant Feynman diagrams are shown in
Fig. \ref{fig9}.

Take the $C_S$ $NN$ contact interaction as an example.
For diagram (a-1), the amplitude is nominally
${\cal O}(Q)$, and reads
\begin{eqnarray} \label{integral1} \nonumber
iM_{a1}&\sim &\hpinn {C_S\over 2}{\sqrt{2} g_A\over F_\pi} \int
{d^Dk\over (2\pi)^D}
{i(S_1\cdot k)\over v_1\cdot (p_1^\prime +k)+i\epsilon}
{i\over v_1\cdot (p_1+k)+i\epsilon}{i\over k^2-m_\pi^2
+i\epsilon}\nonumber\\
&=&- \hpinn C_S{\sqrt{2} g_A\over F_\pi}S_1^\mu
\int_0^\infty {sds}\ \int_0^1 du\  \int {d^Dk\over
(2\pi)^D}\nonumber\\
&&\times {k_\mu\over [k^2+s v_1\cdot k +s (1-u) v_1\cdot p_1^\prime
+us v_1\cdot p_1 +m_\pi^2]^3},
\end{eqnarray}
where $s$ has the dimensions of mass, and where we have
Wick-rotated to Euclidean momenta in the second line. From this
form it is clear that $iM_{a1}\propto S_1\cdot v_1 = 0$. The same
argument holds for diagrams (a-2)-(a-4).

For diagram (b-1), the amplitude reads
\begin{eqnarray} \label{integral2} \nonumber
iM_{b1}&\sim &h_\pi^1 {C_S\over 2}{\sqrt{2} g_A\over F_\pi} \int
{d^Dk\over (2\pi)^D}
{i(S_2\cdot k)\over v_2\cdot (p_2 +k)+i\epsilon}
{i\over v_1\cdot
(p_1^\prime+k)+i\epsilon}{i\over k^2-m_\pi^2
+i\epsilon}     \\  \nonumber
&=&- h_\pi^1 C_S{\sqrt{2} g_A\over F_\pi}S_2^\mu
\int_0^\infty {sds}\ \int_0^1 du\  \int {d^Dk\over
(2\pi)^D}\nonumber\\
&&\times {k_\mu\over [k^2+s v_1\cdot k +s (1-u) v\cdot p_2
+us v\cdot p_1^\prime +m_\pi^2]^3}  =0 \  ,
\end{eqnarray}
where we have used the fact that $v_1=v_2=v=(1, {\vec 0})$
for low-energy $NN$ interaction.
Similarly, (b-2)-(b-4) vanish at ${\cal O}(Q)$.

There remains a third class of diagrams, (c-1)-(c-4). These are
2PR diagrams and their amplitudes
do not vanish at ${\cal O}(Q)$.
For example, the amplitude corresponding to diagram (c-1) reads
\begin{eqnarray} \label{integral3} \nonumber
iM_{c1}\sim h_\pi^1 {C_S\over 2}{\sqrt{2} g_A\over F_\pi} \int
{d^Dk\over (2\pi)^D}
{i(S_2\cdot k)\over v_2\cdot (p_2^\prime-k)+i\epsilon}
{i\over v_1\cdot (p_1^\prime+k)+i\epsilon}{i\over k^2-m_\pi^2
+i\epsilon}.
\end{eqnarray}
However, only the 2PI part of these diagrams should be included.
In other words, the contribution from the two-nucleon intermediate
state should be subtracted from the amplitude. This can be done in
old-fashioned time-ordered perturbation theory. Alternatively, we
may use the following identity to accomplish the subtraction
easily:
\begin{eqnarray}
{i\over -v\cdot k +i\epsilon} =-{i\over v\cdot k +i\epsilon} +2\pi
\delta\left(v\cdot k\right).
\end{eqnarray}
The second term corresponds to the two-nucleon pole,
while the first term is free of the infrared enhancement discussed earlier.
After subtracting the two-nucleon-pole contribution, the modified
amplitude for diagram (c-1) becomes
\begin{eqnarray} \label{integral4} \nonumber
i{\tilde M}_{c1}\sim -h_\pi^1 {C_S\over 2}{\sqrt{2} g_A\over F_\pi} \int
{d^Dk\over (2\pi)^D}
{i(S_2\cdot k)\over v_2\cdot (k-p^\prime_2)+i\epsilon}
{i\over v_1\cdot (p^\prime_1+k)+i\epsilon}{i\over k^2-m_\pi^2
+i\epsilon}=0.
\end{eqnarray}
Similarly, the 2PI parts of diagrams (c-2)-(c-4) vanish.
We see that diagrams
(c-1)-(c-4) can be generated from the PC ${\cal O}(Q^0)$
$C_{S,T}$ contact potential and leading-order PV OPE potential
by iteration in the LS equation.

In summary, the chiral-loop corrections to the PV short-range
potential occur at ${\cal O}(Q^2)$ or higher.

\section*{Appendix D: Derivation of $V^{\rm PV}_{1,\ \rm MR}$}
\label{sec7}

Of course, a consistent calculation
in EFT must include all loop diagrams present
to a given order.
In this appendix we give some details of the
evaluation of the diagrams in Figs. \ref{fig4}, \ref{fig5}, \ref{fig6}.
We use dimensional regularization for simplicity.

Let us consider first the
triangle diagrams in Fig. \ref{fig4}.

\begin{itemize}

\item Flavor-conserving case

In this case the initial and final state on each nucleon line
are the same. {\it i.e.}, a proton remains a proton and a neutron remains
a neutron.
Diagrams (c) and (d) are mirror diagrams of (a) and (b) in Fig.
\ref{fig4}.
We focus on (a) and (b).
The sum of their amplitudes reads
\begin{eqnarray} \label{integ1} \nonumber
iM_{(a)+(b)}&=& -{\sqrt{2} g_A h_\pi^1\over 4 F^3_\pi}
\int {d^Dk\over (2\pi)^D}
{ {\bar N}_1 \tau_3^1 v_1\cdot (2k-q) N_1 {\bar N}_2 1 S_2\cdot (2k-q)
N_2
 \over
v\cdot (p_2+k) (k^2-m_\pi^2) [(k-q)^2-m_\pi^2]
 }   \nonumber\\
&=&-{\sqrt{2} g_A h_\pi^1\over 4 F^3_\pi} \left\{
16\int_0^1dx \int_0^\infty dy \int {d^Dk\over
(2\pi)^D}\right.\nonumber\\
&& \left.\times {(v_1\cdot k) (S_2\cdot k) \over [k^2-y^2-m_\pi^2
-x(1-x){\vec
q}^2]^3}\right.\nonumber \\
&&\left. +8 (S_2\cdot q) \int_0^\infty y dy\int_0^1dx (1-2x)
\int {d^Dk\over (2\pi)^D}\right.\nonumber\\
&&\left.
\times {1 \over [k^2-y^2-m_\pi^2 -x(1-x){\vec q}^2]^3}\right\}  =0.
\end{eqnarray}
After momentum integration the first term contains a factor $v_1\cdot
S_2=0$.
The second term vanishes due to the $x$ integration, since the integrand
is
a total derivative.

\item Flavor-changing case: $n\leftrightarrow p$

In this case the sum of the amplitude for diagram (a) and (b) reads
\begin{eqnarray} \label{integ2} \nonumber
iM_{(a)+(b)}&=& {\sqrt{2} g_A h_\pi^1\over 4 F^3_\pi} \int {d^Dk\over
(2\pi)^D}
{\left(\bar p  v_1\cdot (2k-q) n \right)\left(\bar n  S_2\cdot q p
\right)
 \over
v\cdot (p_2+k) (k^2-m_\pi^2) [(k-q)^2-m_\pi^2]
 }  \\
&=& -i {\sqrt{2} g_A h_\pi^1\over \Lambda^2_\chi F_\pi}  L(q)
\left(\bar  p n\right) \left( \bar n  S_2\cdot q p\right),
\end{eqnarray}
where the function $L(q)$ is defined as
in Eq. (\ref{LandH}).
The sum of the amplitude for diagram (c) and (d) is, likewise,
\begin{eqnarray} \label{integ3}
iM_{(c)+(d)}
= i {\sqrt{2} g_A h_\pi^1\over \Lambda^2_\chi F_\pi}  L(q)
\left(\bar n p \right) \left( \bar p  S_2\cdot q n\right).
\end{eqnarray}
Note that $S_2\cdot q \approx -{1\over 2}{\vec \sigma}_2\cdot {\vec q}$.

Summing the four diagrams  and converting to momentum-space operators we
get
\begin{eqnarray} \label{integ4}
 {L(q)\over 2\sqrt{2}}{g_A h_\pi^1\over   \Lambda^2_\chi F_\pi}
\epsilon^{ab3}
\left( N^\dag \tau^a N   \right) \left( N^\dag \tau^b
{\vec \sigma} \cdot {\vec q} N\right).
\end{eqnarray}

Clearly the sum of triangle diagrams has the same Lorentz, isospin
structure as the $C_6$ contact term.

\end{itemize}

Consider now the crossed-box diagrams in Fig. \ref{fig5}.

\begin{itemize}

\item Flavor-conserving case: $pp\to pp$, $nn\to nn$

In Fig. \ref{fig5},
contributions from diagrams (c) and (d) are equal to (a) and (b).
For $pp\to pp$ the sum of (a) and (b) leads to
\begin{eqnarray}
 i4\sqrt{2}L(q) {g_A^3 h_\pi^1\over   \Lambda^2_\chi F_\pi}
\left( p^\dag [S_1\cdot q, S_1^\mu] p   \right) \left( p^\dag S^2_\mu
p\right).
\end{eqnarray}
Note that, for $pp\to pp$, initial particles are identical.
The operator form will generate both (a), (b) and their mirror diagrams
simultaneously.

For the $nn\to nn$ channel, there is an extra minus sign from PV Yukawa
vertex.
\begin{eqnarray}
\sqrt{2}L(q) {g_A^3 h_\pi^1\over   \Lambda^2_\chi F_\pi}
\epsilon^{ijk} \left( n^\dag q^i \sigma^j  n   \right) \left(
n^\dag \sigma^k n\right).
\end{eqnarray}

Combining both $pp\to pp$ and $nn\to nn$ channels, we get
\begin{eqnarray}   \nonumber
-{L(q)\over \sqrt{2}} {g_A^3 h_\pi^1\over   \Lambda^2_\chi F_\pi}
\epsilon^{ijk} \left( N^\dag\tau_3 q^i \sigma^j  N   \right)
\left( N^\dag \sigma^k N\right) &\\
-{L(q)\over \sqrt{2}} {g_A^3 h_\pi^1\over   \Lambda^2_\chi F_\pi}
\epsilon^{ijk} \left( N^\dag q^i \sigma^j  N   \right) \left(
N^\dag \tau_3\sigma^k N\right).
\end{eqnarray}

\item Flavor changing case: $n\to p, p\to n$

The sum of diagrams (a)-(d) leads to
\begin{eqnarray}
 +{ 3\sqrt{2}\over 16}\left[ -3 L(q) +H(q)\right]
 {g^3_A h_\pi^1\over   \Lambda^2_\chi F_\pi}  \epsilon^{ab3}
\left( N^\dag \tau^a N   \right) \left( N^\dag \tau^b {\vec
\sigma} \cdot {\vec q} N\right),
\end{eqnarray}
where $H(q)$ is defined as in Eq. (\ref{LandH}).

\end{itemize}

Finally, we discuss the box diagrams.  As in Appendix C, we have
to subtract the contribution from the two-nucleon intermediate
state. The corresponding time-ordered diagrams are shown in Fig.
\ref{fig6}. After the subtraction the 2PR part, we find:

\begin{itemize}

\item Flavor-conserving case: $p\to p$ and $n\to n$

For $np\to np$ the sum of all diagrams leads to
\begin{eqnarray}
 i4\sqrt{2}L(q) {g_A^3 h_\pi^1\over   \Lambda^2_\chi F_\pi}
\left( n^\dag [S_1\cdot q, S_1^\mu] n   \right) \left( p^\dag S^2_\mu
p\right).
\end{eqnarray}
Note that, for $pn\to pn$, there is an extra minus sign from
the PV Yukawa vertex,
\begin{eqnarray}
-i4\sqrt{2}L(q) {g_A^3 h_\pi^1\over   \Lambda^2_\chi F_\pi}
\left( p^\dag [S_1\cdot q, S_1^\mu] p   \right) \left( n^\dag S^2_\mu
n\right).
\end{eqnarray}

Combining both channels, we get
\begin{eqnarray}   \nonumber
 {L(q)\over \sqrt{2}} {g_A^3 h_\pi^1\over   \Lambda^2_\chi F_\pi}
\epsilon^{ijk} \left( N^\dag\tau_3 q^i \sigma^j  N   \right)
\left( N^\dag \sigma^k N\right) &\\
-{L(q)\over \sqrt{2}} {g_A^3 h_\pi^1\over   \Lambda^2_\chi F_\pi}
\epsilon^{ijk} \left( N^\dag q^i \sigma^j  N   \right) \left(
N^\dag \tau_3\sigma^k N\right).
\end{eqnarray}

\item Flavor-changing case: $n\leftrightarrow p$

The sum of all diagrams leads to the same result as in the crossed-box
case,
\begin{eqnarray}
 { 3\sqrt{2}\over 16}\left[ -3 L(q) +H(q)\right]
 {g^3_A h_\pi^1\over   \Lambda^2_\chi F_\pi}  \epsilon^{ab3}
\left( N^\dag \tau^a N   \right) \left( N^\dag \tau^b {\vec
\sigma} \cdot {\vec q} N\right).
\end{eqnarray}

\end{itemize}

In summary, the sum of one-loop, TPE diagrams is
\begin{eqnarray}   \nonumber
&& {L(q)\over 2\sqrt{2}}{g_A h_\pi^1\over   \Lambda^2_\chi F_\pi}
\epsilon^{ab3} \left( N^\dag \tau^a N   \right) \left( N^\dag
\tau^b {\vec \sigma} \cdot {\vec q} N\right)\\ \nonumber &+&{
3\sqrt{2}\over 8}\left[ -3 L(q) +H(q)\right]
 {g^3_A h_\pi^1\over   \Lambda^2_\chi F_\pi}  \epsilon^{ab3}
\left( N^\dag \tau^a N   \right) \left( N^\dag \tau^b {\vec
\sigma} \cdot {\vec q}
N\right) \nonumber\\
&-&\sqrt{2}L(q) {g_A^3 h_\pi^1\over   \Lambda^2_\chi F_\pi}
\epsilon^{ijk} \left( N^\dag q^i \sigma^j  N   \right) \left(
N^\dag \tau_3\sigma^k N\right),
\end{eqnarray}
which leads to the medium-range potential (\ref{v3}).

\section*{Appendix E: Illustrative Estimates}

Having a form of the weak parity-violating potential $V^{\rm
PV}(r)$ it is, of course, essential to complete the process by
connecting with the S-matrix---{\it i.e.}, expressing the
phenomenological parameters $\lambda_i,\,\rho_t$ defined in Eq.
(\ref{eq:eff}) in terms of the fundamental
ones---$C_i,\,\tilde{C}_i$ defined in Eq. (\ref{eq:sht}).  This is
a major undertaking and should involve the latest and best $NN$
wave functions such as Argonne V18.  The work is underway, but it
will be some time until this process is completed\cite{gib}.  Even
after this connection has been completed, the results will be
numerical in form.  However, it is very useful to have an analytic
form by which to understand the basic physics of this
transformation and by which to make simple numerical estimates.
For this purpose we shall employ simple phenomenological $NN$ wave
functions, as described below.

Examination of the scattering matrix Eq. (\ref{eq:tpx}) reveals
that the parameters $\lambda_{s,t}$ are associated with the
short-distance component while $\rho_t$ contains contributions
from the both (long-distance) pion exchange as well as short
distance effects. In the former case, since the interaction is
short ranged we can use this feature in order to simplify the
analysis. Thus, we can determine the shift in the deuteron
wavefunction associated with parity violation by demanding
orthogonality with the $^3S_1$ scattering state, which yields,
using the simple asymptotic form of the bound state
wavefunction\cite{khk},\cite{khr}
\begin{equation}
\psi_d(r)=\left[1+\rho_t(\vec{\sigma}_p+\vec{\sigma}_n)\cdot
-i\vec{\nabla} +\lambda_t(\vec{\sigma}_p-\vec{\sigma}_n)\cdot
-i\vec{\nabla})\right] \sqrt{\gamma\over 2\pi}{1\over r}e^{-\gamma
r}\label{eq:deu}
\end{equation}
where $\gamma^2/M=2.23$ MeV is the deuteron binding energy. Now
the shift generated by $V^{\rm PV}(r)$ is found to
be\cite{khk},\cite{khr}
\begin{eqnarray}
\delta\psi_d(\vec{r}) &\simeq&\int
d^3r'G(\vec{r},\vec{r}')V^{\rm PV}(\vec{r}')
\psi_d(r')\nonumber\\
&=&-{M\over 4\pi}\int d^3r'{e^{-\gamma|\vec{r}-\vec{r}'|}
\over |\vec{r}-\vec{r}'|}V^{PV}(\vec{r}')\psi_d(r')\nonumber\\
&\simeq& {M\over 4\pi}\vec{\nabla}\left({e^{-\gamma r}\over
r}\right)\cdot \int
d^3r'\vec{r}'V^{\rm PV}(\vec{r}')\psi_d(r')\label{eq:shr}
\end{eqnarray}
where the last step is permitted by the short range of
$V^{\rm PV}(\vec{r}')$.  Comparing Eqs. (\ref{eq:shr}) and (\ref{eq:deu})
yields then the identification
\begin{equation}
\sqrt{\gamma\over 2\pi}\lambda_t\chi_t\equiv i{M\over
16\pi}\xi_0^\dagger\int d^3r'
(\vec{\sigma}_1-\vec{\sigma}_2)\cdot\vec{r}'V^{\rm PV}
(\vec{r}')\psi_d(r')\chi_t\xi_0 \label{eq:elt}
\end{equation}
where we have included the normalized isospin wave function
$\xi_0$ since the potential involves $\vec{\tau}_1,\vec{\tau}_2.$
When operating on such an isosinglet np state the PV potential can
be written as
\begin{eqnarray}
V^{\rm PV}(\vec{r}')&=&{2\over \Lambda_\chi^3}\left[
(C_1-3C_3)(\vec{\sigma}_1-\vec{\sigma}_2)\cdot(-i\vec{\nabla}f_m(r)+2f_m(r)
\cdot-i\vec{\nabla})\right.\nonumber\\
&+&\left.(\tilde{C}_1-3\tilde{C}_3)(\vec{\sigma}_1\times\vec{\sigma}_2)
\cdot\vec{\nabla}f_m(r)\right]
\end{eqnarray}
where $f_m(r)$ is the Yukawa form
$$f_m(r)={m^2e^{-mr}\over 4\pi r}$$
defined in Eq. (\ref{eq:yuk}).  Using the identity
\begin{equation}
(\vec{\sigma}_1\times\vec{\sigma}_2){1\over 2}
(1+\vec{\sigma}_1\cdot\vec{\sigma}_2)=i(\vec{\sigma}_1
-\vec{\sigma}_2 )
\end{equation}
Eq. \ref{eq:elt} becomes
\begin{eqnarray}
\sqrt{\gamma\over 2\pi}\lambda_t\chi_t&\simeq&{2M\over
16\pi\Lambda_\chi^3}{4\pi\over
3}(\vec{\sigma}_1-\vec{\sigma}_2)^2\chi_t\int_0^\infty
drr^3\nonumber\\
&\times&\left[-2(3C_3-C_1)f_m(r){d\psi_d(r)\over
dr}+(3\tilde{C}_3-3C_3-\tilde{C}_1+C_1)
{df_m(r)\over dr}\psi_d(r)\right]\nonumber\\
&=&\sqrt{\gamma\over 2\pi}\cdot 4\chi_t{1\over 12}{2Mm^2\over
4\pi\Lambda_\chi^3}\left(
{2m(6C_3-3\tilde{C}_3-2C_1+\tilde{C}_1)+\gamma(15C_3-3\tilde{C}_3-5C_1+\tilde{C}
_1)\over
(\gamma+m)^2}\right)\nonumber\\
\quad
\end{eqnarray}
or
\begin{equation}
\lambda_t\simeq {Mm^2\over 6\pi\Lambda_\chi^3}\left(
{2m(6C_3-3\tilde{C}_3-2C_1+\tilde{C}_1)+\gamma(15C_3-3\tilde{C}_3-5C_1+\tilde{C}
_1)\over
(\gamma+m)^2}\right)
\end{equation}

In order to determine the singlet parameter $\lambda_s^{np}$, we
must use the ${}^1S_0$ np-scattering wave function instead of the
deuteron, but the procedure is similar,
yielding\cite{khk},\cite{khr}
\begin{equation}
d_s^{np}(k)\chi_s\equiv i{M\over 48\pi}\xi_1^\dagger\int d^3r'
(\vec{\sigma}_1-\vec{\sigma}_2)\cdot\vec{r}'V^{\rm PV}(\vec{r}')
\psi_{{}^1S_0}(r')\chi_s\xi_1 \label{eq:els}
\end{equation}
and we can proceed similarly.  In this case the potential becomes
\begin{eqnarray}
V^{\rm PV}(\vec{r}')&=&{2\over \Lambda_\chi^3}\left[(C_1+C_3+4C_5)
(\vec{\sigma}_1-\vec{\sigma}_2)\cdot(-i\vec{\nabla}f_m(r)+2f_m(r)
\cdot-i\vec{\nabla})\right.\nonumber\\
&+&\left.(\tilde{C}_1+\tilde{C}_3+4\tilde{C}_5)(\vec{\sigma}_1\times\vec{\sigma}_2)\cdot
\vec{\nabla}f_m(r)\right]
\end{eqnarray}
and Eq. (\ref{eq:els}) is found to have the form
\begin{eqnarray}
d_s^{np}(k)\chi_s&=&{2M\over 48\pi\Lambda_\chi^3}{4\pi\over 3}
(\vec{\sigma}_1-\vec{\sigma}_2)^2\chi_s\int_0^\infty dr{r}^3\left\{
\right.\nonumber\\
&\times&\left.2[C_1+C_3+4C_5]f_m(r){d\psi_{{}^1S_0}(r)\over
dr}\right.\nonumber\\
&+&\left.[C_1+\tilde{C}_1+C_3+\tilde{C}_3+4(C_5+\tilde{C}_5)]
{df_m(r)\over dr}\psi_{{}^1S_0}(r)\right\}\nonumber\\
 &=&- 12\chi_s{1\over 36}{2Mm^2\over
4\pi\Lambda_\chi^3}e^{i\delta_s}\left\{1\over
{(k^2+m^2)^2}\right.\nonumber\\
&\times&\left.\left[\cos\delta_s(4k^2(C_1+C_3+4
C_5)+(C_1+\tilde{C}_1+C_3+\tilde{C}_3+
4(C_5+\tilde{C}_5))(k^2+3m^2)]
\right.\right.\nonumber\\
&+&\left.\left.{2m\over k}\sin\delta_s[(C_1+C_3+4
C_5)(m^2+3k^2)+(C_1+\tilde{C}_1+C_3+\tilde{C}_3+
4(C_5+\tilde{C}_5))m^2)\right] \right\}\nonumber\\
\quad
\end{eqnarray}
which, in the limit as $k\rightarrow 0$, yields the predicted
value for $\lambda_s^{np}$---
\begin{eqnarray}
\lambda_s^{np}&=&-{1\over a_s^{np}}\lim_{k\rightarrow
0}d_s^{np}(k)= {M\over 6\pi a_s^{np}\Lambda_\chi^3}\left\{
3[C_1+\tilde{C}_1+C_3+\tilde{C}_3+
4(C_5+\tilde{C}_5)]\right.\nonumber\\
&-&\left.2ma_s^{np}[2C_1+\tilde{C}_1+2C_3+\tilde{C}_3+
4(2C_5+\tilde{C}_5)]\right\}
\end{eqnarray}
Similarly, we may identify
\begin{eqnarray}
\lambda_s^{pp}&=&-{1\over a_s^{pp}}\lim_{k\rightarrow
0}d_s^{pp}(k)= {M\over 6\pi a_s^{pp}\Lambda_\chi^3}\left\{
3[C_1+\tilde{C}_1+C_2+\tilde{C}_2+C_3+\tilde{C}_3+C_4+\tilde{C}_4-
2(C_5+\tilde{C}_5)]\right.\nonumber\\
&-&\left.2ma_s^{pp}[2C_1+\tilde{C}_1+2C_2+\tilde{C}_2+2C_3+\tilde{C}_3+2C_4
+\tilde{C}_4-
2(2C_5+\tilde{C}_5)]\right\}\nonumber\\
\lambda_s^{nn}&=&-{1\over a_s^{nn}}\lim_{k\rightarrow
0}d_s^{nn}(k)= {M\over 6\pi a_s^{nn}\Lambda_\chi^3}\left\{
3[C_1+\tilde{C}_1-C_2-\tilde{C}_2+C_3+\tilde{C}_3-C_4-\tilde{C}_4-
2(C_5+\tilde{C}_5)]\right.\nonumber\\
&-&\left.2ma_s^{nn}[2C_1+\tilde{C}_1-2C_2-\tilde{C}_2+2C_3+\tilde{C}_3
-2C_4-\tilde{C}_4-
2(2C_5+\tilde{C}_5)]\right\}\nonumber\\
\quad
\end{eqnarray}
In order to evaluate the spin-conserving amplitude $\rho_t$, we
shall assume dominance of the long range pion component. The shift
in the deuteron wave function is given by
\begin{eqnarray}
\delta \psi_d(\vec{r})&=&\xi_0^\dagger\int
d^3r'G_0(\vec{r},\vec{r}')V^{\rm PV}_{\rm LR}(\vec{r}')
\psi_d(r')\nonumber\\
&=&-{M\over 4\pi}\xi_0^\dagger\int
d^3r'{e^{-\gamma|\vec{r}-\vec{r}'|} \over
|\vec{r}-\vec{r}'|}V^{\rm PV}_{\rm LR}(\vec{r}')\psi_d(r')\chi_t\xi_0
\end{eqnarray}
but now with\footnote{Here we have used the identity
\begin{equation}\nonumber
(\vec{\tau}_1\times\vec{\tau}_2)=-i(\vec{\tau_1}-\vec{\tau}_2){1\over
2} (1+\vec{\tau}_1\cdot\vec{\tau}_2)
\end{equation}}
\begin{equation}
V^{\rm PV}_{\rm LR}(\vec{r})={h_\pi g_{\pi NN}\over \sqrt{2}M}{1\over
2}(\tau_1-\tau_2)_z
(\vec{\sigma}_1+\vec{\sigma}_2)\cdot-i\vec{\nabla}w_\pi(r)
\end{equation}
Of course, the meson which is exchanged is the pion so the short
range assumption which permitted the replacement in Eq.
(\ref{eq:shr}) is not valid and we must perform the integration
exactly.  This process is straightforward but tedious\cite{des}.
Nevertheless, we can get a rough estimate by making a ``heavy
pion'' approximation, whereby we can identify the constant
$\rho_t$ via
\begin{equation}
\sqrt{\gamma\over 2\pi}\rho_t\chi_t\approx -i{M\over 32\pi}\int
d^3r' (\vec{\sigma}_1+\vec{\sigma}_2)\cdot\vec{r}'V^{\rm PV}_{\rm LR}
(\vec{r}')\psi_d(r')\chi_t\xi_0
\end{equation}
which leads to\cite{dpl}
\begin{eqnarray}
\sqrt{\gamma\over 2\pi}\rho_t\chi_t&\approx&-{1\over
32\pi}{4\pi\over 3} (\vec{\sigma}_1+\vec{\sigma}_2)^2\chi_t {\hpinn
g_{\pi NN}
\over \sqrt{2}}\int_0^\infty drr^3{df_\pi(r)\over dr}\psi_d(r){1\over
m_\pi^2}\nonumber\\
&=&\sqrt{\gamma\over 2\pi}8\chi_t{1\over 96\pi}{h_\pi g_{\pi
NN}\over \sqrt{2}} {\gamma+2m_\pi\over (\gamma+m_\pi)^2}
\end{eqnarray}
We find then the prediction
\begin{equation}
\rho_t={g_{\pi NN}\over 12\sqrt{2}\pi}{\gamma+2m_\pi\over
(\gamma+m_\pi)^2}\hpinn
\end{equation}

At this point it is useful to obtain rough numerical estimates.
This can be done by use of the numerical estimates given in Table
2.  To make things tractable, we shall use the best values given
therein.  Since we are after only rough estimates and since the
best values assume the DDH relationship---Eq. (\ref{eq:ddhr})
between the tilde- and non-tilde- quantities, we shall express our
results in terms of only the non-tilde numbers---a future complete
evaluation should include the full dependence.  Of course, these
predictions are only within a model, but they has the advantage of
allowing connection with previous theoretical estimates.  In this
way, we obtain the predictions
\begin{eqnarray}
\lambda_t&=&\left[-0.092C_3-0.014C_1\right]m_\pi^{-1}\nonumber\\
\lambda_s^{np}&=&\left[-0.087(C_3+4C_5)-0.037C_1\right]
m_\pi^{-1}\nonumber\\
\lambda_s^{pp}&=&\left[-0.087(C_3+C_4-2C_5)
-0.037(C_1+C_2)\right]m_\pi^{-1}\nonumber\\
\lambda_s^{nn}&=&\left[-0.087(C_3-C_4-2C_5)
-0.037(C_1-C_2)\right]m_\pi^{-1}\nonumber\\
\rho_t&=& 0.346 h_\pi m_\pi^{-1}
\end{eqnarray}
so that, using the best values from Table 2 we estimate
\begin{eqnarray}
\lambda_t&=&-2.39\times 10^{-7}m_\pi^{-1}=-3.41\times
10^{-7}\,\,{\rm fm}\nonumber\\
\lambda_s^{np}&=&-1.12\times 10^{-7}m_\pi^{-1}=-1.60\times
10^{-7}\,\,
{\rm fm}\nonumber\\
\lambda_s^{pp}&=&-3.58\times 10^{-7}\,\,m_\pi^{-1}=-5.22\times
10^{-7}\,\,{\rm fm}\\
\lambda_s^{nn}&=&-2.97\times 10^{-7}\,\,m_\pi^{-1}=-4.33\times
10^{-7}\,\,{\rm fm}\nonumber\\
\rho_t&=&1.50\times 10^{-7}\,\,m_\pi^{-1}=2.14\times
10^{-7}\,\,{\rm fm}
\end{eqnarray}
Again we emphasize that in arriving at the foregoing expressions,
we have used the DDH relationships between the $C_i$ and ${\tilde
C}_i$. In the more general case, one should obtain expressions
containing roughly the linear combinations given in Eqs.
(\ref{eq:lincomb}). A similar caveat applies to the expressions
below.

At this point we note, however, that $\lambda_s^{pp}$ is an order
of magnitude larger than the experimentally determined number, Eq.
(\ref{eq:ppe}).  The problem here is not with the couplings but
with an important piece of physics which has thus far been
neglected---short distance effects. There are two issues here. One
is that the deuteron and $NN$ wave functions should be modified at
short distances from the simple asymptotic form used up until this
point in order to account for finite size effects. The second is
the well-known feature of the Jastrow correlations that suppress
the nucleon-nucleon wave function at short distance.

In order to deal approximately with the short distance properties
of the deuteron wave function, we modify the exponential form to
become constant inside the deuteron radius
$R$\cite{khk},\cite{khr}
\begin{equation}
\sqrt{\gamma\over 2\pi}{1\over r}e^{-\gamma r} \rightarrow
N\left\{\begin{array}{cc}
{1\over R}e^{-\gamma R}& r\leq R\\
{1\over r}e^{-\gamma r}& r>R
\end{array}\right.
\end{equation}
where
$$N=\sqrt{\gamma\over 2\pi}{\exp\gamma R\over \sqrt{1+{2\over 3}\gamma R}}$$
is the modified normalization factor and we use R=1.6 fm.  For the
$NN$ wave function we use\cite{khk},\cite{khr}
\begin{equation}
\psi_{{}^1S_0}(r)=\left\{\begin{array}{cc}
A{\sin\sqrt{p^2+p_0^2}r\over \sqrt{p^2+p_0^2}r}& r\leq r_s\\
{\sin pr\over pr}-{1\over {1\over a_s}+ip}{e^{ipr}\over r}&r>r_s
\end{array}\right.
\end{equation}
where we choose $r_s=2.73$ fm and $p_or_s=1.5$.  The normalization
constant $A(p)$ is found by requiring continuity of the wave
function and its first derivative at $r=r_s$
\begin{equation}
A(p)={\sqrt{p^2+p_0^2}r_s\over \sin\sqrt{p^2+p_0^2}r_s}{\sin pr_s-
pa_s\cos pr_s\over pr_s(1+ipa_s)}
\end{equation}
As to the Jastrow correlations we multiply the wave function by
the simple phenomenological form\cite{rib}
\begin{equation}
\phi(r)=1-ce^{-dr^2},\quad{\rm with}\quad c=0.6,\quad d=3\,\,{\rm
fm}^{-2}
\end{equation}
With these modifications we find the much more reasonable values
for the constants $\lambda_s^{{pp},{np}}$ and $\lambda_t$
\begin{eqnarray}
\lambda_s^{pp}&=&\left[-0.011(C_3+C_4-2C_5)-0.004
(C_1+C_2)\right]m_\pi^{-1}\nonumber\\
\lambda_s^{nn}&=&\left[-0.011(C_3-C_4+2C_5)-0.004
(C_1-C_2)\right]m_\pi^{-1}\nonumber\\
\lambda_s^{np}&=&\left[-0.011(C_3+4C_5)-0.004C_1\right]m_\pi^{-1}\nonumber\\
\lambda_t&=&\left[-0.019C_3-0.0003C_1\right]m_\pi^{-1}
\label{eq:cal}
\end{eqnarray}
Using the best values from Table 2 we find then the benchmark
values
\begin{eqnarray}
\lambda_s^{pp}&=&-4.2\times 10^{-8}m_\pi^{-1}= -6.1\times
10^{-8}\,\,{\rm fm}
\nonumber\\
\lambda_s^{nn}&=&-3.6\times 10^{-8}m_\pi^{-1}= -5.3\times
10^{-8}\,\,{\rm fm}
\nonumber\\
\lambda_s^{np}&=&-1.3\times 10^{-8}m_\pi^{-1}= -1.9\times
10^{-8}\,\,{\rm fm}
\nonumber\\
\lambda_t&=&-4.7\times 10^{-8}m_\pi^{-1}=-6.7\times
10^{-8}\,\,{\rm fm}
\end{eqnarray}
Since $\rho_t$ is a long distance effect, we use the same value as
calculated previously as our benchmark number
\begin{equation}
\rho_t=1.50\times 10^{-7}\,\,m_\pi^{-1}=2.14\times 10^{-7}\,\,{\rm
fm}
\end{equation}

Obviously the value of $\lambda_s^{pp}$ is now in much better
agreement with the experimental value Eq. (\ref{eq:ppe}).  Of
course, our rough estimate is no substitute for a reliable state
of the art wave function evaluation.  This has been done recently
by Carlson {\rm et al.}  and yields, using the Argonne V18
wavefunctions\cite{car}
\begin{equation}
\lambda_s^{pp}=\left[-0.008(C_3+C_4-2C_5)
-0.003(C_1+C_2)\right]m_\pi^{-1}
\end{equation}
in reasonable agreement with the value calculated in Eq.
(\ref{eq:cal}). Similar efforts should be directed toward
evaluation of the remaining parameters using the best modern wave
functions.

We end our brief discussion here, but clearly this was merely a
simplistic model calculation. It is important to complete this
process by using the best contemporary nucleon-nucleon wave
functions with the most general EFT potential developed above, in
order to allow the best possible restrictions to be placed on the
unknown counter-terms.



\begin{thebibliography}{99}

\bibitem{few}
See, {\it e.g.},
S.C. Pieper and R.B. Wiringa,
Ann. Rev. Nucl. Part. Sci. {\bf 51}, 53 (2001);
B.R. Barrett, P. Navr\'atil, W.E. Ormand, and J.P. Vary,
Acta Phys. Polon. {\bf B33}, 297 (2002).

\bibitem{eft}
See, {\it e.g.},
P.F. Bedaque and U. van Kolck,
Ann. Rev. Nucl. Part. Sci. {\bf 52}, 339 (2002).

\bibitem{wei}
S. Weinberg,
Phys. Lett. {\bf B251}, 288 (1990);
Nucl. Phys. {\bf B363}, 3 (1991).

\bibitem{ray}
C. Ord\'o\~nez and U. van Kolck,
Phys. Lett. {\bf B291}, 459 (1992);
C. Ord\'o\~nez, L. Ray, and U. van Kolck,
Phys. Rev. {\bf C53}, 2086 (1996);
N. Kaiser, R. Brockmann, and W. Weise,
Nucl. Phys. {\bf A625}, 758 (1997);
N. Kaiser, S. Gerstendorfer, and W. Weise,
Nucl. Phys. {\bf A637}, 395 (1998);
N. Kaiser,
Phys. Rev. {\bf C65}, 017001 (2002) and references therein.

\bibitem{3Npot}
U. van Kolck,
Phys. Rev. {\bf C49}, 2932 (1994);
J.L. Friar, D. H\"uber, and U. van Kolck,
Phys. Rev. {\bf C59}, 53 (1999).

\bibitem{fits}
E. Epelbaum, W. Gl\"ockle, and U.-G. Mei{\ss}ner, Nucl. Phys. {\bf
A671}, 295 (2000) and [arxiv:nucl-th/0405048]; D.R. Entem and R.
Machleidt, Phys. Rev. {\bf C68}, 041001 (2003).

\bibitem{chiralfew}
E. Epelbaum, A. Nogga, W. Gl\"ockle, H. Kamada, U.-G. Mei{\ss}ner,
and H. Wita\l a,
Phys. Rev. {\bf C66}, 064001 (2002).

\bibitem{aleph}
U. van Kolck,
Nucl. Phys. {\bf A645}, 273 (1999).

\bibitem{crs}
J.-W. Chen, G. Rupak, and M.J. Savage,
Nucl. Phys. {\bf A653}, 386 (1999).

\bibitem{3stooges}
P.F. Bedaque and U. van Kolck,
Phys. Lett. {\bf B428}, 221 (1998);
P.F. Bedaque, H.-W. Hammer, and U. van Kolck,
Phys. Rev. {\bf C58}, 641 (1998);
F. Gabbiani, P.F. Bedaque, and H.W. Grie{\ss}hammer,
Nucl. Phys. {\bf A675}, 601 (2000).

\bibitem{triton}
P.F. Bedaque, H.-W. Hammer, and U. van Kolck,
Nucl. Phys. {\bf A676}, 357 (2000);
P.F. Bedaque, G. Rupak, H.W. Grie{\ss}hammer, and H.-W. Hammer,
Nucl. Phys. {\bf A714}, 589 (2003).

\bibitem{npd}
T.-S. Park, D.-P. Min, and M. Rho,
Nucl. Phys. {\bf A596}, 515 (1996);
J.-W. Chen, G. Rupak, and M.J. Savage,
Phys. Lett. {\bf B464}, 1 (1999);
G. Rupak, Nucl. Phys. {\bf A678}, 405 (2000);
T.-S. Park, K. Kubodera, D.-P. Min, and M. Rho,
Phys. Lett. {\bf B472}, 232 (2000).

\bibitem{tan}
N. Tanner,
Phys. Rev. {\bf 107}, 1203 (1957).

\bibitem{par}
C.S. Wu et al.,
Phys. Rev. {\bf 105}, 1413 (1957).

\bibitem{mic}
F.C. Michel,
Phys. Rev. {\bf B133}, 329 (1964).

\bibitem{ope}
B.H.J. McKellar,
Phys. Lett. {\bf B26}, 107 (1967).

\bibitem{oth}
E. Fischbach,
Phys. Rev. {\bf 170}, 1398 (1968);
D. Tadic,
Phys. Rev. {\bf 174}, 1694 (1968);
W. Kummer and M. Schweda,
Acta Phys. Aust. {\bf 28}, 303 (1968);
B.H.J. McKellar and P. Pick,
Phys. Rev. {\bf D7}, 260 (1973).

\bibitem{pir}
H.J. Pirner and D.O. Riska,
Phys. Lett. {\bf B44}, 151 (1973).

\bibitem{ddh}
B. Desplanques, J.F. Donoghue, and B.R. Holstein,
Ann. Phys. (NY) {\bf 124}, 449 (1980).

\bibitem{wilburn}
W.S. Wilburn and J.D. Bowman,
Phys. Rev. {\bf C57}, 3425 (1998).

\bibitem{ana}
W.C. Haxton, C.P. Liu, and M.J. Ramsey-Musolf,
Phys. Rev. {\bf C65}, 045502 (2002);
W.C. Haxton and C.E. Wieman,
Ann. Rev. Part. Nucl. Sci. {\bf 51}, 261 (2001).

\bibitem{jerry}
G.A. Miller,
Phys. Rev. {\bf C67}, 042501(R) (2003).

\bibitem{str}
See, {\it e.g.},
D.H. Beck and B.R. Holstein,
Int. J. Mod. Phys. {\bf E10}, 1 (2001);
D.H. Beck and R.D. McKeown,
Ann. Rev. Nucl. Part. Sci. {\bf 51}, 189 (2001).

\bibitem{dan}
G.S. Danilov,
Phys. Lett. {\bf 18}, 40 (1965).

\bibitem{dm}
B. Desplanques and J. Missimer,
Nucl. Phys. {\bf A300}, 286 (1978);
B. Desplanques,
Phys. Rept. {\bf 297}, 2 (1998).

\bibitem{bar}
G. Barton,
Nuovo Cim. {\bf 19}, 512 (1961).

\bibitem{bhh}
B.R. Holstein,
Phys. Rev. {\bf D23}, 1618 (1981).

\bibitem{dz}
V.M. Dubovik and S.V. Zenkin,
Ann. Phys. (NY) {\bf 172}, 100 (1986).

\bibitem{fcdh}
G.B. Feldman, G.A. Crawford, J. Dubach, and B.R. Holstein,
Phys. Rev. {\bf C43}, 863 (1991).

\bibitem{atomic}
C.S. Wood et al.,
Science {\bf 275}, 1759 (1997).

\bibitem{kaplan}
D.B. Kaplan and M.J. Savage,
Nucl. Phys. {\bf A556}, 653 (1993).

\bibitem{zhu}
S.-L. Zhu, S. Puglia, B.R. Holstein, and M.J. Ramsey-Musolf,
Phys. Rev. {\bf D63}, 033006 (2001).

\bibitem{anapole1} 
S.-L. Zhu, S. Puglia, B.R. Holstein, and M.J. Ramsey-Musolf,
Phys. Rev. {\bf D62}, 033008 (2000).

\bibitem{anapole2} 
C.M Maekawa and U. van Kolck,
Phys. Lett. {\bf B478}, 73 (2000);
C.M Maekawa, J.S. Veiga, and U. van Kolck,
Phys. Lett. {\bf B488}, 167 (2000).

\bibitem{bs1} 
P.F. Bedaque and M.J. Savage,
Phys. Rev. {\bf C62}, 018501 (2000).

\bibitem{bs2} 
J.-W. Chen, T.D. Cohen, and C.W. Kao,
Phys. Rev. {\bf C64}, 055206 (2001).

\bibitem{cj1} 
J.-W. Chen and X. Ji,
Phys. Rev. Lett. {\bf 86}, 4239 (2001);
Phys. Lett. {\bf B501}, 209 (2001).

\bibitem{cj2} 
S.-L. Zhu, S. Puglia, B.R. Holstein, and M.J. Ramsey-Musolf,
Phys. Rev. {\bf C64}, 035502 (2001);
S.-L. Zhu, C.M. Maekawa, B.R. Holstein, and M.J. Ramsey-Musolf,
Phys. Rev. Lett. {\bf 87}, 201802 (2001);
S.-L. Zhu, C.M. Maekawa, G. Sacco, B.R. Holstein, and M.J. Ramsey-Musolf,
Phys. Rev. {\bf D65}, 033001 (2002).

\bibitem{ss1}
D.B. Kaplan, M.J. Savage, R.P. Springer and M.B. Wise,
Phys. Lett. {\bf B449}, 1 (1999).

\bibitem{ss2}
M.J. Savage and R.P. Springer,
Nucl. Phys. {\bf A644}, 238 (1998);
            {\bf A657}, 457 (1999);
            {\bf A686}, 413 (2001).

\bibitem{radcappionful}
C.H. Hyun, T.-S. Park, and D.-P. Min,
Phys. Lett. {\bf B516}, 321 (2001).

\bibitem{towards}
S.R. Beane, P.F. Bedaque, M.J. Savage, and U. van Kolck,
Nucl. Phys. {\bf A700}, 377 (2002);
S.R. Beane and M.J. Savage,
Nucl. Phys. {\bf A713}, 148 (2003);
            {\bf A717}, 91 (2003).

\bibitem{pp}
A.R. Berdoz et al.,
Phys. Rev. {\bf C68}, 034004 (2003);
Phys. Rev. Lett. {\bf 87}, 272301 (2001).

\bibitem{hf}
K.S. Krane et al.,
Phys. Rev. {\bf C4}, 1906 (1971).

\bibitem{la}
V.W. Yuan et al.,
Phys. Rev. {\bf C44}, 2187 (1991);
Y. Masuda et al.,
Nucl. Phys. {\bf A504}, 269 (1989);
V.P. Alfimenko et al.,
Nucl. Phys. {\bf A398}, 93 (1983).

\bibitem{f1}
E.G. Adelberger et al.,
Phys. Rev. {\bf C27}, 2833 (1983).

\bibitem{f2}
K. Elsener et al.,
Nucl. Phys. {\bf A461}, 579 (1987);
Phys. Rev. Lett. {\bf 52}, 1476 (1984).

\bibitem{n1}
K.A. Snover et al.,
Phys. Rev. Lett. {\bf 41}, 145 (1978).

\bibitem{n2}
E.D. Earle et al.,
Nucl. Phys. {\bf A396}, 221 (1983).

\bibitem{f3}
C.A. Barnes et al.,
Phys. Rev. Lett. {\bf 40}, 840 (1978).

\bibitem{f4}
M. Bini et al.,
Phys. Rev. Lett. {\bf 55}, 795 (1985).

\bibitem{f5}
G. Ahrens et al.,
Nucl. Phys. {\bf A390}, 496 (1982).

\bibitem{f6}
S.A. Page et al.,
Phys. Rev. {\bf C35}, 1119 (1987).

\bibitem{hol89}
B.R. Holstein,
{\em Weak Interactions in Nuclei}, World Scientific, Singapore (1989).

\bibitem{ah}
E. Adelberger and W.C. Haxton,
Ann. Rev. Nucl. Part. Sci. {\bf 35}, 501 (1985).

\bibitem{hh}
W. Haeberli and B.R. Holstein,
in {\em Symmetries and Fundamental Interactions in Nuclei,''}
ed. W.C. Haxton and E.M. Henley, World Scientific, Singapore (1995).

\bibitem{wphysica}
S. Weinberg,
Physica {\bf 96A}, 327 (1979).

\bibitem{gl}
J. Gasser and H. Leutwyler,
Ann. Phys. (NY) {\bf 159}, 142 (1984);
Nucl. Phys. {\bf B250}, 465 (1985).

\bibitem{rocco} R. Schiavilla and J. Carlson, private communication.

\bibitem{bhk}
V.R. Brown, E.M. Henley and F.R. Krejs,
Phys. Rev. {\bf C9}, 935 (1974).

\bibitem{oka}
T. Oka,
Prog. Theor. Phys. {\bf 66}, 977 (1981).

\bibitem{mil}
D.E. Driscoll and G.A. Miller,
Phys. Rev. {\bf C39}, 1951 (1989).

\bibitem{bon}
P.D. Evershiem et al.,
Phys. Lett. {\bf B256}, 11 (1991).

\bibitem{psi}
S. Kistryn et al.,
Phys. Rev. Lett. {\bf 58}, 1616 (1987);
R. Balzer et al.,
Phys. Rev. {\bf C30}, 1409 (1984).

\bibitem{lan}
J.M. Potter et al.,
Phys. Rev. Lett. {\bf 33}, 1307 (1974);
D.E. Nagle et al.,
in ``3rd Intl. Symp. on High Energy Physics with
Polarized Beams and Targets'', AIP Conf. Proc. {\bf 51}, 224 (1978).

\bibitem{moresin}
J. Lang et al.,
Phys. Rev. Lett. {\bf 54}, 170 (1985).

\bibitem{lansce}
M. Snow et al.,
Nucl. Inst. and Meth. {\bf 440}, 729 (2000).

\bibitem{nist}
D. Markov, private communication.

\bibitem{athens}
C. Papanicholas, private communication.

\bibitem{duke} H. Weller,
private communication.

\bibitem{snow}
M. Snow, private communication.

\bibitem{v18}
R.B. Wiringa, V.G.J. Stoks, and R. Schiavilla,
Phys. Rev. {\bf C51}, 38 (1995).

\bibitem{riad}
R. Suleman and S. Kowalski, Jefferson Laboratory proposal PR02-006.

\bibitem{bogdan} B. Wojtsekhowski, Jefferson Laboratory Letter of Intent LOI-00-002.

\bibitem{georgi}
H. Georgi,
Phys. Lett. {\bf B240}, 447 (1990).

\bibitem{jenkins}
E. Jenkins and A.V. Manohar,
Phys. Lett. {\bf B255}, 558 (1991);
            {\bf B259}, 353 (1991).

\bibitem{reparam}
M. Luke and A.V. Manohar,
Phys. Lett. {\bf B286} (1992) 348.

\bibitem{hol}
B.R. Holstein,
Phys. Rev. {\bf D60}, 114030 (1999).

\bibitem{kong}
X. Kong and F. Ravndal,
Nucl. Phys. {\bf A656}, 421 (1999).

\bibitem{radcappionless}
M.J. Savage,
Nucl. Phys. {\bf A695}, 365 (2001).

\bibitem{prezeau}
G. Prezeau, M.J. Ramsey-Musolf, and P. Vogel, {\tt hep-ph/0303205}.

\bibitem{ijmpe}
V. Bernard, N. Kaiser, and U.-G. Mei{\ss}ner,
Int. J. Mod. Phys. {\bf E1}, 561 (1992).

\bibitem{bho}
B.R. Holstein,
Phys. Lett. {\bf B244}, 83 (1990).

\bibitem{barrysdeltas}
T.R. Hemmert, B.R. Holstein, and J. Kambor,
J. Phys. {\bf G24}, 1831 (1998).

\bibitem{nadia}
N. Fettes et al., Annals Phys. 283, 273 (2000); Erratum-ibid. 288,
249 (2001).

\bibitem{gm}
A. Manohar and H. Georgi,
Nucl. Phys. {\bf B234}, 189 (1984).

\bibitem{luke92}
M.E. Luke and A.V. Manohar,
Phys. Lett. {\bf B286}, 348 (1992).

\bibitem{sca}
J.F. Donoghue, E. Golowich, and B.R. Holstein,
Phys. Rev. {\bf D30}, 587 (1984).

\bibitem{KSW}
D.B. Kaplan, M.J. Savage, and M.B. Wise,
Phys. Lett. {\bf B424}, 390 (1998);
Nucl. Phys. {\bf B534}, 329 (1998).

\bibitem{FMS}
S. Fleming, T. Mehen, and I.W. Stewart,
Nucl. Phys. {\bf A677}, 313 (2000).

\bibitem{nijmegen}
M.C.M. Rentmeester, R.G.E. Timmermans, J.L. Friar, and J.J. de Swart,
Phys. Rev. Lett. {\bf 82}, 4992 (1999).

\bibitem{largeN}
E. Witten, Nucl. Phys. {\bf B160}, 57 (1979).

\bibitem{egpr}
G. Ecker, J. Gasser, A. Pich, and E. de Rafael,
Nucl. Phys. {\bf B321}, 311 (1989).

\bibitem{Hoh76}
G. H\"ohler et al.,
Nucl. Phys. {\bf B114}, 505 (1976).

\bibitem{Stoks}
V.G.J. Stoks, R.A.M. Klomp, C.P. F. Terheggen, and J. J. de Swart,
Phys. Rev. {\bf C49}, 2950 (1994).

\bibitem{BjD}
J.D. Bjorken and S. Drell,
{\sl Relativistic Quantum Fields}, McGraw-Hill, New York (1965).

\bibitem{bonn}
G.E. Brown and R. Machleidt,
Phys. Rev. {\bf C50}, 1731 (1986).

\bibitem{speth}
G. Janssen, K. Holinde, and J. Speth,
Phys. Rev. {\bf C54}, 2218 (1996).

\bibitem{zhu5}
Shi-Lin Zhu,
Phys. Rev. {\bf C59}, 435 (1999); 3455 (1999).

\bibitem{des}
B. Desplanques,
Nucl. Phys. {\bf A335}, 147 (1980).

\bibitem{gib}
B.F. Gibson, V.R. Brown, R. Schiavilla private communications.

\bibitem{khk}
Here we follow the approach of
I.B. Khriplovich and R.V. Korkin,
Nucl. Phys. {\bf A690}, 610 (2001).

\bibitem{khr}
I.B. Khriplovich,
Phys. At. Nucl. {\bf 64}, 516 (2001).

\bibitem{car}
J. Carlson, R. Schiavilla, V.R. Brown, and B.F. Gibson,
Phys. Rev. {\bf C65} 035502 (2002).

\bibitem{dpl}
B. Desplanques,
Phys. Lett. {\bf B512}, 305 (2001).

\bibitem{rib}
D.O Riska and G.E. Brown, Phys. Lett. {\bf B38}, 193 (1972).

\end{thebibliography}
\end{document}